\documentclass[
  ,draft            
  ,numberedheadings 
  ]
  {aipproc}

\usepackage{epsfig}
\usepackage{epsf}

\layoutstyle{6x9}

\begin{document}



\def\beq{\begin{equation}}
\def\eeq{\end{equation}}
\def\be{\begin{equation}}
\def\ee{\end{equation}}

\def\iomn{i\omega_n}
\def\iom{i\omega}
\def\iom#1{i\omega_{#1}}
\def\c#1#2#3{#1_{#2 #3}}
\def\cdag#1#2#3{#1_{#2 #3}^{+}}
\def\epsk{\epsilon_{{\bf k}}}
\def\Ga{\Gamma_{\alpha}}
\def\Seff{S_{eff}}
\def\dinf{$d\rightarrow\infty\,$}
\def\T{\mbox{Tr}\,}
\def\t{\mbox{tr}\,}
\def\cG0{{\cal G}_0}
\def\cG{{\cal G}}
\def\cU{{\cal U}}
\def\cS{{\cal S}}
\def\divnum{\frac{1}{N_s}}
\def\vac{|\mbox{vac}\rangle}
\def\intR{\int_{-\infty}^{+\infty}}
\def\intb{\int_{0}^{\beta}}
\def\spinup{\uparrow}
\def\spindown{\downarrow}
\def\bra{\langle}
\def\ket{\rangle}

\def\ka{{\bf k}}
\def\vk{{\bf k}}
\def\vq{{\bf q}}
\def\vQ{{\bf Q}}
\def\vr{{\bf r}}
\def\q{{\bf q}}
\def\R{{\bf R}}
\def\vR{{\bf R}}
\def\kp{\bbox{k'}}
\def\a{\alpha}
\def\b{\beta}
\def\d{\delta}
\def\D{\Delta}
\def\e{\varepsilon}
\def\ed{\epsilon_d}
\def\ef{\epsilon_f}
\def\g{\gamma}
\def\G{\Gamma}
\def\l{\lambda}
\def\L{\Lambda}
\def\o{\omega}
\def\ph{\varphi}
\def\s{\sigma}
\def\chib{\overline{\chi}}
\def\et{\widetilde{\epsilon}}
\def\hn{\hat{n}}
\def\hnu{\hat{n}_\uparrow}
\def\hnd{\hat{n}_\downarrow}

\def\hc{\mbox{h.c}}
\def\Im{\mbox{Im}}

\def\est{\varepsilon_F^*}
\def\uc2{$U_{c2}$}
\def\uc1{$U_{c1}$}

\def\v2o3{V$_2$O$_3$\,}
\def\vcr2o3{(V$_{1-x}$Cr$_x$)$_2$O$_3$\,}
\def\vo2{VO$_2$\,}
\def\casr{Ca$_{1-x}$Sr$_x$VO$_3$\,}
\def\srvo3{SrVO$_3$\,}
\def\cavo3{CaVO$_3$\,}
\def\latio3{LaTiO$_3$\,}
\def\lasrtio3{La$_{1-x}$Sr$_x$TiO$_3$\,}
\def\ytio3{YTiO$_3$\,}
\def\nisse{NiS$_{2-x}$Se$_x$\,}
\def\ceal3{CeAl$_3$\,}
\def\naxcoo2{Na$_x$CoO$_2$\,}
\def\kappacl{$\kappa$-(BEDT-TTF)$_{2}$Cu[N(CN)$_{2}$]Cl\,}

\def\cT{{\cal T}}
\def\cd{c^\dagger}
\def\ad{a^\dagger}
\def\tmu{\widetilde{\mu}}
\def\dS{\Delta\Sigma\,}
\def\fG{G({\bf r},{\bf r'};i\omega)\,}
\def\fW{W({\bf r},{\bf r'};i\omega)\,}
\def\eV{\,\mbox{eV}\,}
\def\etal{{\it et al.\,}}

\title[Dynamical Mean-Field Theory]{Strongly Correlated Electron Materials:\\
Dynamical Mean-Field Theory\\ and Electronic Structure}

\author{Antoine Georges}{
address={Centre de Physique Th\'eorique,
Ecole Polytechnique, 91128 Palaiseau Cedex, France}}
                                                                                         
\begin{abstract}
These are introductory lectures to some aspects of the physics of
strongly correlated electron systems.
I first explain the main reasons for strong
correlations in several classes of materials.
The basic principles of dynamical mean-field theory
(DMFT) are then briefly reviewed.
I emphasize the formal analogies with
classical mean-field theory and density functional theory, through
the construction of free-energy functionals of
a local observable.
I review the application of DMFT to the Mott transition,
and compare to recent spectroscopy and transport experiments.
The key role of the quasiparticle coherence scale, and
of transfers of spectral weight between low- and intermediate or high
energies is emphasized.
Above this scale, correlated metals enter an incoherent regime with
unusual transport properties. The recent combinations of DMFT with electronic
structure methods are also discussed, and illustrated by
some applications to transition metal oxides and f-electron materials.
                                                                                         
\end{abstract}
                                                                                         
\maketitle

\tableofcontents


\section{Introduction: why strong correlations ?}
\label{strong}

\subsection{Hesitant electrons: delocalised waves or localised particles ?}

The physical properties of electrons in many solids can be described,
to a good approximation, by assuming an independent particle
picture. This is particularly successful when one deals with broad
energy bands, associated with a large value of the kinetic energy. In such
cases, the (valence) electrons are highly {\it itinerant}: they are delocalised
over the entire solid. The typical time spent near a specific atom in the crystal
lattice is very short. In such a situation, valence electrons are well described
using a {\it wave-like picture}, in which individual wavefunctions are calculated from an
effective one-electron periodic potential.

For some materials however, this physical picture suffers from severe limitations and
may fail altogether. This happens when valence electrons spend a larger time
around a given atom in the crystal lattice, and hence have a tendency towards
{\it localisation}. In such cases, electrons tend to ``see each other'' and the effects
of statistical correlations between the motions of individual electrons become
important. An independent particle description will not be appropriate, particularly
at short or intermediate time scales (high to intermediate energies).
A {\it particle-like picture} may in fact be more appropriate than a wave-like one
over those time scales, involving wavefunctions localised around specific
atomic sites.
Materials in which electronic correlations are significant are generally
associated with moderate values of the bandwidth (narrow bands). The small kinetic
energy implies a longer time spent on a given atomic site. It also implies that
the ratio of the Coulomb repulsion energy between electrons and the available
kinetic energy becomes larger. As a result delocalising the valence
electrons over the whole solid may become less favorable energetically.
In some extreme cases, the balance may even become unfavorable, so that the
corresponding electrons will remain localised. In a naive picture, these electrons
sit on the atoms to which they belong and refuse to move.
If this happens to all the electrons close to the Fermi level, the solid
becomes an insulator. This insulator is difficult to understand in the wave-like
language: it is not caused by the absence of available one-electron states caused by
destructive interference in $\vk$-space, resulting in a band-gap, as in conventional
band insulators. It is however very easy to understand in real space (thinking of
the solid as made of individual atoms pulled closer to one another in
order to form the crystal lattice). This mechanism was understood
long ago~\cite{mott_1949,mott_mit_book} by Mott (and
Peierls), and such insulators are therefore called
{\it Mott insulators} (Sec.~\ref{sec:mott}.
In other cases, such as f-electron materials, this electron localisation affects only
part of the electrons in the solid (e.g the ones corresponding to the f-shell),
so that the solid remains a (strongly correlated) metal.

The most interesting situation, which is also the one which is hardest to handle
theoretically, is when the localised character on short time-scales and the itinerant
character on long time-scales coexist. In such cases, the electrons ``hesitate''
between being itinerant and being localised. This gives rise to a number of
physical phenomena, and also results in several possible instabilities of the
electron gas which often compete, with very small energy differences between them.
In order to handle such situations theoretically, it is necessary to think
both in $\vk$-space
and in real space, to handle both the particle-like and the wave-like character
of the electrons and, importantly, to be able to describe physical
phenomena on {\it intermediate energy scales}. For example, one needs to
explain how long-lived (wave-like) quasiparticles may eventually emerge at
low energy/temperature
in a strongly correlated metal while at higher energy/temperature, only incoherent
(particle-like) excitations are visible. It is the opinion of the author that, in many
cases, understanding these intermediate energy scales and the associated
coherent/incoherent
crossover is the key to the intriguing physics often observed in correlated metals.
In these lectures, we discuss a technique,
the dynamical mean-field theory (DMFT), which is able to (at least partially)
handle this problem.
This technique has led to significant progress in our understanding of
strong correlation physics, and allows for a quantitative description of
many correlated materials\cite{georges_review_dmft,pruschke_jarrell_review}.
Extensions and generalisations of this technique are
currently being developed in order to handle the most difficult/mysterious situations
which cannot be tamed by the simplest version of DMFT.

\subsection{Bare energy scales}

\paragraph{Localised orbitals and narrow bands}

In practice, strongly correlated materials are generally associated with
partially filled d- or f- shells. Hence, the suspects are materials
involving:
\begin{itemize}
\item Transition metal elements (particularly from the 3d-shell from Ti to Cu,
and to a lesser extent 4d from Zr to Ag).
\item Rare earth (4f from Ce to Yb) or actinide elements (5f from Th to Lw)
\end{itemize}
To this list, one should also add molecular (organic) conductors with
large unit cell volumes in which
the overlap between molecular orbitals is weak.

What is so special about d- and f- orbitals (particularly 3d and 4f) ?
Consider the atomic wavefunctions of the 3d shell in a 3d transition
metal atom (e.g Cu). There are no atomic
wavefunctions with the same value $l=2$ of the angular momentum quantum number,
but lower principal quantum number $n$ than $n=3$ (since one must have $l\leq\,n-1$).
Hence, the 3d wavefunctions are orthogonal to all the $n=1$ and $n=2$ orbitals just because
of their angular dependence, and the radial part needs not have nodes or
extend far away from the nucleus.
As a result, the 3d-orbital wave functions are confined
more closely to the nucleus than for s or p states of comparable
energy. The same argument applies to the 4f shell in rare earths. It also
implies that the 4d wavefunctions in the 4d transition metals or the 5f ones in
actinides will be more extended (and hence that these materials are expected to display,
on the whole, weaker correlation effects than 3d transition metals,
or the rare earth, respectively).

Oversimplified as it may be, these qualitative arguments at least tell us that
a key energy scale in the problem is the degree of overlap between
orbitals on neighbouring atomic sites. This will control the bandwidth and
the order of magnitude of the kinetic energy. A simple estimate of this
overlap is the matrix element:
\beq
t\,_{\vR\vR'}^{LL'}\,\sim\, \int d\vr\, \chi^*_L(\vr-\vR)\,
\frac{\hbar^2\nabla^2}{2m}\,\chi_{L'}(\vr-\vR')
\label{eq:def_hop}
\eeq
In the solid, the wavefunction $\chi_L(\vr-\vR)$ should be thought of as a Wannier-like
wave function centered on atomic site $\vR$. In narrow band systems, typical values
of the bandwith are a few electron-volts.

\paragraph{Coulomb repulsion and the Hubbard $U$}
Another key parameter is the typical strength of the Coulomb repulsion
between electrons sitting in the most localized orbitals.
The biggest repulsion is associated with electrons with opposite spins
occupying the same orbital: this is the Hubbard repulsion which we can
estimate as:
\beq
U\,\sim\,\int d\vr d\vr' |\chi_L(\vr-\vR)|^2\,U_{s}(\vr-\vr')\,|\chi_L(\vr'-\vR)|^2
\label{eq:hubbard_U}
\eeq
In this expression, $U_s$ is the interaction between electrons {\it including screening
effects} by other electrons in the solid.
Screening is a very large effect: if we were to estimate (\ref{eq:hubbard_U}) with the unscreened
Coulomb interaction $U(\vr-\vr')=e^2/|\vr-\vr'|$, we would typically obtain values in the range
of tens of electron-volts. Instead, the screened value of $U$ in correlated materials is typically
a few electron-volts. This can be comparable to the kinetic energy for narrow bandwiths, hence the
competition between localised and itinerant aspects. Naturally, other matrix elements (e.g
between different orbitals, or between different sites) are important for a realistic
description of materials (see the last section of these lectures).

In fact, a precise description of screening in solids is a rather difficult problem. An
important point is, again, that this issue crucially {\it depends on energy scale}. At
very low energy, one should observe the fully screened value, of order a few \eV's, while
at high energies (say, above the plasmon energy in a metal) one should observe the
unscreened value, tens of \eV's.
Indeed, the screened effective interaction $W(\vr,\vr';\omega)$ as estimated e.g from the
RPA approximation, is a strong function of frequency
(see e.g Ref.~\cite{springer_womega_1998_prb,ferdi_uomega_2004_condmat} for
an ab-initio GW treatment in the case of Nickel).
As a result, using an energy-independent parametrization
of the on-site matrix elements of the Coulomb interaction such as (\ref{eq:hubbard_U})
can only be appropriate for a description restricted to
low- enough energies~\cite{ferdi_uomega_2004_condmat}.
The Hubbard interaction can only be given a precise meaning in a solid, over
a large enery range, if it is made energy-dependent. I shall come back to this issue in
the very last section of these lectures (Sec.~\ref{sec:abinitio_dmft}).

\paragraph{The simplest model hamiltonian}
From this discussion, it should be clear that the simplest model in which
strong correlation physics can be discussed is that of a
lattice of single-level ``atoms'', or equivalently of a
single
tight-binding band (associated with Wannier orbitals centered on the sites of
the crystal lattice), retaining only the on-site interaction term between electrons
with opposite spins:
\begin{equation}
H = - \sum_{\vR\vR',\sigma} t_{\vR\vR'}\,c^{+}_{\vR\sigma} c_{\vR'\sigma} \,+\,
\e_0\sum_{\vR\sigma} n_{\vR\sigma}\,+\, U\sum_{\vR} n_{\vR\uparrow} n_{\vR\downarrow}
\label{eq:hubbard}
\end{equation}
The kinetic energy term is diagonalized in a single-particle basis of Bloch's wavefunctions:
\beq
H_0 = \sum_{\vk\sigma} \e_\vk c^{+}_{\vk\sigma} c_{\vk\sigma}\,\,\,;\,\,\,
\e_\vk \equiv \sum_{\vR'}\,t_{\vR\vR'}\,e^{i\vk\cdot(\vR-\vR')}
\eeq
with e.g for nearest-neighbour hopping on the simple cubic lattice in d-dimensions:
\beq
\e_\vk\,=\,-2t \sum_{\mu=1}^{d} \cos (k_\mu a)
\eeq
In the absence of hopping, we have, at each site, a single atomic level and
hence four possible quantum states:
$|\,0\ket,|\,\uparrow\ket,|\,\downarrow\ket$ and $|\,\uparrow\downarrow\ket$ with
energies $0\,,\,\e_0$ and  $U+2\e_0$, respectively.

Eq.~(\ref{eq:hubbard}) is the famous Hubbard model~\cite{hubbard_1,hubbard_2,hubbard_3}.
It plays in this field the same role than that played by
the Ising model in statistical mechanics: a laboratory for testing physical ideas, and
theoretical methods alike.
Simplified as it may be, and despite the fact that it already has a 40-year old history,
we are far from having explored all the physical phenomena contained in this model, let
alone of being able to reliably calculate with it in all parameter ranges !

\subsection{Examples of strongly correlated materials}

In this section, I give a few examples of strongly correlated
materials. The discussion emphasizes a few key points but is
otherwise very brief. There are many useful references related to this
section, e.g~\cite{varma_giam,imada_mit_review,mott_mit_book,
tsuda_oxides_book,harrison_es_book}.

\subsubsection{Transition metals}

\begin{center}
\begin{figure}
\includegraphics[width=12cm]{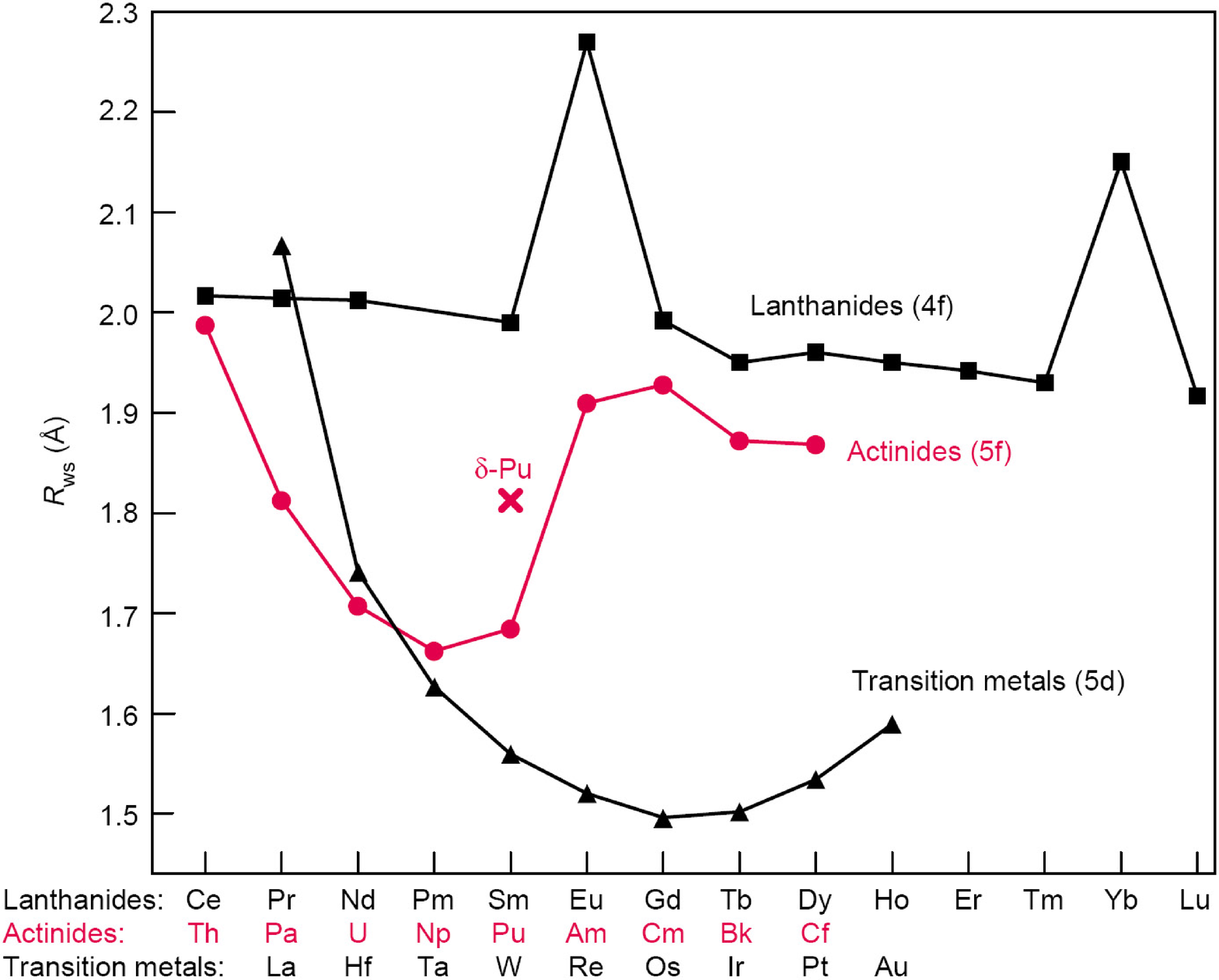}
\caption{Experimental Wigner-Seitz Radius of Actinides, Lanthanides, and
5d Transition Metals. The equilibrium volume of the primitive unit cell is given
by $V=4\pi\,R_{WS}^3/3$.
Elements that lie on top of each other have the same number of valence electrons.
The volume of the transition metals has a roughly parabolic shape, indicating
delocalised 5d electrons. The volumes of the lanthanides remain roughly
constant, indicating localised 4f electrons.
The volumes of the light actinides decrease with
increasing atomic number, whereas the volumes of the late actinides behaves
similarly to that of the lanthanides.
From Ref.~\cite{wills_losalamos_2000}}
\label{fig:unitcell_volume}
\end{figure}
\end{center}

In 3d transition metals, the 4s orbitals have lower energy than
the 3d and are therefore filled first. The 4s orbitals extend
much further
from the nucleus, and thus overlap strongly. This holds the atoms
sufficiently far apart so that the d- orbitals have a {\it small
direct overlap}.
Nevertheless, d-orbitals extend much further from
the nucleus than the ``core'' electrons (corresponding to shells which are
deep in energy below the Fermi level).
As a result, throughout the 3d series of transition metals (and even more so
in the 4d series), d-electrons
do have an itinerant character, giving rise to quasiparticle bands.
That this is the case is already clear from
a very basic property of the material, namely how the
equilibrium unit- cell volume depends on the element as one moves along
the 3d series (Fig.~\ref{fig:unitcell_volume}).
The unit-cell volume has a very
characteristic, roughly parabolic, dependence. A simple model of a narrow band
being gradually filled, introduced long ago by
Friedel~\cite{friedel_metals} accounts for this parabolic dependence
(see also \cite{mcmahan_collapse_review,wills_losalamos_2000}).
Because the states at the bottom of the
band are bonding-like while the states at the top of the band are anti-bonding like,
the binding energy is maximal (and hence the equilibrium volume is
minimal) for a half-filled shell. Instead, if the d-electrons were
localised we would expect little contribution of the d-shell
to the cohesive energy of the solid, and the equilibrium volume should
not vary much along the series.
Screening is relatively efficient in transition metals because
the 3d band is not too far in energy from the 4s band.
The latter plays the
dominant role in screening the Coulomb interaction
(crudely speaking, one has to consider
the following charge transfer
process between two neighbouring atoms: $3d^n4s + 3d^n4s
\rightarrow 3d^{n-1}4s^2 + 3d^{n+1}$,
see e.g \cite{anisimov_calcU_1991_prb} for further discussion).
For all these reasons (the band not being extremely narrow,
screening being efficient), electron correlations do have
important physical effects for 3d transition metals, but not
extreme ones like localisation. Magnetism of these metals below
the Curie temperature, but also the existence of fluctuating
local moments in the paramagnetic phase are exemples of such
correlation effects. Band structure calculations based
on DFT-LDA methods overestimate the width of the occupied
d-band (by about 30\% in the case of nickel). Some features
observed in spectroscopy experiments (such as the (in)famous 6\eV
satellite in nickel) are also signatures of correlation effects, and
are not reproduced by standard electronic structure calculations.

\subsubsection{Transition metal oxides}

\begin{figure}
\includegraphics[height=10 cm]{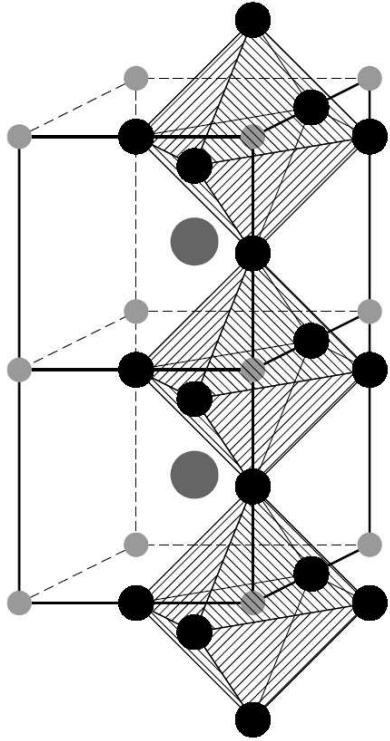}
\caption{The cubic perovskite structure, e.g of the compound \srvo3.
Transition-metal atoms (V) - small grey spheres- are at the
center of oxygen octahedra (dark spheres). Sr atoms are the larger
spheres in-between planes. From Ref.~\cite{maiti_phd}.}
\label{fig:perovskite}
\end{figure}

In transition metal compounds (e.g oxides or chalcogenides),
the direct overlap between d-orbitals is generally so small
that d-electrons can only move through hybridisation with the
ligand atoms (e.g oxygen 2p-bands).
For example, in the cubic perovskite structure shown on
Fig.~\ref{fig:perovskite}, each transition-metal atom is ``encaged'' at the
center of an octahedron made of six oxygen atoms.
Hybridisation leads to the formation of bonding and
antibonding orbitals.
An important energy scale is the
{\it charge-transfer energy} $\Delta=\e_d-\e_p$, i.e the
energy difference between the
average position of the oxygen and transition metal bands.
When $\Delta$ is large as compared to the overlap integral
$t_{pd}$, the bonding orbitals have mainly oxygen character
and the antibonding ones mainly transition-metal character.
In this case, the effective metal-
to- metal hopping can be estimated as
$t_{eff}\sim\,t_{pd}^2/\Delta$, and is therefore
quite small.

The efficiency of screening in transition-metal oxides
depends crucially on the relative position of the 4s and 3d band.
For 3d transition metal monoxides MO with M to the right of
Vanadium, the 4s level is much higher in energy than 3d, thus
leading to poor screening and large values of $U$. This, in
addition to the small bandwidth and relatively large $\Delta$,
leads to dramatic correlation
effects, turning the system into a Mott insulator (or rather, a charge-
transfer insulator, see below), in spite of the incomplete
filling of the d-band.
The Mott phenomenon plays a key role in the physics of transition- metal
oxides, as discussed in detail later in these lectures
(see Sec.~\ref{sec:mott} and Fig.~\ref{fig:fujimori_map}).

\paragraph{Crystal field splitting}
The 5-fold (10-fold with spin) degeneracy of the d- orbitals in
the atom is lifted in the solid, due to the influence of the
electric field created by neighbouring atoms, i.e the ligand
oxygen atoms in transition metal oxides.
For a transition metal ion in an octahedral environment (as in
Fig.~\ref{fig:perovskite}), this results in a three-fold group of
states ($t_{2g}$) which is lower in energy and a doublet ($e_g$)
higher in energy. Indeed, the
$d_{xy}, d_{yz}, d_{zx}$ orbitals forming the $t_{2g}$ multiplet
do not point towards the ligand atoms, in contrast to the states
in the $e_g$ doublet ($d_{x^2-y^2}$, $d_{3z^2-r^2}$). The latter
therefore lead to a higher cost in Coulomb repulsion energy.
For a crystal with perfect cubic symmetry, the $t_{2g}$ and $e_g$
multiplets remain exactly degenerate, while a lower symmetry of the
crystal lattice
lifts the degeneracy further. For a tetrahedral environment of the
transition-metal ion, the opposite situation is found,
with $t_{2g}$ higher in energy than $e_g$.
In transition metals, the energy scale associated
with crystal- field splitting is typically
much smaller than the bandwith.
This is not so in transition-metal oxides, for which
these considerations become essential. In some materials, such as
e.g \srvo3 and the other $d^1$ oxides studied in Sec.~\ref{sec:esc_d1oxides} of these
lectures, the energy bands emerging from the $t_{2g}$ and $e_g$ orbitals
form two groups of bands well separated in energy.

\paragraph{Mott- and charge-transfer insulators}
There are two important considerations, which are responsible for
the different physical properties of the ``early'' (i.e
involving Ti, V, Cr, ...)
and ``late'' (Ni, Cu) transition- metal oxides:

- whether the Fermi level falls within the $t_{2g}$ or $e_g$  multiplets,

- what is the relative position of oxygen ($\e_p$) and transition-metal
($\e_p$) levels ?

For those compounds which correspond to an octahedral environment:

\begin{itemize}
\item In early transition-metal oxides,
$t_{2g}$ is partially filled, $e_g$ is
empty. Hence, the hybridisation with ligand is very weak (because
$t_{2g}$ orbitals point away from the 2p oxygen orbitals). Also,
the d-orbitals are much higher in energy than the 2p orbitals of
oxygen. As a result, the charge-transfer energy $\Delta =
\e_d-\e_p$ is large, and the bandwidth is small. The local d-d
Coulomb repulsion $U_{dd}$ is a smaller scale than $\Delta$ but
it can be larger than the bandwidth ($\sim t_{pd}^2/\Delta$):
this leads to Mott insulators.

\item For late transition-metal oxides, $t_{2g}$
is completely filled, and the Fermi level lies within $e_g$. As a
result, hybridisation with the ligand is stronger. Also, because of
the greater electric charge on the nuclei, the attractive
potential is stronger and as a result, the Fermi level moves
closer to the energy of the 2p ligand orbitals. Hence, $\Delta$ is
a smaller scale than $U_{dd}$ and controls the energy cost of
adding an extra electron. When this cost becomes larger than the
bandwidth, insulating materials are obtained, often called
``charge transfer insulators''~\cite{fujimori_minami_1984_prb,zsa_1985_prl}.
The mechanism is not
qualitatively different than the Mott mechanism, but
the insulating gap is set by the scale $\Delta$ rather than
$U_{dd}$ and separates the oxygen band from a d-band
rather than a lower and upper Hubbard bands having both d-character.
\end{itemize}
%
%

\paragraph{The p-d model} The single-band Hubbard model is easily extended
in order to take into account both transition-metal and oxygen orbitals in a
simple modelisation of transition-metal oxides.
The key terms to be retained are\footnote{For simplicity, the hamiltonian is written
in the case where only one d-band is relevant, as e.g for cuprates.}:
\begin{equation}
H_{pd} = - \sum_{\vR\vR',\sigma} t_{pd}\,(d^{+}_{\vR\sigma} p_{\vR'\sigma}+h.c) \,+\,
\e_d\sum_{\vR\sigma} n^d_{\vR\sigma}+\e_p\sum_{\vR'\sigma} n^p_{\vR'\sigma}
\,+\,U_{dd}\sum_{\vR} n^d_{\vR\uparrow} n^d_{\vR\downarrow}
\label{eq:pd_model}
\end{equation}
to which one may want to add other terms, such as: Coulomb repulsions $U_{pp}$ and
$U_{pd}$ or direct oxygen-oxygen hoppings $t_{pp}$.

\subsubsection{f-electrons: rare earths, actinides and their compounds}
\label{sec:felec}

A distinctive character of the physics of rare-earth metals (lanthanides)
is that the 4f electrons
tend to be localised rather than itinerant (at ambiant pressure).
As a result, the f-electrons contribute contribute little to the cohesive
energy of the solid, and the unit-cell volume depends very weakly on the
filling of the 4f shell (Fig.~\ref{fig:unitcell_volume}). Other electronic
orbitals do form bands which cross the Fermi level however, hence the metallic
character of the lanthanides.
When pressure is applied, the f-electrons become increasingly itinerant. In fact,
at some critical pressure, some rare-earth metals (mots notably Ce and Pr)
undergo a sharp first-order
transition which is accompanied by a discontinuous drop of the equilibrium unit-cell
volume. Cerium is a particularly remarkable case, with a volume drop of
as much as $15\%$ and the same crystal symmetry (fcc) in the low-volume ($\alpha$) and
high-volume ($\gamma$) phase. In other cases, the transition corresponds to a change
in crystal symmetry, from a lower symmetry phase at low pressure to a higher symmetry
phase at high pressure. For a recent review on the volume-collapse transition
of rare earth metals, see Ref.~\cite{mcmahan_collapse_review}.

The equilibrium volume of actinide (5f) metals display behaviour which is
intermediate between transition metals and rare earths. From the beginning of
the series (Th) until Plutonium (Pu), the volume has an approximately parabolic
dependence on the filling of the f-shell, indicating delocalised 5f electrons.
From Americium onwards, the volume has a much weaker dependence on the number of f-electrons,
suggesting localised behaviour. Interestingly, plutonium is right on the verge
of this delocalisation to localisation transition. Not surprisingly then, plutonium is,
among all actinide metals, the one which has the most complex phase diagram and
which is also the most difficult to describe using conventional electronic structure
methods (see \cite{wills_losalamos_2000,kotliar_savrasov_mott_felectrons_2003} for recent
reviews). This will be discussed further in the last section of these lectures.
This very brief discussion of rare-earths and actinide compounds is meant to illustrate
the need for methods able to deal simultaneously with the itinerant and localised
character of electronic degrees of freedom.
\begin{center}
\begin{figure}
\includegraphics[width=10 cm]{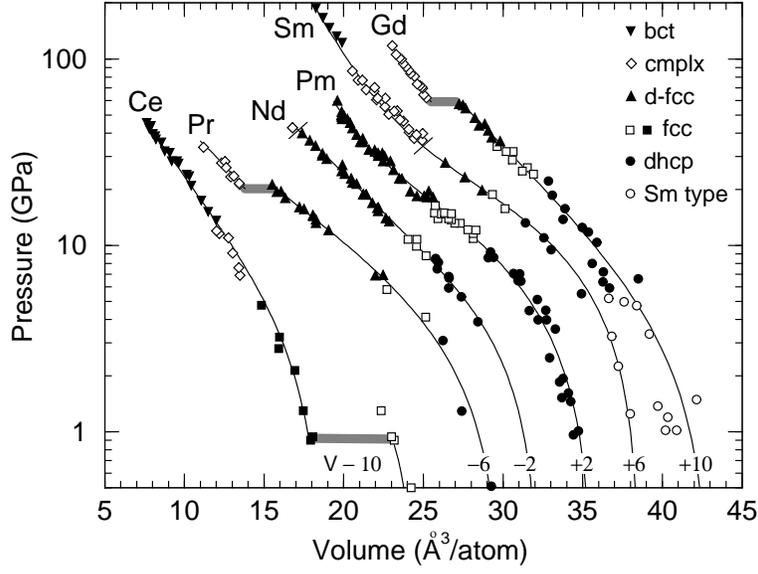}
\caption{Pressure volume data for the rare earths. Structures are
identified, with ``cmplx'' signifiying a number of complex,
low-symmetry structures.  The volume collapse transitions are marked by
the wide hatched lines for Ce, Pr, and Gd, while lines perpendicular to
the curves denote the d-fcc to hP3 symmetry change in Nd and Sm.  The
curves are guides to the eye.  Note that the data and curves have been
shifted in volume by the numbers (in \AA$^3$/atom) shown at the bottom
of the figure. Figure and caption reproduced from Ref.~\cite{mcmahan_collapse_review}.
}
\label{fig:volume_collapse}
\end{figure}
\end{center}

The physics of strong electronic correlations becomes even more apparent for
f-electron materials which are compounds involving rare-earth (or actinide) ions
and other atoms, such as e.g \ceal3. A common aspect of such compounds is the
formation of quasiparticle bands with extremely large effective masses (and hence
large values of the low-temperature specific heat coefficient $\gamma=C/T$), up to
a thousand time the bare electron mass ! Hence the term ``heavy-fermion'' given to
these compounds: for reviews, see e.g \cite{coleman_lectures_vietri_2002,hewson_book}.
The origin of these large effective
masses is the weak hybridization between the very localised f-orbitals and the rather
broad conduction band associated with the metallic ion. At high temperature/energy, the
f-electron have localised behaviour (yielding e.g local magnetic moments and a Curie law for
the magnetic susceptibility). At low temperature/energy, the conduction electrons screen the
local moments, leading to the formation of quasiparticle bands with mixed f- and conduction
electron character (hence a large Fermi surface encompassing both f- and conduction electrons).
The low-temperature susceptibility has a Pauli form and the low-energy physics is, apart
from some specific compounds, well described by Fermi liquid theory. This screening process,
the Kondo effect, is associated with a very low energy coherence scale,
the (lattice~\cite{burdin_exhaustion_prl}-) Kondo temperature,
considerably renormalised
as compared to the bare electronic energy scales.

\paragraph{The periodic Anderson model}

The simplest model hamiltonian appropriate for f-electron materials is the Anderson
lattice or periodic Anderson model. It retains the f-orbitals associated with the rare-earth
or actinide atoms at each lattice site, as well as the relevant conduction electron degrees
of freedom which hybridise with those orbitals. In the simplest form, the hamiltonian reads:
\begin{equation}
H_{PAM}\,=\, \sum_{\vk\sigma} \e_\vk\,c^\dagger_{\vk\sigma}c_{\vk\sigma} \,+\,
\sum_{\vk\sigma\,m} (V_{\vk}\,c^\dagger_{\vk\sigma}f_{m\vk\sigma} + h.c\,)
+\,\e_f\sum_{\vR\sigma\,m}\,n^f_{\vR\sigma\,m}
\,+\,U\,\sum_{\vR} \left(\sum_{\sigma\,m}n^f_{\vR\sigma\,m}\right)^2
\label{eq:pam}
\end{equation}
Depending on the material considered, other terms may be necessary for increased realism,
e.g an orbital dependent f-level $\e_{fm}$, hybridisation $V_{\vk\,m}$
or interaction matrix $U_{mm'}^{\sigma\sigma'}$ or a
direct f-f hopping $t_{ff}$.

\section{Dynamical Mean-Field Theory at a glance}
\label{dmft_glance}

Dealing with strong electronic correlations is a notoriously difficult
theoretical problem. From the physics point of view, the difficulties come mainly
from the wide range of energy scales involved (from the bare electronic energies,
on the scale of electron-Volts, to the low-energy physics on the scale of Kelvins)
and from the many competing orderings and instabilities associated with
small differences in energy.

It is the opinion of the author that, on top of the essential
guidance from physical intuition and phenomenology, the
development of quantitative techniques is essential in order to
solve the key open questions in the field (and also in order to
provide a deeper understanding of some ``classic'' problems, only
partially understood to this day).

In this section, we explain the basic principles of Dynamical Mean-Field Theory (DMFT).
This approach has been developed over the
last fifteen years and has led to some significant advances in our understanding
of strong correlations. In this section, we explain the basic principles of this approach
in a concise manner. The Hubbard model is taken as an example.
For a much more detailed presentation, the reader is referred to the available
review articles~\cite{georges_review_dmft,pruschke_jarrell_review}.

\subsection{The mean-field concept, from classical to quantum}

\begin{center}
\begin{figure}
\includegraphics[width=10cm]{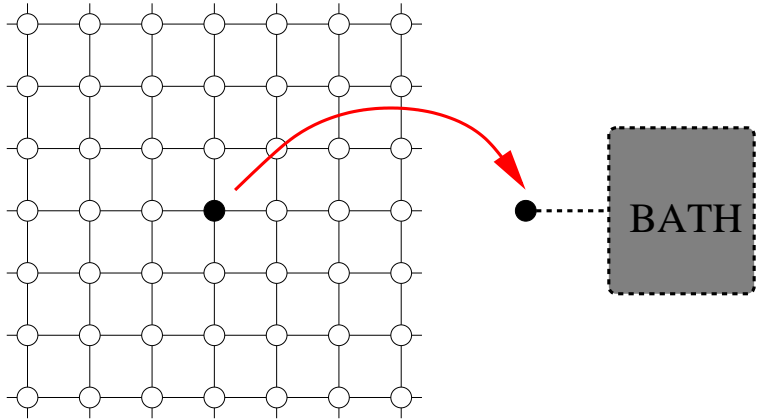}
\caption{Mean-field theory replaces a lattice model by a single site coupled to a
self-consistent bath.}
\label{fig:mft}
\end{figure}
\end{center}

Mean-field theory approximates a lattice problem with many
degrees of freedom by a {\it single-site effective problem} with less degrees
of freedom. The underlying physical idea is that the dynamics at a given site
can be thought of as the interaction of the local degrees of freedom at this site
with an external bath created by all other degrees of freedom on other sites
(Fig.~\ref{fig:mft}).
\paragraph{Classical mean-field theory}
The simplest illustration of this idea is for the Ising model:
\begin{equation}
H = -\sum_{(ij)} J_{ij}S_{i}S_{j} - h\, \sum_{i} S_{i}
\label{ising}
\end{equation}
Let us focus on the thermal average of the magnetization on each lattice site:
$m_i=\bra S_i\ket$. We consider an equivalent problem of {\it independent spins}:
\begin{equation}
H_{eff} = - \sum_i h^{eff}_i\,S_{i}
\label{effising}
\end{equation}
in which the (Weiss) effective field is chosen in such a way that the
value of $m_i$ is accurately reproduced. This requires:
\beq
\beta h^{eff}_i = \tanh^{-1} m_i
\eeq
Let us consider, for definiteness, a ferromagnet with nearest-neighbour couplings
$J_{ij}=J\,>0$.
The mean-field theory approximation (first put forward by Pierre Weiss, under
the name of ``molecular field theory'') is that $h_i^{eff}$ can be approximated by the
thermal average of the local field seen by the spin at site $i$, namely:
\begin{equation}
h_i^{eff} \simeq h + \sum_{j} J_{ij}\, m_j = h + z J m
\label{weiss}
\end{equation}
where $z$ is the connectivity of the lattice, and translation
invariance has been used ($J_{ij}=J$ for n.n sites, $m_i=m$).
This leads to a self-consistent equation for the magnetization:
\begin{equation}
m\,=\,\mbox{tanh}\,(\beta h + z \beta J m)
\label{mfising}
\end{equation}
We emphasize that replacing the problem of interacting spins
by a problem of non-interacting ones in a effective bath is not an approximation,
as long as we use this equivalent model for the only purpose of
calculating the local magnetizations. The approximation is made when relating
the Weiss field to the degrees of freedom on neighbouring sites, i.e in the
{\it self-consistency condition} (\ref{weiss}). We shall elaborate further on
this point of view in the next section, where exact energy functionals will
be discussed.
The mean-field approximation becomes {\it exact} in the limit
where the connectivity $z$ of the lattice becomes large.
It is quite intuitive indeed that the neighbors of a given site can be
treated globally as an external bath when their number becomes large,
and that the spatial fluctuations of the local field become negligible.

\paragraph{Generalisation to the quantum case: dynamical mean-field theory}
This construction can be extended to quantum many-body systems.
Key steps leading to this quantum generalisation where: the introduction of
the limit of large lattice coordination for interacting fermion models
by Metzner and Vollhardt~\cite{metzner_vollhardt} and the mapping onto a
self-consistent quantum impurity
by Georges and Kotliar~\cite{georges_kotliar_dmft}, which established the
DMFT framework\footnote{See also the later work in Ref.~\cite{jarrell_1992_qmc_prl}, and
Ref.~\cite{georges_review_dmft} for an extensive list of references.}.

I explain here the DMFT construction on the simplest example of the
Hubbard model\footnote{The energy $\e_0$ of the
single-electron atomic level has been introduced in this section for the sake of
pedagogy. Naturally, in the single band case, everything depends only on the energy
$\e_0-\mu$ with respect to the global chemical potential so that one can
set $\e_0=0$}:
\begin{equation}
H = - \sum_{ij,\sigma} t_{ij}\, c^{\dagger}_{i\sigma} c_{j\sigma} +
U\sum_{i} n_{i\uparrow} n_{i\downarrow} +\e_0\,\sum_{i\sigma} n_{i\sigma}
\label{hubbard}
\end{equation}
As explained above, it describes a collection of single-orbital ``atoms'' placed at the
nodes $\vR_i$ of a periodic lattice. The orbitals overlap from site to site, so that
the fermions can hop with an amplitude $t_{ij}$.
In the absence of hopping, each ``atom'' has 4 eigenstates:
$|0\ket,|\uparrow\ket,|\downarrow\ket$ and $\uparrow\downarrow\ket$ with
energies $0\,,\,\e_0$ and  $U+2\e_0$, respectively.

The key quantity on which DMFT focuses is the {\it local} Green's function at
a given lattice site:
\beq
\label{eq:def_glocal}
G^\sigma_{ii}(\tau-\tau')\,\equiv\,
- \bra T c_{i\sigma}(\tau)\cd_{i\sigma}(\tau')\ket
\eeq
In classical mean-field theory, the local magnetization $m_i$ is represented as
that of a single spin on site $i$ coupled to an effective Weiss field. In a
completely analogous manner, we shall introduce a representation of the
local Green's function as that of a {\it single atom coupled to an effective bath}.
This can be described by the hamiltonian of an Anderson impurity model
\footnote{Strictly speaking, we have a collection of independent impurity models,
one at each lattice site. In this section, for simplicity, we assume a phase with
translation invariance and focus on a particular site of the lattice (we therefore
drop the site index for the impurity orbital $\cd_\sigma$). We also assume a paramagnetic
phase. The formalism easily generalizes to phases with long-range order
(i.e translational and/or spin-symmetry breaking)~\cite{georges_review_dmft}}:
\begin{equation}
H_{AIM} = H_{atom} + H_{bath} + H_{coupling}
\end{equation}
in which:
\begin{eqnarray}
&H_{atom} = U\, n^c_{\uparrow} n^c_{\downarrow}
+(\e_0-\mu)\,(n^c_{\spinup}+n^c_{\spindown})\nonumber \\
&H_{bath} = \sum_{l\sigma} \et_l\, \ad_{l\sigma} a_{l\sigma}\nonumber \\
&H_{coupling} =\sum_{l\sigma} V_{l}\,(\ad_{l\sigma}c_{\sigma}+
\cd_{\sigma}a_{l\sigma})
\label{am}
\end{eqnarray}
In these expressions, a set of non-interacting fermions
(described by the $a^\dagger_l$'s) have been introduced, which are the degrees of freedom
of the effective bath acting on site $\R_i$. The $\et_l$ and $V_l$'s
are parameters which should be chosen in such
a way that the c-orbital (i.e impurity) Green's function of (\ref{am}) coincides with the
local Green's function of the lattice Hubbard model under consideration. In fact, these
parameters enter only through the hybridisation function:
\beq
\Delta(\iomn) = \sum_l \frac{|V_l|^2}{\iomn-\et_l}
\eeq
This is easily seen when the effective on-site problem is recast in a form
which does not explicitly
involves the effective bath degrees of freedom. However, this requires the use of an
effective action functional integral formalism rather than a simple hamiltonian
formalism. Integrating out the bath degrees of freedom one obtains the effective
action for the impurity orbital only under the form:
\begin{equation}
S_{eff}=
- \int^{\beta}_{0} d\tau \int^{\beta}_{0} d\tau'\, \sum_{\sigma}
c^{+}_{\sigma}(\tau) {\cal G}_0^{-1}(\tau-\tau') c_{\sigma}(\tau')
+U\,\int^{\beta}_{0}d\tau\, n_{\uparrow}(\tau)n_{\downarrow}(\tau)
\label{Seff}
\end{equation}
in which:
\beq
{\cal G}_0^{-1}(\iomn) = \iomn+\mu-\e_0 -\Delta(\iomn)
\label{eq:def_g0}
\eeq
This local action represents the effective dynamics of the local site
under consideration: a fermion is created on this site
at time $\tau$ (coming from the
"external bath", i.e from the other sites of the lattice)
and is destroyed at time $\tau'$ (going back to
the bath). Whenever two fermions (with opposite spins) are present
at the same time, an energy cost $U$ is included.
Hence this effective action describes the fluctuations between
the 4 atomic states $|0\ket,|\uparrow\ket,|\downarrow\ket,\uparrow\downarrow\ket$
induced by the coupling to the bath.
We can interpret ${\cal G}_0(\tau-\tau')$ as the quantum generalisation
of the Weiss effective field in the classical case.
The main difference with the classical case is that this
``dynamical mean-field''
is a {\it function of energy} (or time) instead of a single number.
This is required in order to take full account of
local quantum fluctuations, which is the main purpose of DMFT.
${\cal G}_0$ also plays the role of a bare Green's function for the effective
action $\Seff$, but it should {\it not be confused} with the non-interacting ($U=0$)
local Green's function of the original lattice model.

\begin{center}
\begin{figure}
\includegraphics[width=10 cm]{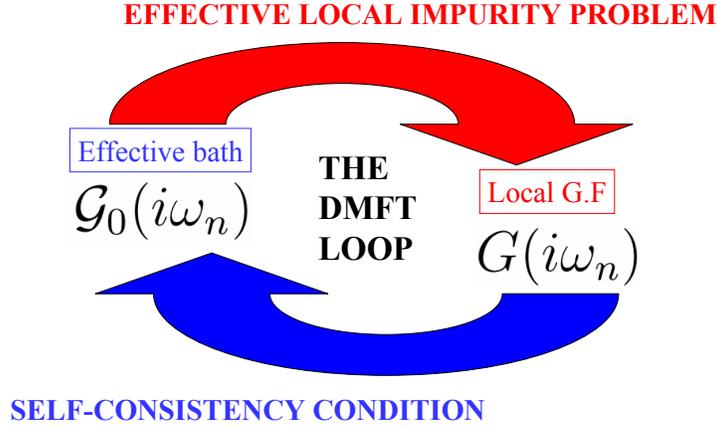}
\caption{The DMFT iterative loop. The following procedure
is generally used in practice: starting from an initial guess for $\cG_0$, the
impurity Green's function $G_{imp}$ is calculated by using an appropriate solver for the
impurity model (top arrow). The impurity self-energy is also calculated from
$\Sigma_{imp}=\cG_0^{-1}(\iomn)-G_{imp}^{-1}(\iomn)$.
This is used in order to obtain the on-site
Green's function of the lattice model by performing a $\vk$-summation (or
integration over the free d.o.s):
$G_{loc}=\sum_\vk [\iomn+\mu-\e_\vk-\Sigma_{imp}(\iomn)]^{-1}$.
An updated Weiss function is then obtained as
${\cal G}_{0,new}^{-1}=G_{loc}^{-1}+\Sigma_{imp}$,
which is injected again into the impurity solver (bottom arrow).
The procedure is iterated until convergence is reached.}
\label{fig:dmft_loop}
\end{figure}
\end{center}

At this point, we have introduced the quantum generalisation of the Weiss effective field
and have represented the local Green's function $G_{ii}$ as that of a single atom
coupled to an effective bath. This can be viewed as an {\it exact representation}, as
further detailed in Sec.~\ref{sec:func}. We now have to generalise to the quantum case
the mean-field {\it approximation} relating the Weiss function to $G_{ii}$ (in the classical
case, this is the self-consistency relation (\ref{mfising})).
The simplest manner in which this can be explained - but perhaps not the more
illuminating one conceptually (see Sec.~\ref{sec:func} and
\cite{georges_review_dmft,georges_windsor_dmft})-
is to observe that, in the effective impurity model (\ref{Seff}), we can define a
local self-energy from the interacting Green's
function
$G(\tau-\tau') \equiv -<Tc(\tau) c^{+}(\tau')>_{\Seff}$
and the Weiss dynamical mean-field as:
\begin{eqnarray}\nonumber
\Sigma_{imp}(i\omega_n)\,\equiv &{\cal G}_0^{-1}(i\omega_n)-G^{-1}(i\omega_n)\\
=&\iomn+\mu-\e_0-\Delta(\iomn)-G^{-1}(i\omega_n)
\label{eq:def_sig_imp}
\end{eqnarray}
Let us, on the other hand, consider the
self-energy of the original lattice model,
defined as usual from the full Green's function
$G_{ij}(\tau-\tau') \equiv -<Tc_{i,\sigma}(\tau) c_{j,\sigma}^{+}(\tau')>$ by:
\beq
G({\ka},i\omega_{n}) \,=\,
\frac{1}{i\omega_{n} + \mu -\epsilon_0- \epsilon_{\ka} -\Sigma(\ka,i\omega_{n})}
\label{eq:latticeG}
\eeq
in which $\epsk$ is the Fourier transform of the hopping integral, i.e the
dispersion relation of the non-interacting tight-binding band:
\beq
\epsilon_{\ka}\equiv \sum_{j} t_{ij} e^{i \ka.(\R_i-\R_j)}
\eeq
We then make the approximation that the lattice self-energy coincides with the
impurity self-energy. In real-space, this means that we neglect all non-local
components of $\Sigma_{ij}$ and approximate the on-site one by $\Sigma_{imp}$:
\begin{equation}
\Sigma_{ii}\simeq\Sigma_{imp}\,\,\,,\,\,\,\Sigma_{i\neq j} \simeq 0
\end{equation}
We immediately see that this is a consistent approximation only provided it leads
to a unique determination of the local (on-site) Green's function, which by construction
is the impurity-model Green's function. Summing (\ref{eq:latticeG}) over $\vk$
in order to obtain the on-site component $G_{ii}$ of the
the lattice Green's function, and using
(\ref{eq:def_sig_imp}), we arrive at the self-consistency
condition\footnote{Throughout these notes, the sums over momentum are
normalized by the volume of the Brillouin zone, i.e $\sum_\vk 1 =1$}:
\beq
\sum_{\vk}\frac{1}{\Delta(\iomn)+G(\iomn)^{-1}-\e_\vk} \,=\,G(\iomn)
\label{eq:scc1}
\eeq
Defining the non-interacting density of states:
\beq
D(\e)\,\equiv\,\sum_\vk\,\delta(\e-\e_\vk)
\label{eq:def_free_dos}
\eeq
this can also be written as:
\beq
\int d\e\,\frac{D(\e)}{\Delta(\iomn)+G(\iomn)^{-1}-\e} \,=\,G(\iomn)
\label{eq:scc2}
\eeq
This {\it self-consistency condition} relates, for each frequency, the dynamical
mean-field $\Delta(\iomn)$ and the local Green's function $G(\iomn)$. Furthermore,
$G(\iomn)$ is the interacting Green's function of the effective impurity model
(\ref{am}) -or (\ref{Seff})-. Therefore, we have a closed set of equations that
fully determine in principle the two functions $\Delta,G$ (or ${\cal G}_0$,$G$)).
In practice, one will use an {\it iterative procedure}, as represented on
Fig.~\ref{fig:dmft_loop}. In many cases, this iterative procedure converges to
a unique solution independently of the initial choice of $\Delta(\iomn)$. In some
cases however, more than one stable solution can be found (e.g close to the Mott
transition, see section below). The close analogy between the classical mean-field
construction and its quantum (dynamical mean-field) counterpart is summarized in
Table~\ref{table:dmft}.
%
\begin{center}
\begin{table}
\begin{tabular}{|c|c|c|} \hline
Quantum Case & Classical Case &  \\ \hline
$-\sum_{ij\sigma} t_{ij} c^{+}_{i\sigma}c_{j\sigma} +
\sum_i H_{atom}(i)$ &
$H =-\sum_{(ij)} J_{ij} S_{i}S_{j}-h\sum_{i}S_{i}$ &
Hamiltonian \\ \hline
$G_{ii}(\iomn)=-<c^{+}_i(i\omega_{n})c_i(i\omega_{n})>$ &
$m_i = <S_{i}>$ & Local Observable \\ \hline
$H_{eff} = H_{atom} + \sum_{l\sigma} \et_l a^{+}_{l\sigma} a_{l\sigma} + $&
$H_{eff} = -h_{eff}\,S $ &
Effective single-site \\
$+ \sum_{l\sigma} V_{l}(a^{+}_{l\sigma}c_{\sigma}+h.c)$ & & Hamiltonian\\ \hline
$\Delta\,(\iomn) = \sum_l \frac{|V_l|^2}{\iomn-\et_l}$ & $h_{eff}$ & Weiss function/Weiss field \\
${\cal G}_0^{-1}(\iomn) \equiv \iomn + \mu - \Delta(\iomn)$ & & \\ \hline
$\sum_{\vk}[\Delta(\iomn)+G(\iomn)^{-1}-\epsilon_{\vk}]^{-1} = G(\iomn)$ &
$ h_{eff} = \sum_j J_{ij}m_j +h$ & Self-consistency relation \\ \hline
\end{tabular}
\caption{Correspondance between the mean-field theory of a classical
system and the dynamical mean-field theory of a quantum system.}
\label{table:dmft}
\end{table}
\end{center}

\subsection{Limits in which DMFT becomes exact}
\label{sec:dmft_exact_limits}

\paragraph{Two simple limits: non-interacting band and isolated atoms}
It is instructive to check that the DMFT equations yield the exact answer in two simple
limits:
\begin{itemize}
\item In the {\it non-interacting
limit} $U=0$, solving (\ref{Seff}) yields $G(\iomn)={\cal G}_0(\iomn)$ and
$\Sigma_{imp}=0$. Hence, from (\ref{eq:scc1}),
$G(i\omega_n)=\sum_\vk 1/(i\omega_n+\mu-\e_0-\e_\vk)$
reduces to the free on-site Green's function. DMFT is trivially exact in
this limit since the self-energy is not only $\vk$-independent but vanishes
altogether.
\item In the {\it atomic limit} $t_{ij}=0$, one just has a collection
of independent atoms on each site and $\e_\vk=0$.
Then (\ref{eq:scc1}) implies $\Delta(\iomn)=0$: as expected, the
dynamical mean-field vanishes since the atoms are isolated.
Accordingly, the self-energy only has on-site components, and hence DMFT is
again exact in this limit.
The Weiss field reads
${\cal G}_0^{-1}=\iomn+\mu-\e_0$, which means that the action $\Seff$ simply corresponds to the
quantization of the atomic hamiltonian $H_{atom}$. This yields:
\begin{eqnarray}\nonumber
&G(i\omega_n)_{atom}=\frac{1-n/2}{i\omega_n+\tmu}
+\frac{n/2}{i\omega_n+\tmu-U}\\
&\Sigma(\iomn)_{atom} = \frac{nU}{2}
+ \frac{n/2(1-n/2)U^2}{\iomn+\tmu-(1-n/2)U}
\end{eqnarray}
with $\tmu\equiv\mu-\e_0$ and
$n/2=(e^{\beta\tmu}+e^{\beta(2\tmu-U)})/(1+2e^{\beta\tmu}+e^{\beta(2\tmu-U)})$.
\end{itemize}
Hence, the dynamical mean-field approximation is exact in the two limits of
the non-interacting band and of isolated atoms, and provides an interpolation
in between. This interpolative aspect is a key to the success of this approach
in the intermediate coupling regime.

\paragraph{Infinite coordination}
The dynamical mean-field approximation becomes exact in the limit where the
connectivity $z$ of the lattice is taken to infinity. This is also true of the
mean-field approximation in classical statistical mechanics. In that case,
the exchange coupling between nearest-neighbour sites must be scaled as:
$J_{ij}=J/z$ (for $J_{ij}$'s of uniform sign),
so that the Weiss mean-field $h_{eff}$ in (\ref{weiss})
remains of order one. This also insures that the
entropy and internal energy per site remain finite and hence preserves the
competition which is essential to the physics of magnetic ordering.
In the case of itinerant quantum systems~\cite{metzner_vollhardt},
a similar scaling must be made
on the hopping term in order to maintain the balance between
the kinetic and interaction energy.
The nearest-neighbour hopping amplitude must be scaled as:
$t_{ij}=t/\sqrt{z}$. This insures that the non-interacting
d.o.s $D(\e)=\sum_\vk\delta(\e-\e_\vk)$ has a non-trivial limit
as $z\rightarrow\infty$. Note that it also insures that the superexchange
$J_{ij}\propto t_{ij}^2/U$ scales as $1/z$, so that magnetic ordering
is preserved with transition temperatures of order unity. In practice,
two lattices are often considered in the $z=\infty$ limit:
\begin{itemize}
\item The d-dimensional cubic lattice with $z=2d\rightarrow\infty$ and
$\e_\vk=-2t\sum_{p=1}^{d}\mbox{cos}(k_p)/\sqrt{z}$. In this case
the non-interacting d.o.s becomes a Gaussian:
$D(\epsilon) = \frac{1}{t\sqrt{2\pi}}\exp-(\frac{\epsilon^{2}}{2t^{2}})$

\item The Bethe lattice (Cayley tree) with
coordination $z\rightarrow\infty$ and nearest-neighbor hopping
$t_{ij}=t/\sqrt{z}$. This corresponds to a semicircular d.o.s:
$D(\epsilon) = \frac{2}{\pi D} \sqrt{1-(\epsilon/D)^2}$ with a
half-bandwidth $D=2t$. In this case, the self-consistency condition
(\ref{eq:scc1}) can be inverted explicitly in order to relate the
dynamical mean-field to the local Green's function as:
$\Delta(\iomn)=t^2 G(\iomn)$.

\end{itemize}
Apart from the intrinsic interest of solving strongly correlated fermion models in the
limit of infinite coordination, the fact that the DMFT equations become exact
in this limit is important since it guarantees, for example, that exact
constraints (such as causality of the self-energy, positivity of the
spectral functions, sum rules such as the Luttinger theorem or the f-sum rule)
are preserved by the DMFT approximation.

\subsection{Important topics not reviewed here}

There are several important topics related to the DMFT framework, which I have
not included in these lecture notes. Some of them were covered in the lectures,
but extensive review articles are available in which these topics are
at least partially described.

This is a brief list of such topics:

\paragraph{DMFT for ordered phases}
The DMFT equations can easily be extended to study phases with long-range order,
calculate critical temperatures for ordering as well as phase diagrams,
see e.g~\cite{georges_review_dmft}.

\paragraph{Response and correlation functions in DMFT}
Response and correlation functions can be expressed in terms
of the lattice Green's functions, and of the impurity model vertex functions,
see e.g~\cite{georges_review_dmft,pruschke_jarrell_review}.
Note that momentum-dependence enters, through the lattice Green's function.

\paragraph{Physics of the Anderson impurity model}
Understanding the various possible fixed points of quantum impurity models
is important for gaining physical intuition when solving lattice models within
DMFT. See Ref.~\cite{hewson_book} for a review and references on the Anderson
impurity model. It is important to keep in mind that, in contrast to the common
situation in the physics of magnetic impurities or mesoscopics, the effective
conduction electron bath in the DMFT context has significant energy-dependence.
Also, the self-consistency condition can drive the effective impurity
model from one kind of low-energy behaviour to another, depending on
the range of parameters (e.g close to
the Mott transition, see Sec.~\ref{sec:mott}).

\paragraph{Impurity solvers}
Using reliable methods for calculating the impurity Green's function and
self-energy is a key step in solving the DMFT equations. A large numbers
of ``impurity solvers'' have been implemented in the DMFT
context\footnote{Some early versions of numerical codes are available at:
http://www.lps.ens.fr/$\sim$krauth},
including: the quantum Monte Carlo (QMC)
method~\cite{jarrell_1992_qmc_prl}
(see also~\cite{rozenberg_mott_qmc,georges_krauth_mott_prl}),
based on the Hirsch-Fye algorithm~\cite{hirsch_fye},
adaptative exact diagonalisation or projective schemes
(see~\cite{georges_review_dmft} for a review and references),
the Wilson numerical renormalisation group (NRG, see e.g \cite{bulla_mott_NRG}
and references therein). Approximation schemes have also proven useful,
when used in appropriate regimes,
such as the ``iterated perturbation theory''
approximation (IPT, ~\cite{georges_kotliar_dmft,kajueter_ipt_1996_prl}),
the non-crossing approximation (NCA, see \cite{pruschke_jarrell_review} for
references) and
various extensions~\cite{florens_rotors_imp_2002_prb},
as well as schemes interpolating between high and
low energies~\cite{oudovenko_solver_2004}.

\paragraph{Beyond DMFT}
DMFT does capture ordered phases, but does not take into account the
coupling of short-range spatial correlations (let alone long-wavelength)
to quasiparticle properties, in the absence of ordering. This is a key aspect
of some strongly correlated materials (e.g cuprates, see the concluding section of
these lectures), which requires
an extension of the DMFT formalism.
Two kinds of extensions have been explored:
\begin{itemize}

\item $\vk$-dependence of the self-energy can be reintroduced by considering
cluster extensions of DMFT, i.e a small cluster of sites (or coupled atoms) into
a self-consistent bath. Various embedding schemes have been
discussed~\cite{georges_review_dmft,schiller_cluster_1996_prl,hettler_dca_1998_prb,
kotliar_cdmft,biermann_chain_2001_prl,biroli_cluster,onoda_finiteT_mott} and I will not attempt a review of this
very interesting line of research here. One of the key questions is whether
such schemes can account for a strong variation of the quasiparticle properties
(e.g the coherence scale) along the Fermi surface.

\item Extended DMFT (E-DMFT~\cite{si_smith_edmft_1996,kajueter_phd,sengupta_georges_metallic_sg,
smith_si_edmft_2000}) focuses on two-particle local observables, such as
the local spin or charge correlation functions, in addition to the local
Green's function of usual DMFT. For applications to electronic structure, see
Sec.~\ref{sec:abinitio_dmft}.

\end{itemize}
\section{Functionals, local observables, and interacting systems}
\label{sec:func}

In this section\footnote{This section is based in part on Ref.~\cite{georges_windsor_dmft}},
I would like to discuss a theoretical framework which
applies quite generally to interacting systems. This framework reveals
common concepts underlying different theories such as:
the Weiss mean-field theory (MFT) of a classical magnet, the density functional theory (DFT)
of the inhomogeneous electron gas in solids, and the dynamical mean-field theory (DMFT) of
strongly correlated electron systems.
The idea which is common to these diverse theories is the construction of
{\it a functional of some local quantity} (effective action) by the Legendre transform method.
Though exact in principle, it requires in practice that
the exact functional is approximated in some manner.
This method has a wide range of applicability in
statistical mechanics, many-body physics and field-theory~\cite{fukuda_ptps_121_1996}.
The discussion will be (hopefully) pedagogical, and for this reason
I will begin with the example of a classical magnet. For a somewhat more detailed
presentation, see Ref.~\cite{georges_windsor_dmft}.

There are common concepts underlying all these constructions (cf. Table),
as will become clear below, namely:
\begin{itemize}
\item i) These theories focus on a specific {\it local quantity}: the local magnetization in MFT, the
local electronic density in DFT, the local Green's function (or spectral density) in DMFT.

\item ii) The original system of interest is replaced by an {\it equivalent system}, which
is used to provide a representation of the selected quantity: a single spin in an effective field
for a classical magnet, free electrons in an effective one-body potential in DFT,
a single impurity Anderson model within DMFT.
The effective parameters entering this equivalent problem define {\it generalized
Weiss fields} (the Kohn-Sham potential in DFT, the effective hybridization within DMFT), which are
self-consistently adjusted. I note that the associated equivalent system can be a non-interacting
(one-body) problem, as in MFT and DFT, or a fully interacting many-body problem (albeit simpler
than the original system) such as in DMFT and its extensions.

\item iii) In order to pave the way between the real problem of interest and the equivalent model, the method
of coupling constant integration will prove to be very useful in constructing (formally)
the desired functional using the Legendre transform method. The coupling constant can be either the
coefficient of the interacting part of the hamiltonian (which leads to a non-interacting equivalent
problem, as in DFT), or in front of the non-local part of the hamiltonian (which leads in general to a
local, but interacting, equivalent problem such as in DMFT).
\end{itemize}

\noindent
Some issues and questions are associated with each of these points:

\begin{itemize}

\item i) While the theory and associated functional primarily aims at calculating the selected
local quantity, it always come with the possibility of determining some more general object. For example,
classical MFT aims primarily at calculating the local magnetization, but it can be used to
derive the Ornstein-Zernike expression of the correlation function between different sites. Similarly,
DFT aims at the local density, but Kohn-Sham orbitals can be {\it interpreted} (without a firm
formal justification) as one-electron
excitations. DMFT produces a local self-energy which one may interpret as the lattice self-energy from
which the full k-dependent Green's function can be reconstructed. In each of these cases, the precise
status and interpretation of these additional quantities can be questioned.

\item ii) I emphasize that the choice of an equivalent representation of the local quantity has
nothing to do with subsequent approximations made on the functional.
The proposed equivalent system is in fact an exact representation
of the problem under consideration (for the sake of calculating the selected local quantity).
It does raise a {\it representability} issue, however: is
it always possible to find values of the generalised Weiss field which will lead to a specified
form of the local quantity, and in particular to the exact form associated with the specific system
of interest? For example: given the local electronic density $n(x)$ of a specific solid, can one always
find a Kohn-Sham effective potential such that the one-electron local density obtained by solving
the Schr\"{o}dinger equation in that potential coincides with $n(x)$ ? Or, in the context of DMFT: given
the local Green's function of a specific model, can one find a hybridisation function such that it can be viewed
as the local Green's function of the specified impurity problem ?

\item iii) There is also a stability issue of the exact functional: is the equilibrium value
of the local quantity a minimum ?
More precisely, one would like to show that negative eigenvalues of the
stability matrix correspond to true physical instabilities of the system.
I will not seriously investigate this issue in this lecture
(for a discussion within DMFT, where it is still quite open, see \cite{chitra_bk}).

\end{itemize}

\begin{center}
\begin{table}
\begin{tabular}{|c|c|c|c|}
\hline
 Theory & MFT &  DFT  &   DMFT  \\
 \hline
Quantity & Local magnetization $m_i$ & Local density $n(x)$ &
Local GF $G_{ii}(\omega)$\\ \hline
Equivalent & Spin in  & Electrons in  &
Quantum  \\
system & effective field &  effective potential &
impurity model \\ \hline
Generalised & Effective & Kohn-Sham & Effective \\
Weiss field &  local field & potential & hybridisation \\ \hline
\end{tabular}
\caption{Comparison of theories based on functionals of a local observable}
\label{table:func}
\end{table}
\end{center}

\subsection{The example of a classical magnet}

For the sake of pedagogy, I will consider in this section the simplest
example on which the above ideas can be made concrete: that of a classical
Ising magnet with hamiltonian
\be
H = -\sum_{ij} J_{ij} S_i S_j
\ee
\paragraph{Construction of the effective action}

We want to construct a functional $\Gamma[m_i]$ of a
{\it preassigned} set of local magnetizations $m_i$,
such that minimizing this functional yields the
equilibrium state of the system. This functional is of course the
Legendre transform of the free-energy with respect to a set of
local magnetic fields.
To make contact with the field-theory literature, I note that $\beta\,\Gamma$
is generally called the {\it effective action} in this context.
I will give a formal construction of this functional, following a
method due to Plefka \cite{plefka} and Yedidia and myself
\cite{georges_yedidia_mft}. Let us introduce a varying coupling constant
$\a\in[0,1]$, and define:
\be
H_{\a}\equiv \a\,H = \sum_{ij} \alpha J_{ij}\, S_i S_j
\ee
Introducing local Lagrange multipliers $\l_i$, we consider the functional:
\be
\Omega[m_i,\l_i;\a]\equiv-{1\over\beta} \ln \T \,
e^{-\b H_{\a} + \b\sum_i\lambda_i (S_i-m_i)} = F[\l_i]+\sum_i \l_i m_i
\ee
Requesting stationarity of this functional with respect to the $\l_i$'s amounts to
impose that, {\it for all values of $\a$}, $\bra S_i\ket$ coincides with the
preassigned local magnetization $m_i$. The equations $m_i=\bra S_i\ket$ which expresses
the magnetization as a function of the sources $\lambda_i$ can then be inverted to
yield the $\l_i$'s as functions of
the $m_j$'s and of $\a$:
\begin{equation}
\label{eq:inversion}
\bra S_i \ket_{\l,\a} = m_i \rightarrow
\l_i = \l_i[m_j;\a]
\end{equation}
(The average $\bra\cdots\ket_{\l,\a}$ in this equation is with respect
to the Boltzmann weight appearing in the above definition of $\Omega$,
including $\lambda_i$'s and $\a$).
The Lagrange parameters can then be substituted into $\Omega$ to obtain
the $\a$-dependent Legendre transformed functional:
\be
\Ga[m_i] = \Omega[m_i,\l_i[m_j,\a]] = F\left[\l_i[m]\right] + \sum_i \l_i[m] m_i
\ee
Of course, the functional we are really interested in is that of the
original system with $\a=1$, namely:
\be
\Gamma [m_i] \equiv \Gamma_{\a=1}[m_i]
\ee
Let us first look at the non-interacting limit $\a=0$ for which
the explicit expression of $\Omega$ is easily obtained as:
\be
\Omega_0 = \sum_i \left(-{1\over\b}\ln \cosh \b\l_i+m_i\l_i\right)
\ee
Varying in the $\lambda$'s yields:
\be
\label{weissIsing}
\tanh \b\l_i^{(\a=0)} = m_i
\ee
and finally:
\be
\label{A0Ising}
\Gamma_{\a=0}[m_i] = {1\over\b} \sum_i \left(
{{1+m_i}\over{2}}\ln{{1+m_i}\over{2}} +
{{1-m_i}\over{2}}\ln{{1-m_i}\over{2}}\right)
\ee
The $\a=0$ theory defines the {\it equivalent problem} that
we want to use in order to deal with the
original system. Here, it is just a theory of
{\it independent spins in a local effective field}. The expression
(\ref{A0Ising}) is simply the entropy term corresponding to independent Ising spins
for a given values of the local magnetizations.

The value taken by the Lagrange multiplier in the equivalent system,
$\l_i^{\a=0}$ (denoted $\l_i^0$ in the following), must be interpreted as
the {\it Weiss effective field}. We note that, in this simple example, there is an explicit and very simple
relation (\ref{weissIsing}) between the Weiss field and $m_i$, so that one can work
{\it equivalently} in terms of either quantities. Also, because of the
simple form of (\ref{weissIsing}), {\it representability} is trivially satisfied:
given the actual values of the magnetizations $m_i$'s ($\in [-1,1]$) at equilibrium
for the model under consideration, one
can always represent them by the Weiss fields
$\beta h_i^{eff}=\mbox{arctanh} \,m_i$.

To proceed with the construction of $\Gamma$, we use a coupling constant
integration and write:
\be
\G[m_i;\a=1] = \G_0[m_i] + \int_0^{1} d\a {{d\G_{\a}}\over{d\a}}[m_i]
\ee
It is immediate that, because of the constraint $\bra (S_i-m_i)\ket=0$:
\be
{{d \Ga}\over{d\a}} = \bra H \ket_{\a,\l[\a]}
= -\sum_{ij} J_{ij} \bra S_iS_j\ket_{\a,\l[\a,m]}
\ee
In this expression, the correlation must be viewed as a functional of the
local magnetizations (thanks to the inversion formula (\ref{eq:inversion})).
Introducing the connected correlation function:
\be
g^c_{ij}[\{m_k\};\a] \equiv \bra (S_i-m_i)(S_j-m_j)\ket_{\a,\l[\a,m]}
\ee
we obtain:
\be
\label{eq:diff_gamma_Ising}
{{d \Ga}\over{d\a}} = -\sum_{ij} J_{ij} m_i m_j
-\sum_{ij} J_{ij} \,g^c_{ij}[\{m_k\};\a]
\ee
So that finally, one obtains the formal expression for
$\G[m_i]\equiv \G_{\a=1}[m_i]$:
\be\label{eq:exact_Gamma_Ising}
\G[m_i] = \G_0[m_i] - \sum_{ij} J_{ij} m_i m_j -
\sum_{ij} J_{ij} \int_0^1 d\a g^c_{ij}[m_k;\a]
\equiv
\G_0 + E_{MF} + \G_{corr}
\end{equation}
In this expression, $g^c$ denotes the connected correlation function
for a given value of the coupling constant, {\it expressed as a functional of
the local magnetisations}.

Hence, the {\it exact functional} $\G$ appears as a sum of three contributions:
\begin{itemize}
\item The part associated with the equivalent system (corresponding here
to the entropy of constrained but otherwise free spins)
\item The mean-field energy $\sum_{ij} J_{ij} m_i m_j$
\item  A contribution from correlations which contains all corrections beyond mean-field
\end{itemize}
As explained in the next section, there is a direct analogy between this and
the various contributions to the density functional within DFT (kinetic energy,
Hartree energy and exchange-correlation).

I note in passing that one can derive a closed equation for the exact functional,
which reads (see \cite{georges_windsor_dmft} for a derivation):

%
%
\be
\label{eq:exact_int_1_Ising}
\G_\a[m_i] = \G_0[m_i] - \a\sum_{ij} J_{ij} m_i m_j - {1\over\beta}
\sum_{ij} J_{ij} \int_0^{\a} d\a'
\left[ {{\d^2 \G_{\a'}}\over{\d m_k\d m_l}} \right]^{-1}_{ij}
\ee
This equation fully determines in principle the effective
action functional. However, in order to use it in practice, one generally has
to start from a limit in which the functional is known explicitly, and expand around
that limit.
For example, an expansion around the high-temperature limit
yields systematic corrections to mean-field
theory~\cite{georges_yedidia_mft,georges_windsor_dmft}.
This equation is closely related~\cite{georges_windsor_dmft} to the Wilson-Polchinsky equation
\cite{polchinsky} for the effective
action (after a Legendre transformation: see also \cite{ledou_schehr}), which
can be taken as a starting point for a renormalisation group analysis by starting
from the local limit and expanding in the ``locality''
(see e.g \cite{chauve_ledou_exact_RG,ledou_schehr}.


\paragraph{Equilibrium condition and stability}

The physical values of the magnetisations at equilibrium are obtained
by minimising $\G$, which yields:
\be
\label{EqIsing}
m_i^* = \tanh\left(\b\sum_jJ_{ij}m_j^* -\b {{\d \G_{corr}}\over{\d m_i}}
\right)
\ee
and the Weiss field takes the following value:
\be
\label{WeissEqIsing}
h_i^{eff}\,\equiv\,(\l_i^0)^* = \sum_j J_{ij}\, m_j^* - {{\d \G_{corr}}\over{\d m_i}}|^*
\ee
This equation is a {\it self-consistency condition} which determines the Weiss
field in terms of the local magnetizations on all other sites. Its physical
interpretation is clear: $h_i^{eff}$ is the
true (average) local field seen by site $i$. It is equal to the sum of two terms: one
in which all spins are treated as independent, and a correction due to correlations.

The stability of the functional around equilibrium is controlled by the
fluctuation matrix:
\be
{{\d^2 \G}\over{\d m_i\d m_j}} = {{\d\l_i^0}\over{\d m_j}} -J_{ij}
+ {{\d^2 \G_{corr}}\over{\d m_i\d m_j}}
\ee
At equilibrium, this is nothing else than the inverse of the
susceptibility (or correlation function) matrix:
\be
{{\d^2 \G}\over{\d m_i\d m_j}} \equiv
(\chi^{-1})_{ij} =
(\chi^{-1}_0)_{ij} - J_{ij} + {{\d^2 A_{corr}}\over{\d m_i\d m_j}}
\ee
with:
\be
(\chi^{-1}_0)_{ij} = {{1}\over{\b(1-m_i^2)}} \delta_{ij}
\ee
Hence, our functional does satisfy a stability criterion as defined in the introduction:
a negative eigenvalue of this matrix (i.e of $\chi(\vec{q})$) would correspond to a
physical instability of the system.
Note that at the simple mean-fied level, we recover the RPA formula for the
susceptibility: $(\chi^{-1})_{ij} =
(\chi^{-1}_0)_{ij} - J_{ij}$.

\paragraph{Mean-field approximation and beyond}

Obviously, this construction of the {\it exact} Legendre transformed free energy, and
the exact equilibrium condition (\ref{EqIsing})
has formal value, but concrete applications require some further approximations to be
made on the correlation term $\G_{corr}$.
The simplest such approximation is just to neglect $\G_{corr}$ altogether. This is
the familiar Weiss mean-field theory:
\be\label{eq:MFT_Ising}
\G_{MFT} = {1\over\b} \sum_i \left(
{{1+m_i}\over{2}}\ln{{1+m_i}\over{2}} +
{{1-m_i}\over{2}}\ln{{1-m_i}\over{2}}\right) -\sum_{ij}
J_{ij} m_i m_j
\ee
For a ferromagnet (uniform positive $J_{ij}$'s),
this approximation becomes {\it exact in the limit of infinite coordination} of the lattice.

The formal construction above is a useful guideline when
trying to improve on the mean-field approximation.
I emphasize that, within the present approach, {\it it is
the self-consistency condition (\ref{WeissEqIsing}) (relating the Weiss field to
the environment)
{\it that needs to be corrected}}, while the equation $m_i=\tanh\beta h_i^{eff}$ is
attached to our choice of equivalent system and will be always valid.
For example, in \cite{plefka,georges_yedidia_mft} it was shown how to construct
$\G_{corr}$ by a systematic high-temperature expansion in $\beta$. This expansion can
be conveniently generated by iterating the exact equation (\ref{eq:exact_int_1_Ising}).
It can also be turned
into an expansion around the limit of infinite coordination \cite{georges_yedidia_mft}.
The first contribution to $\G_{corr}$ in this expansion
appears at order $\beta$ (or $\alpha^2$) and reads:
\begin{equation}\label{eq:Onsager}
\G_{corr}^{(1)} \,=\,-\frac{\beta}{2}\sum_{ij}\,J_{ij}^2\,(1-m_i^2)(1-m_j^2)
\end{equation}
This is a rather famous correction to mean-field theory, known as the ``Onsager reaction term''.
For spin glass models ($J_{ij}$'s of random sign), it is crucial to include this term
even in the large connectivity limit. The corresponding equations for the equilibrium
magnetizations are those derived by Thouless, Anderson and Palmer~\cite{thouless_TAP}.


\subsection{Density functional theory}
\label{sec:dft}

In this section, I explain how density-functional theory
\footnote{I actually consider the finite-temperature extension of DFT \cite{mermin_DFT_finiteT}}
(DFT) \cite{hohenberg_kohn,kohn_sham} can be derived along very similar lines.
This section borrows from the work of Fukuda et al.
\cite{fukuda_ptp_92_1994,fukuda_ptps_121_1996} and
of Valiev and Fernando \cite{valiev_pla_1997}. For a recent pedagogical review
emphasizing this point of view, see \cite{argaman_dft}.
For detailed reviews of the DFT formalism,
see e.g~\cite{dreizler_gross_dft,jones_gunnarsson_rmp}.

Let us consider the inhomogeneous electron gas of a solid,
with hamiltonian:
\beq
H = -\sum_i \frac{1}{2} \nabla^2_i + \sum_i v(\vr_i) +
\frac{1}{2} \sum_{i\neq j} U(\vr_i-\vr_j)
\eeq
in which $v(x)$ is the external potential due to
the nuclei and $U(x-~x')$ ($=e^2/|x-x'|$) is the
electron-electron interaction. (I use conventions in which $\hbar=m=1$).
Let us write this hamiltonian in second- quantized form, and again introduce a
coupling-constant parameter $\a$ (the physical case is $\a=1$):
\begin{equation}
H_\a = -\frac{1}{2}\int dx\,\psi^\dagger\nabla^2\psi+\,
\int dx\, v(x)\hn(x) + {\a\over 2} \int dx\,dx'\,
\hn(x) U(x-x') \hn(x')
\end{equation}
We want to construct the free energy functional of the system
while constraining
the average density to be equal to some specified function $n(x)$.
In complete analogy with the previous section, we
introduce a Lagrange multiplier function $\lambda(x)$, and
consider
\footnote{Note that
I chose in this expression a different sign convention for $\lambda$ than
in the previous section, and also that $\T$ denotes the full many-body
trace over all $N$-electrons degrees of freedom.}:
\begin{equation}
\Omega_{\a}[n(x),\lambda(x)] \equiv -{1\over\b}
\ln \T \exp{\left(-\beta H_\a
+ \b\int dx\, \lambda(x) (n(x)-\hn(x))\right)}
\end{equation}
A functional of {\it both} $n(x)$ and $\lambda(x)$.
As before, stationarity in $\lambda$ insures that:
\begin{equation}
\langle \hn(x)\rangle_{\lambda,\a} = n(x)\,\,\rightarrow\,\,
\lambda(x) = \lambda_{\a}[n(x)]
\label{eql}
\end{equation}
This will be used to eliminate $\lambda(x)$ in terms of $n(x)$
and construct the functional of $n(x)$ only:
\begin{equation}
\G_{\a}[n(x)] \equiv \Omega_{\a}[n(x),\lambda_{\a}[n(x)]]
\end{equation}

\subsubsection{Equivalent system: non-interacting electrons in an effective potential}

Again, I first look at the non-interacting case $\a=0$.
Then we have to solve
a one-particle problem in an $x$-dependent external potential.
This yields:
\begin{equation}
\label{OmegaDFT0}
\Omega_0[n[x],\l[x]] =
- \t\ln [\iomn -\hat{t} - \hat{v} -\hat{\l}] -
\int dx\, \l(x)n(x)
\end{equation}
In this equation, $\t$ denotes the trace over the degrees of freedom of a single electron,
$\iomn$ is the usual Matsubara frequency, and $\hat{t}\equiv -\nabla^2/2$, $\hat{v}$,
$\hat{\lambda}$ are the one-body operators corresponding to the kinetic
energy, external potential and $\lambda(x)$ respectively. The identity
$\ln det = \t\ln$ has been used.

Minimisation with respect to $\lambda(x)$ yields the following relation
between $\lambda^0$ and $n(x)$:
\be
\label{WeissDFT}
{1\over\b}\sum_n
\bra x| {{1}\over{\iomn-\hat{t}-\hat{v}-\hat{\lambda}_0}} |x\ket \,=\,
n(x)
\ee
This defines the functional $\lambda_0[n(x)]$, albeit in a somewhat
implicit manner. This is directly analogous to Eq.(\ref{weissIsing}) defining
the Weiss field in the
Ising case (but in that case, this equation was easily invertible).
If we want to be more explicit, what we have to do is solve the
one-particle Schrodinger equation:
\be
\label{Schro}
\left(-\frac{1}{2}\Delta + v_{KS}(x) \right)\phi_l(x) = \e_l\phi_l(x)
\ee
where the {\it effective one-body potential} (Kohn-Sham potential)
is {\it defined} as:
\be
v_{KS}(x) \equiv v(x) + \lambda^0(x)
\ee
It is convenient to construct the associated resolvent:
\be
\label{G_DFT}
R(x,x';\iomn) = \sum_l {{\phi_l(x)\phi_l^*(x')}\over{
\iomn-\e_l}}
\ee
and the relation (\ref{WeissDFT}) now reads:
\be
\sum_l |\phi_l(x)|^2 f_{FD}(\e_l) = n(x)
\ee
in which $f_{FD}$ is the Fermi-Dirac distribution.

This relation expresses the local density in an interacting many-particle
system as that of a one-electron problem {\it in an effective potential}
defined by (\ref{WeissDFT}). In so doing, the effective
one-particle wave functions and energies (Kohn-Sham orbitals) have been introduced,
whose relation to the original system (and in particular their interpretation as
excitation energies) is far from obvious (see e.g \cite{jones_gunnarsson_rmp}).
There is, for example, no fundamental justification in identifying the resolvent
(\ref{G_DFT}) with the true one-electron Green's function of the interacting system.
The issue of {\it representability} (i.e whether an effective potential can always
be found given a density profile $n(x)$) is far from being as obvious as in the
previous section, but has been established on a rigorous
basis~\cite{chayes_dft,chayes_ruskai_dft}.

To summarize, the non-interacting functional $\G_0[n(x)]$ reads:
\beq
\label{A0DFT}
\G_0[n(x)] =
- \t\ln [\iomn -\hat{t} - \hat{v} -\hat{\l_0}[n]] -
\int dx\, \l^0[x;n] n(x)
\eeq
which can be rewritten as:
\beq
\G_0[n(x)] = = -\frac{1}{\b} \sum_l \ln\left[1+e^{-\b\e_l[n]}\right] -\int dx\, v_{KS}(x) n(x)
+ \int dx\, v(x) n(x)
\eeq
in which $\lambda_0$ and $v_{KS}$ are viewed as a functional of $n(x)$, as
detailed above.

In the limit of zero temperature ($\beta\to\infty$), this reads:
\beq
\G_0[n(x),T=0] = \sum_{l}^{\prime} \e_l -\int dx\, v_{KS}(x) n(x)
+ \int dx\, v(x) n(x)
\eeq
in which the sum is over the N occupied Kohn-Sham states. We note that it contains extra
terms beyond the ground-state energy of the KS equivalent system
(see also Sec.~\ref{sec:energy_lda+dmft}).

We also note that $\G_0$ is not a very explicit functional of $n(x)$.
It is a somewhat more explicit functional of $\lambda_0(x)$ (or equivalently
of the KS effective potential $v_{KS}(x)$) so that it is often
more convenient to think in terms of this quantity directly.
At any rate, in order to evaluate $\G_0$ for a specific density profile or effective
potential one must solve the Schr\"{o}dinger equation for KS orbitals and eigenenergies.
This is a time-consuming task for realistic three-dimensional potentials and
practical calculations would be greatly facilitated if a more explicit accurate expression
for $\Gamma_[n(x)]$ would be available~\footnote{see e.g the lecture notes by K.Burke:
http://dft.rutgers.edu/kieron/beta/index.html}.

\subsubsection{The exchange-correlation functional}

We turn to the interacting theory, and use the coupling constant integration
method (see \cite{harris_dft} for its use in DFT):
\be
\G[n(x)] = \G[n(x);\a=0] + \int_0^1 d\a {{d \G_{\a}}\over{d\a}}
\ee
Similarly as before:
\be
{{d \G_{\a}}\over{d\a}} =
\bra \hat{U} \ket_{\lambda,\a} =  {1\over 2}
\int dx dx'\, U(x-x') \bra\hn(x)\hn(x')\ket_{\lambda,\a}
\ee
Separating again a Hartree (mean-field) term, we get:
\be
\label{eq:total_functional}
\G[n(x)] = \G_0[n(x)] + E_{Hartree}[n(x)] + \G_{xc}[n(x)]
\ee
with:
\be
E_{Hartree}[n(x)] = {1\over 2} \int dx dx'\, U(x-x') n(x) n(x')
\ee
and $\G_{xc}$ is the correction-to mean field term (the exchange-correlation
functional):
\be\label{eq:xc_functional}
\G_{xc}[n(x)]
= {1\over 2} \int dx dx'\, U(x-x') \int_0^1d\a g^c_{\a}[n;x,x']
\ee
In which:
\be
g^c_{\a}[n;x,x'] \equiv \bra(\hn(x)-n(x))(\hn(x')-n(x'))\ket_{\lambda_{\a}[n],\a}
\ee
is the (connected) density-density correlation function, expressed
as a functional of the local density, for a given value of the coupling $\a$.

It should be emphasized that the exchange-correlation functional $\Gamma_{xc}$
is {\it independent} of the specific form of the crystal potential $v(x)$: it is
a {\it universal functional} which depends only on the form of the
inter-particle interaction $U(x-~x')$~!
To see this, we first observe that, because $\G[n(x)]$ is the Legendre
transform of the free energy with respect to the one-body potential,
we can easily relate the functional in the presence of the crystal potential $v(x)$
to that of the homogeneous electron gas (i.e with $v=0$):
\be
\G [n(x)] = \G_{HEG} [n(x)] + \int dx\, v(x) n(x)
\ee
Since this relation is also obeyed for the non-interacting system
(see Eq.~(\ref{A0DFT})), and using $\G=\G_0+\G_H+\G_{xc}$, we see that
the functional form of $\G_{xc}$ is independent of $v(x)$. It is the same for all solids, and
also for the homogeneous electron gas.

I finally note that an exact relation can again be derived for the
density functional (or alternatively the exchange-correlation functional)
by noting that:
\be
\beta g^c_{\a}[n;x,x'] = \left[\frac{\d\G_{\a}}{\d n(x) \d n(y)}\right]^{-1}_{xx'}
\ee
Inserting this relation into (\ref{eq:total_functional},\ref{eq:xc_functional}),
one obtains:
\be
\G_\a[n] = \G_0[n] + \a E_H[n] +
\frac{1}{2}\int dx dx'\, U(x-x') \int_0^{\a} d\a'
\left[\frac{\d\G_{\a}}{\d n(x) \d n(y)}\right]^{-1}_{xx'}
\ee
in complete analogy with (\ref{eq:exact_int_1_Ising}).
For applications of this exact functional equation,
see e.g~\cite{khodel_dft_physrep,amusia_dft}. Analogies with the exact renormalization
group approach (see previous section) might suggest further use of this relation
in the DFT context.

\subsubsection{The Kohn-Sham equations}

Let us now look at the condition for equilibrium. We vary $\G[n(x)]$, and
we note that, as before, the terms originating from the variation
$\d\l ^0/\d n(x)$ cancel because of the relation (\ref{WeissDFT}).
We thus get:
\be
{{\d \G}\over{\d n(x)}} = -\l_0(x) +
\int dx'\, U(x-x') n(x') + {{\d \G_{xc}}\over{\d n(x)}}
\ee
so that the equilibrium density $n^*(x)$ is determined by:
\be
\label{WeissEqDFT}
\l^0(x)^* =
\int dx'\, U(x-x') n^*(x') + {{\d \G_{xc}}\over{\d n(x)}}|_{n=n^*}
\ee
which equivalently specifies the KS potential at equilibrium as:
\be
\label{KSpot}
v_{KS}^{*}(x) = v(x) +
 \int dx'\, U(x-x') n^*(x') + {{\d \G_{xc}}\over{\d n(x)}}|_{n=n^*}
\ee
Equation (\ref{WeissEqDFT}) is the precise analog of
Eq.(\ref{WeissEqIsing}) determining the Weiss field in the Ising case,
and $v_{KS}^{*}$ is the true effective potential seen by an electron
at equilibrium, in a one-electron picture. Together with (\ref{Schro}), it
forms the fundamental (Kohn-Sham) equations of the DFT approach. To summarize,
the expression of the total energy ($T=0$) reads:
\beq
\G[n(x),T=0] = \sum_{l}^{\prime} \e_l -\int dx\, v_{KS}(x) n(x)
+ \int dx\, v(x) n(x) \,+\,\G_{xc}\left[n(x)\right]
\eeq

Concrete applications of the DFT formalism require an approximation to be made
on the exchange-correlation term. The celebrated {\it local density approximation}
(LDA) reads:
\be
\G_{xc}\left[n(x)\right]|_{LDA} = \int dx\, n(x)\,\epsilon_{xc}^{HEG}[n(x)]
\ee
in which $\epsilon_{xc}^{HEG}(n)$ is the exchange-correlation energy density of
the {\it homogeneous} electron gas, for an electron density $n$.
Discussing the reasons for the successes of this approximation (as well as its
limitations) is quite beyond the scope of these lectures. The interested reader is
referred e.g to \cite{jones_gunnarsson_rmp,argaman_dft}.

Finally, we observe that DFT satisfies the stability properties discussed in the
introduction, since $\d ^2 \G/\d n(x)\d n(x')$ is the inverse of
the density-density response function
($\vq$-dependent compressibility). A negative eigenvalue would correspond
to a charge ordering instability.


\subsection{Exact functional of the local Green's function,
and the Dynamical Mean-Field Theory approximation}

In this section, I would like to explain how the concepts of the previous
sections provide a broader perspective on the dynamical mean field approach
to strongly correlated fermion systems. In contrast to DFT which focuses on
ground-state properties (or thermodynamics), the goal of DMFT
(see \cite{georges_review_dmft}
for a review) is to address excited states by focusing on the {\it local
Green's function} (or the {\it local spectral density}).
Thus, it is natural to formulate this approach in terms of a functional of the local
Green's function. This point of view has been recently emphasized by Chitra and Kotliar
\cite{chitra_local} and by the author in Ref.~\cite{georges_windsor_dmft}.

I describe below how such an {\it exact functional} can be formally constructed
for a correlated electron model (irrespective, e.g of dimensionality), hence leading
to a {\it local Green's function (or local spectral density)
functional theory}.
I will adopt a somewhat different viewpoint than in \cite{chitra_local},
by taking the {\it atomic limit} (instead of the non-interacting limit) as a reference
system. This leads naturally to represent the exact local Green's function
as that of a quantum impurity model, with a suitably chosen hybridisation
function. There is no approximation involved in this mapping (only a
representability assumption). This gives a general value to the impurity model mapping of
Ref.\cite{georges_kotliar_dmft}.
Dynamical mean field theory as usually implemented can then be viewed
as a {\it subsequent approximation}
made on the non-local
contributions to the exact functional (e.g. the kinetic energy).

For the sake of simplicity, I will take the Hubbard model as an example throughout this section.
The hamiltonian is decomposed as:
\begin{equation}
H_\a = U\,\sum_{i} n_{i\uparrow} n_{i\downarrow} -
\a\,\sum_{ij,\sigma} t_{ij}\, c^{+}_{i\sigma} c_{j\sigma}
\label{hubbard}
\end{equation}
I emphasize that the varying coupling constant $\a\in[0,1]$ has been introduced in front
of the hopping term, which is the non-local term of this hamiltonian, and {\it not} in front
of the interaction. When dealing with a more general hamiltonian, we would similarly decompose
$H=H_{\mbox{loc}}+\a H_{\mbox{non-loc}}$.

\subsubsection{Representing the local Green's function by a quantum impurity model}

In order to constrain the local Green's function
$\bra c_i(\tau)c_i^+(\tau') \ket$ to take a specified value $G(\tau-\tau')$, we
introduce conjugate sources (or Lagrange multipliers) $\Delta(\tau-~\tau')$ and consider
\footnote{In this section, I will divide the free energy functional by the number $N_s$
of lattice sites (restricting myself for simplicity to an homogeneous system)}:
\begin{eqnarray}
\label{eq:omegafull}
\nonumber
\Omega_\a[G(\omega),\D(\o)] \equiv
&&-\frac{1}{N_s\beta}\ln \int Dc Dc^+
\exp \{ \int_0^{\b}d\tau (\sum_{i\sigma} c_{i\sigma}^+(-\partial_{\tau}+\mu)c_{i\sigma}
-H_\a[c,c^+])+\\
&&+\int_0^{\b}\int_0^{\b}d\tau d\tau'\sum_{i\sigma}
\D (\tau-\tau') [G(\tau-\tau')-c_{i\sigma}^+(\tau)c_{i\sigma}(\tau')]\}
\end{eqnarray}
Inverting the relation $G=G_\a[\Delta]$ yields $\Delta=\D_\a[G]$, and a functional
of the local Green's function is obtained as $\G_\a[G] = \Omega_\a\left[G,\D_\a[G]\right]$.
This is the Legendre transform of the free energy with respect to the local source
$\Delta$.

I would like to emphasize that this construction is quite different from the
Baym-Kadanoff formalism, which considers a functional of all the components of the lattice
Green's function $G_{ij}$, not only of its local part $G_{ii}$.
The Baym-Kadanoff
approach also gives interesting insights into the DMFT
construction~\cite{georges_review_dmft,chitra_bk},
and will be considered at a later stage in these lectures.

Consider first the $\a=0$ case, in which the hamiltonian is purely local (atomic limit).
Then, we have
to consider a local problem defined by the action:
\begin{eqnarray}\nonumber
S_{imp}=
&&-\int^{\beta}_{0} d\tau \int^{\beta}_{0} d\tau' \sum_{\sigma}
c^{+}_{\sigma}(\tau)
\left[(-\partial_{\tau}+\mu)\delta(\tau-\tau')-\D_0(\tau-\tau')\right] c_{\sigma}(\tau')\\
&&+U\,\int^{\beta}_{0}d\tau\, n_{\uparrow}(\tau)n_{\downarrow}(\tau)
\label{Simp}
\end{eqnarray}
Hence, the local Green's function $G(\iomn)$ is represented as that of a quantum impurity problem
(an Anderson impurity problem in the context of the Hubbard model):
\be\label{Weiss_DMFT}
G\,=\,G_{imp}\,[\D_0]
\ee
As before, $\D_0$ plays the role of a Weiss field (analogous to the effective field for a magnet, or
to the
KS effective potential in DFT). Formally, this Weiss field specifies \cite{georges_kotliar_dmft}
the effective bare Green's function
of the impurity action (\ref{Simp}):
\be
{\cal G}_0^{-1}(\iomn) = \iomn+\mu-\D_0(\iomn)
\ee
There are however two important new aspects here:
\begin{itemize}
\item i) The Weiss function $\D_0$ is a {\it dynamical} (i.e frequency dependent) object. As a result
the local equivalent problem (\ref{Simp}) is not in Hamiltonian form but involves
retardation
\item ii) The equivalent local problem is not a one-body problem, but involves local interactions.
\end{itemize}
We note that, as in DFT, the explicit inversion of (\ref{Weiss_DMFT}) is not possible in general.
In practice, one needs a (numerical or approximate) technique to solve the quantum impurity problem
(an {\it ``impurity solver''}), and one can use an iterative procedure. Starting from some initial condition
for $\D_0$ (or $\cG0$) , one computes the interacting Green's function $G_{imp}$, and the associated self-energy
$\Sigma_{imp}\equiv {\cal G}_0^{-1}-G_{imp}^{-1}$. One then updates $\cG0$ as:
$\cG0^{new}=[\Sigma_{imp}+G^{-1}]^{-1}$, where $G$ is the specified value of the local Green's
function.

\subsubsection{Exact functional of the local Green's function}

We proceed with the construction of the exact functional of the local Green's function,
by coupling constant integration (starting from the atomic limit).

At $\a=0$ (decoupled sites, or infinitely separated atoms), we have
\footnote{In this formula and everywhere below, $\T\,$ denotes
$\frac{1}{\b}\sum_n$, with possibly a convergence factor~$e^{\iomn 0+}$.} :
\be
\Omega_0[\D_0,G]= F_{imp}[\D_0]-\T\,(G\D_0)
\ee
where $F_{imp}$ is the free energy of the local quantum impurity model viewed as
a functional of the hybridisation function. By formal inversion
$\D_0=\D_0[G]$:
\be
\G_0[G] = F_{imp}\left[\D_0[G]\right]-\T\,\left(G\D_0[G]\right)
\ee
We then observe that (since the $\a$-derivatives of the Lagrange multipliers do
not contribute because of the stationarity of $\Omega$):
\be
\frac{d\Gamma_\a}{d\a}\,=\,\bra H_{\mbox{non-loc}} \ket
\ee
which, for the Hubbard model, reduces to the kinetic energy:
\be
\frac{d\Gamma_\a}{d\a}\,=\,\bra \hat{T} \ket =
- \divnum\sum_{ij} t_{ij} \bra c^+_i c_j \ket |_G \,=\,
\T\,\divnum\sum_{\vk} \e_{\vk}\,G_{\a}(\vk,\iomn)|_G
\ee
In this expression, the lattice Green's function $G_{\a}(\vk,\iomn)$ should be
expressed, for a given $\a$, as a functional of the local Green's function $G$.

This leads to the following formal expression of the exact functional $\G[G]=\G_{\a=1}[G]$:
\beq
\label{exact_Gamma_G}
\G [G] \,=\,F_{imp}\left[\D_0[G]\right]-\T\,\left(G\D_0[G]\right) + {\cal T}[G]
\eeq
in which $\cT[G]$ is the kinetic energy functional (evaluated while keeping $G_{ii}=G$ fixed):
\beq
\cT[G]\,=\,\int_0^{1} d\a \divnum\sum_{ij} t_{ij} \bra c^+_i c_j \ket |_G
\,=\,
\int_0^{1} d\a \T\,\divnum\sum_{\vk} \e_{\vk}\,G_{\a}(\vk,\iomn)|_G
\eeq
The condition $\delta\Gamma/\delta G=0$ determines the actual value of the local Green's function
at equilibrium as (using $\delta\Gamma_0/\delta G = - \Delta_0$):
\beq
\label{eq:exact_scc}
\Delta_0[G(\iomn)] = \frac{\delta\cT[G]}{\delta G(\iomn)}
\eeq
We recall that the generalized Weiss function (hybridization) and $G$ are, by construction,
related by (\ref{Weiss_DMFT}):
\beq
\label{eq:imp_solver}
G=G_{imp}[\Delta_0]
\eeq
Equations (\ref{eq:exact_scc},\ref{eq:imp_solver}) (together with the definition of the
impurity model, Eq.~\ref{Simp})) are the key equations of dynamical mean-field theory,
viewed as an exact approach.
The cornerstone of this approach~\cite{georges_kotliar_dmft} is that, in order to obtain
the local Green's function,
one has to solve an impurity model (\ref{Simp}),
submitted to the self-consistency condition (\ref{eq:exact_scc}) relating the hybridization
function $\Delta_0$ to $G(\iomn)$ itself.
I emphasize that, since $\Gamma[G]$ is an exact functional, this construction is
completely general: it is
valid for the Hubbard model in arbitrary
dimensions and on an arbitrary lattice.

Naturally, using it in practice requires a concrete approximation
to the kinetic energy functional $\cT[G]$
(similarly, the DFT framework is only practical once an approximation to
$\Gamma_{xc}$ is used, for example the LDA).
The DMFT {\it approximation} usually employed is described below. In fact, it might be useful
to employ a different terminology and call ''local spectral density functional theory''
(or ``local impurity functional theory'') the exact framework, and DMFT the
subsequent approximation commonly made in $\cT[G]$.

\subsubsection{A simple case: the infinite connectivity Bethe lattice}

It is straightforward to see that the formal expression for the kinetic energy functional
$\cT[G]$ simplifies into a simple closed expression for
the Bethe lattice with connectivity $z$, in the limit $z\rightarrow\infty$.
In fact, a closed form can be given on an arbitrary lattice in the limit of
large dimensions, but this is a bit more tedious and we postpone it to the next
section.

In the limit of large connectivity, the hopping must be scaled as:
$t_{ij}=t/\sqrt{z}$ \cite{metzner_vollhardt}.
Expanding the kinetic energy functional in (\ref{exact_Gamma_G}) in powers of
$\a$, one sees that only the term of order $\a$ remains in the $z=\infty$
limit thanks to the tree-like geometry, namely:
\be
\a \sum_{ijkl} t_{ij}t_{kl} \bra c^+_i c_j c^+_k c_l\ket_{\a=0}
=\a \sum_{ij} t_{ij}^2 \T\, G^2 = \a (z N_s)\,\frac{t^2}{z}\, \T\, G^2
\ee
So that, integrating over $\a$, one obtains $\cT[G]=t^2\T\,G^2/2$ and finally:
\be
\label{Gamma_Bethe}
\Gamma_{Bethe, z=\infty}[G] = F_{imp}\left[\D_0[G]\right]-\T\,\left(G\D_0[G]\right) +
\frac{t^2}{2} \T\,G^2
\ee
This functional is similar (although different in details) to the one recently
used by Kotliar \cite{kotliar_landau_functional_mott} in a Landau analysis of the
Mott transition within DMFT.

The self-consistency condition (\ref{eq:exact_scc}) that finally determines both the local Green's function and
the Weiss field (through an iterative solution of the impurity model) thus reads in this case:
\be\label{sc_Bethe}
\D_0[G,\iomn] = t^2\,G(\iomn)
\ee

\subsubsection{DMFT as an approximation to the kinetic energy
functional.}

Now, I will show that the usual form of DMFT \cite{georges_review_dmft}
(for a general non-interacting dispersion $\e_k$)
corresponds to a very simple approximation of the kinetic energy term $\cT[G]$
in the exact functional $\Gamma[G]$.
Consider the one-particle Green's function $G_\a(\vk,\iomn)$ associated with the action
(\ref{eq:omegafull}) of the Hubbard model,
in the presence of the source term $\Delta_\alpha$ and for an arbitrary
coupling constant. We can define a self-energy associated with this Green's function:
\be\label{true_G_alpha}
G_\a(\vk,\iomn) = \frac{1}{\iomn+\mu-\D_\a[\iomn]-\a\e_{\vk}-
\Sigma_{\a}[\vk,\iomn]}
\ee
The self-energy $\Sigma_a$ is in general a $\vk$-dependent object, except obviously
for $\a=0$ in which all sites are decoupled into independent impurity models.
The DMFT approximation consists in replacing $\Sigma_a$ for arbitrary $\a$
by the impurity model self-energy
$\Sigma_0$ (hence depending only on frequency), at least for the purpose of
calculating $\cT[G]$. Hence:
\be\label{lattice_G_alpha}
G_\a(\vk,\iomn)|_{DMFT} = \frac{1}{\iomn+\mu-\D_\a[\iomn;G]-\a\e_{\vk}-
\Sigma_{\a=0}[\iomn,G]}
\ee
With:
\be\label{Sigma_alpha}
\Sigma_{\a=0}[G;\iomn]\equiv \cG0^{-1}-G^{-1}=\iomn+\mu-\D_0[\iomn,G]-G^{-1}
\ee
Summing over $\vk$, one then expresses the local Green's function in terms of the
hybridisation as:
\be
G(\iomn) = \int d\e \frac{D(\e)}{\zeta-\a\e} =
\frac{1}{\a} \widetilde{D}
\left(\frac{\zeta}{\a}\right)
\ee
With $\zeta\equiv \iomn+\mu-\D_\a -\Sigma_0=\D_0-\D_a+G^{-1}$.
In this expression, $D(\e)=\divnum\sum_{\vk}\delta(\e-\e_{\vk})$ is the non-interacting
density of states, and $\widetilde{D}(z) = \int d\e \frac{D(\e)}{z-\e}$ its Hilbert transform.
Introducing the inverse function such that $\widetilde{D}[R(g)]=g$, we can invert the relation
above to obtain the hybridisation function as a functional of the local $G$ for $U=0$:
\be
\label{inversion_U=0}
\D_\a[\iomn;G] = G^{-1}+\D_0[G] - \a R\left[\a G \right]
\ee
So that the lattice Green's function is also expressed as a functional of $G$ as:
\be
\label{eq:lattice_Green}
G_\a(\vk,\iomn) = \frac{1}{\a R(\a G)-\a\e_{\vk}}
\ee
Inserting this into (\ref{exact_Gamma_G}), we can evaluate the kinetic energy:
\be
\divnum\sum_{\vk} \e_{\vk}G_\a (\vk) =
\frac{1}{\a}\int d\e \frac{\e\,D(\e)}{R(\a G)-\e}
=\frac{1}{\a} \left[-1 + \a G R(\a G) \right]
\ee
and hence the DMFT approximation to $\cT[G]$:
\begin{equation}
\cT_{DMFT}[G] \,=\,\int_0^{1} d\a \T\,\left[ G(\iomn) R \left(\a G(\iomn)\right) -
\frac{1}{\a} \right]
\eeq
So that the total functional reads, in the DMFT approximation:
\begin{eqnarray}\nonumber
\Gamma_{DMFT}[G]\,=\,&&F_{imp}\left[\D_0[G]\right]-\T\,\left(G\D_0[G]\right) +\\
&&+\int_0^{1} d\a \T\,\left[ G(\iomn) R \left(\a G(\iomn)\right) -
\frac{1}{\a} \right]
\end{eqnarray}
In the case of an infinite-connectivity Bethe lattice, corresponding to a
semi-circular d.o.s of width $4t$, one has:
$R[g]=t^2g+1/g$, so that the result (\ref{Gamma_Bethe}) is recovered from this general
expression.
I note that the DMFT approximation to the functional $\cT[G]$
is completely independent of the interaction strength $U$.

The equilibrium condition (\ref{eq:exact_scc}) $\d\G/\d G=0$ thus reads
\footnote{When deriving this equation, it is useful to note that
$R(\a G)+\a G R'(\a G)= \partial_\a [\a R(\a G)]$.}
, in the DMFT approximation~\cite{georges_review_dmft}:
\be
\D_0[\iomn,G]|_{DMFT} = R[G(\iomn)]-\frac{1}{G(\iomn)}
\ee
This can be rewritten in a more familiar form, using (\ref{Sigma_alpha}):
\be
G(\iomn) = \int d\e \frac{D(\e)}{\iomn+\mu-\Sigma_{imp}(\iomn)}\,\,\,,\,\,\,
\mbox{with:}\,\, \Sigma_{imp} = \cG0^{-1}-G^{-1}
\ee
The self-consistency condition is equivalent to the condition $\D_{\a=1}[G]=0$,
as expected from the fact that $\D_{\a=1}=\d\G/\d G$. Hence, within the DMFT approximation,
the lattice Green's function is obtained by setting $\a=1$ into (\ref{lattice_G_alpha}):
\be
G(\vk,\iomn)|_{DMFT} = \frac{1}{\iomn+\mu-\e_{\vk}-\Sigma_{imp}(\iomn)}
\ee

\subsection{The Baym-Kadanoff viewpoint}
\label{sec:BK}

Finally, let me briefly mention that the DMFT approximation can also be formulated
using the more familiar Baym-Kadanoff functional. In contrast to the previous
section, this is a functional of {\it all components} $G_{ij}$ of the lattice
Green's function, not only of the local one $G_{ii}$.
The Baym-Kadanoff functional is defined as:
\begin{equation}
\Omega_{BK}[G_{ij},\Sigma_{ij}]\,=\,
-\t\ln\left[(\iomn+\mu)\delta_{ij}-t_{ij}-\Sigma_{ij}(\iomn)\right]-\t[\Sigma\cdot\,G]+
\Phi_{LW}[\{G_{ij}\}]
\label{eq:omega_BK}
\end{equation}
Variation with respect to $\Sigma_{ij}$ yields the usual Dyson's equation relating
the Green's function and the self-energy. The Luttinger-Ward functional $\Phi_{LW}$
has a simple diagrammatic definition as the sum of all skeleton diagrams in the
free-energy. Variation with respect to $G_{ij}$ express the self-energy as a total derivative
of this functional:
\begin{equation}
\Sigma_{ij}(i \omega_n)=\frac{\delta\Phi}{\delta G_{ij}(i\omega_n)}
\label{eq:sigma_lw}
\end{equation}
The DMFT approximation amounts to approximate the Luttinger-Ward functional
by a functional which is the sum of that of {\it independent atoms}, retaining only
the dependence over the local Green's function, namely:
\begin{equation}
\Phi^{DMFT}_{LW} = \sum_i \Phi_{imp}\,[G_{ii}]
\label{eq:phi_dmft}
\end{equation}
An obvious consequence is that the self-energy is site-diagonal:
\begin{equation}
\Sigma_{ij}(\iomn)=\delta_{ij}\,\Sigma(\iomn)
\end{equation}
Eliminating $\Sigma_{ii}$ amounts to do a Legendre transformation
with respect to $G_{ii}$, and therfore leads to a different expression of the
local DMFT functional introduced in the previous section~\cite{chitra_local}:
\beq
\Gamma_{DMFT}[G_{ii}]\,=\,
-\t\ln\left[(\iomn+\mu-\frac{\delta\Phi_{imp}}{\delta G_{ii}})\delta_{ij}-t_{ij}\right]
-\t[\frac{\delta\Phi}{\delta G_{ii}}\cdot\,G_{ii}]+\sum_i \Phi_{imp}\,[G_{ii}]
\eeq
The Baym-Kadanoff formalism is useful for total energy calculations, and will
be used in Sec.~\ref{sec:energy_lda+dmft}.

%
%
%



%
%
%
%
%
%
\section{The Mott metal-insulator transition}
\label{sec:mott}

\subsection{Materials on the verge of the Mott transition}

Interactions between electrons can be responsible for the
insulating character of a material, as realized early on by Mott~\cite{mott_1949,mott_mit_book}.
The Mott mechanism plays a key role in the physics of strongly
correlated electron materials.
Outstanding examples \cite{mott_mit_book,imada_mit_review} are
transition-metal oxides (e.g superconducting cuprates),
fullerene compounds, as well as organic conductors\footnote{The Mott phenomenon may also be
partly responsible for the localization of
f-electrons in some rare earth and actinides {\it metals},
see \cite{johansson_1974_mott_philmag,skriver_actinides_1978,skriver_americium_1980,
savrasov_kotliar_pu_prl_2000,savrasov_kotliar_pu_nature_2001}
and \cite{wills_losalamos_2000,kotliar_savrasov_mott_felectrons_2003} for recent reviews.}.
Fig.~\ref{fig:fujimori_map} illustrates this in the case of transition metal oxides with
perovskite structure ABO$_3$~\cite{fujimori_diagram_1992}.
\begin{center}
\begin{figure}
\includegraphics[width=15cm]{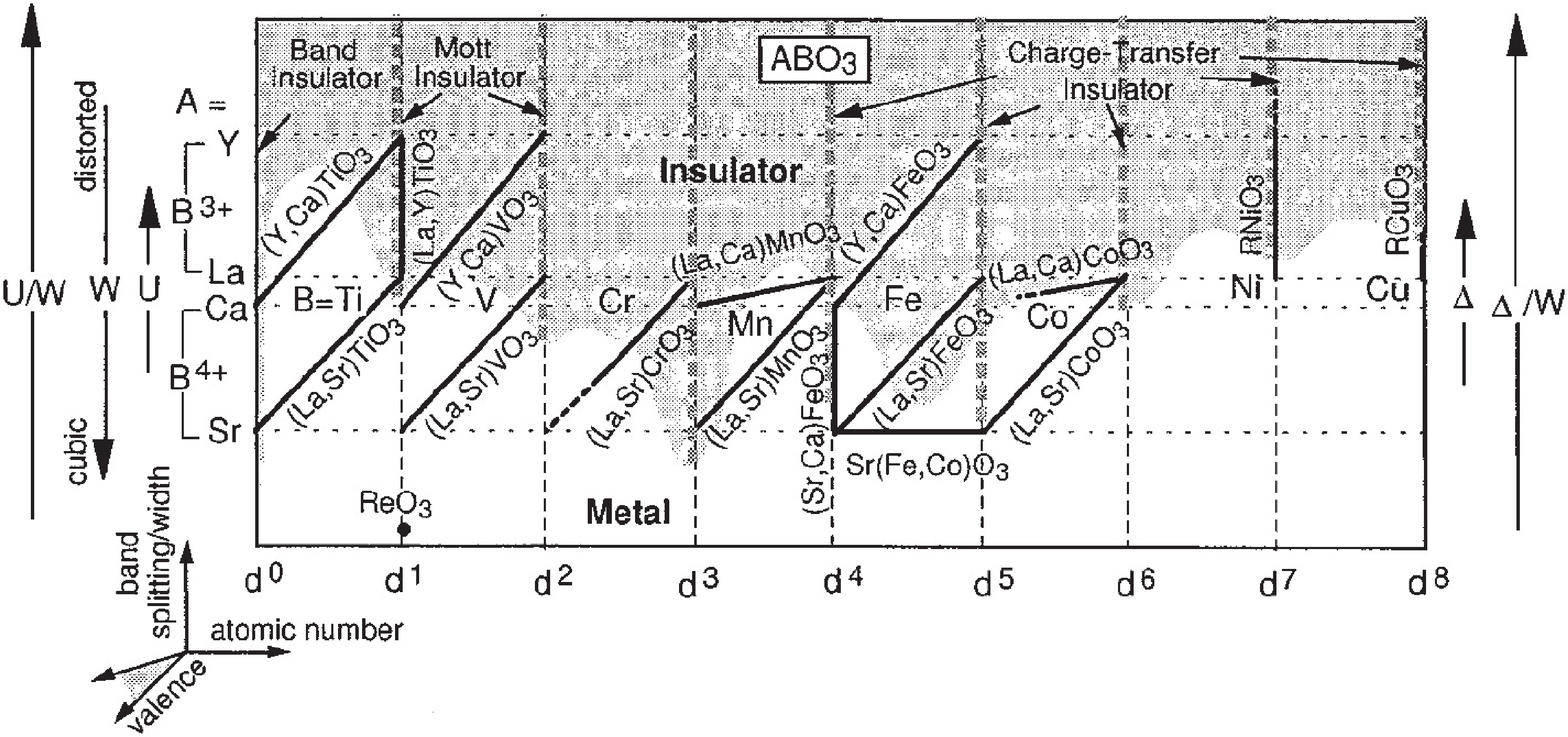}
\caption{This diagram (due to A.Fujimori~\cite{fujimori_diagram_1992}, see also
\cite{imada_mit_review}) can be viewed as a
map of the vast territory of transition- metal compounds with perovskite structure ABO$_3$.
Varying the transition metal ion B corresponds to gradual filling of the 3d-shell.
Different substitutions on the A-site can be made (A= Sr,Ca and A=La,Y are mainly considered in
this diagram). This allows to change either the valence of the transition metal ion (doping), or
the structural parameters in an isoelectronic manner. The shaded region corresponds to insulating
compounds, while the unshaded one corresponds to metals. This illustrates the key role of the
Mott phenomenon in the physics of transition-metal oxides.
}
\label{fig:fujimori_map}
\end{figure}
\end{center}

A limited number of materials are poised right on the verge of this electronic
instability. This is the case, for example, of V$_2$O$_3$, NiS$_{2-x}$Se$_x$ and of quasi two-dimensional
organic conductors of the $\kappa$-BEDT family. These materials are particularly
interesting for the fundamental investigation of the Mott transition, since they offer the
possibility of going from one phase to the other by varying some external parameter
(e.g chemical composition,temperature, pressure,...).
Varying external pressure is definitely
a tool of choice since it allows to sweep continuously from the insulating phase
to the metallic phase (and back). The phase diagrams of (V$_{1-x}$ Cr$_{x}$)$_2$O$_3$ and
of $\kappa$-(BEDT-TTF)$_{2}$Cu[N(CN)$_{2}$]Cl under pressure are displayed
in Fig.~\ref{fig:materials}.
\begin{figure}
\begin{tabular}{cc}
\includegraphics[width=8 cm]{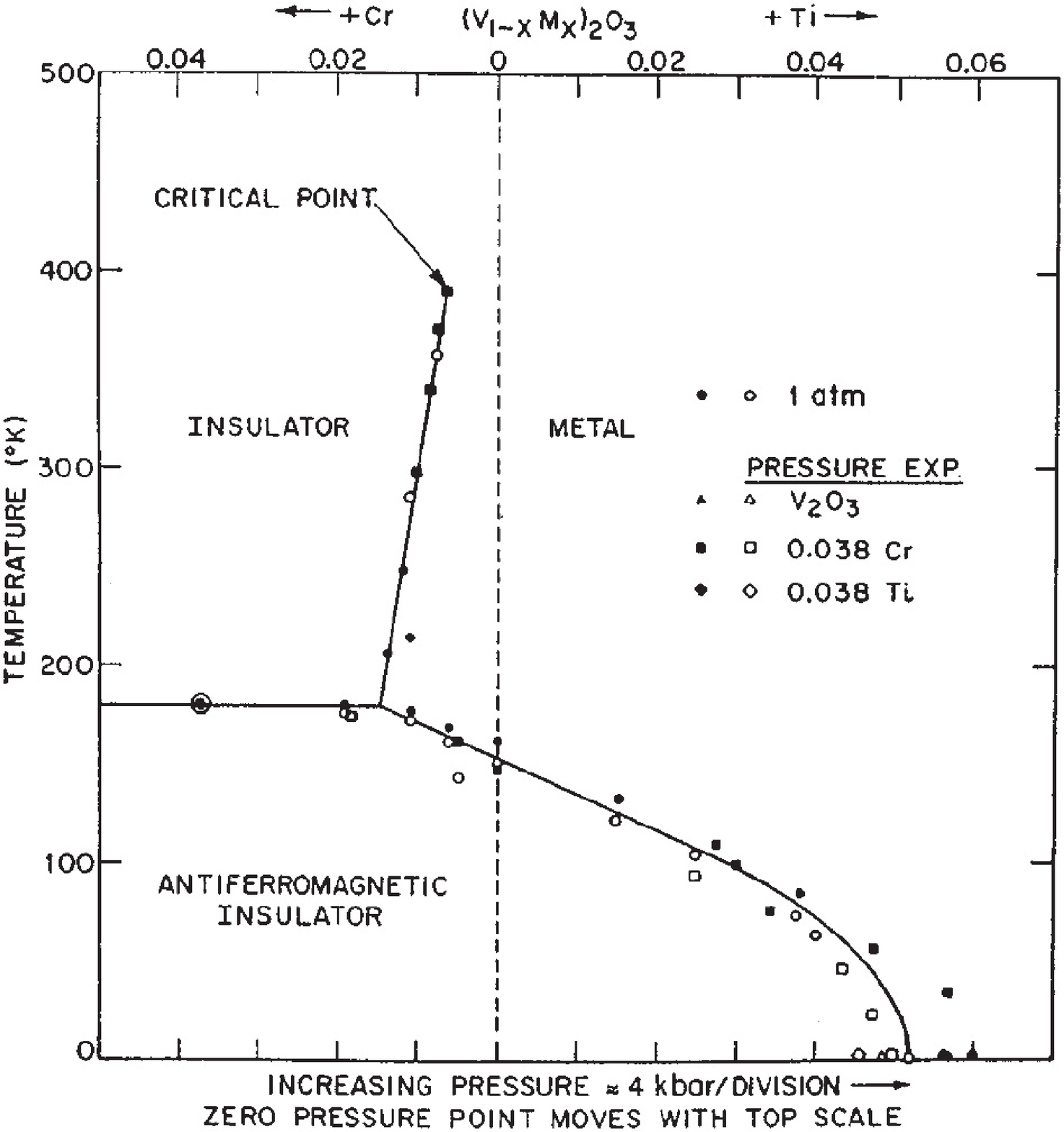} &
\includegraphics[width=8 cm]{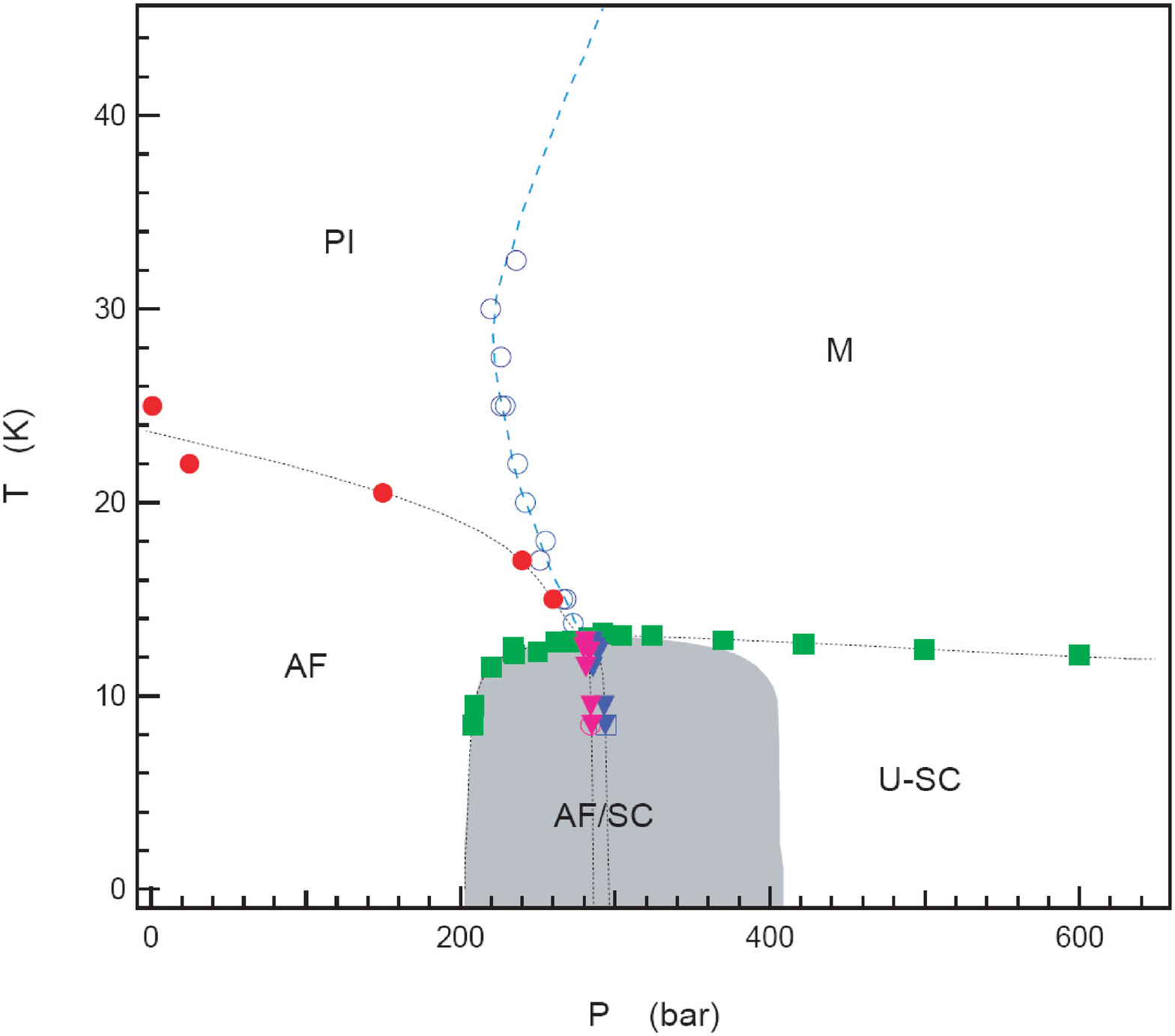}
\end{tabular}
\caption{
Left: Phase diagram of (V$_{1-x}$ Cr$_{x}$)$_2$O$_3$ as a
function of either Cr-concentration $x$ or pressure (after\cite{McWhan73}).
Increasing $x$ by $1\%$
produces similar effects than {\it decreasing} pressure by $\sim 4$kbar, for this material.
Right: Phase diagram of $\kappa$-(BEDT-TTF)$_{2}$Cu[N(CN)$_{2}$]Cl as a function of
pressure (after \cite{Lef00}).
}
\label{fig:materials}
\end{figure}
There is a great similarity between the high-temperature part of the phase diagrams of these materials,
despite very different energy scales. At low-pressure they are {\it paramagnetic} Mott insulators, which
are turned into metals as pressure is increased. Above a critical temperature $T_c$
(of order $\sim 450$K for the oxide compound and $\sim 40$K for the organic one), this
corresponds to a smooth crossover. In contrast, for $T<T_c$ a first-order transition is observed, with a
discontinuity of all physical observables (e.g resistivity). The first order transition line
ends in a second order critical endpoint at $(T_c,P_c)$. We observe that in both cases, the
critical temperature is a very small fraction of the bare electronic
energy scales (for \v2o3 the half-bandwidth is of order $0.5-1$~eV, while it is of
order $2000$~K for the organics).

There are also some common features between the low-temperature part of the phase diagram
of these compounds, such as the fact that the paramagnetic Mott insulator orders into an
antiferromagnet as temperature is lowered. However, there are also striking differences: the
metallic phase has a superconducting instability for the organics, while this is not the case for
V$_2$O$_3$. Also, the magnetic transition is only superficially similar\,: in the case of
\v2o3, it is widely believed to be accompanied (or even triggered) by orbital
ordering\cite{bao_v2o3} (in contrast to NiS$_{2-x}$Se$_x$\cite{kotliar_V2O3_NiS}), and
as a result the transition is first-order. In general, there is a higher degree of universality
associated with the vicinity of the Mott critical endpoint than in the low-temperature region,
in which long-range order takes place in a material- specific manner.

Mott localization into a paramagnetic insulator implies a high spin entropy, which must therefore
be quenched in some way as temperature is lowered. An obvious possibility is magnetic
ordering, as in these two materials. In fact, a Mott transition between a paramagnetic Mott insulator
and a metallic phase is only observed in those compounds where magnetism is sufficiently {\it frustrated}
so that the transition is not preempted by magnetic ordering. This is indeed the case in both
compounds discussed here: \v2o3 has competing ferromagnetic and antiferromagnetic exchange constants,
while the two-dimensional layers in the organics have a triangular structure.
Another possibility is that the entropy is quenched through a Peierls instability (dimerization),
in which case the Mott insulator can remain paramagnetic (this is the case, for example, of VO$_2$).
Whether it is possible to stabilize a paramagnetic Mott insulator down to $T=0$ without breaking
spin or translational symmetries is a fascinating problem, both theoretically and from the
materials point of view
(for a recent review on resonating valence bond phases in frustrated quantum magnets, see
e.g~\cite{misguich_review} and \cite{sachdev_mott_review_2004}).
The compound
$\kappa$-(BEDT-TTF)$_2$Cu$_2$(CN)$_3$ may offer~\cite{shimizu_spinliquid}
a realization of such a spin-liquid
state (presumably through a combination of strong frustration and
strong charge fluctuations \cite{imada_2003_spinliquid}),
but this behaviour is certainly more the exception than the rule.

\subsection{Dynamical mean-field theory of the Mott transition}

Over the last decade, a detailed theory of the strongly correlated metallic
state, and of the Mott transition itself has emerged,
based on the
{\it dynamical mean-field theory} (DMFT). We refer to \cite{georges_review_dmft} for a review and
an extensive list of original
references~\cite{georges_kotliar_dmft,rozenberg_mott_qmc,georges_krauth_mott_prl,georges_krauth_mott_prb,
rozenberg_mott_prb,laloux_mott_prb}
We now review some key features of this theory.

\paragraph{Quasiparticle coherence scale}
In the metallic state, Fermi-liquid theory applies below a low energy scale $\est$,
which can be interpreted as the coherence-scale for quasiparticles
(i.e long-lived quasiparticles exist only for energies and temperature smaller than
$\est$). This low-energy coherence scale
is given by $\est\sim Z D$ (with $D$ the half-bandwith, also equal to the Fermi energy
of the non-interacting system at half-filling) where $Z$ is the quasiparticle weight.
In the strongly correlated metal close to the transition, $Z\ll 1$, so that $\est$ is
strongly reduced as compared to the bare Fermi energy.

\paragraph{Three peaks in the d.o.s: Hubbard bands and quasiparticles}
In addition to low-energy quasiparticles (carrying a fraction $Z$ of the spectral weight),
the one-particle spectrum of
the strongly correlated metal contains high-energy excitations
carrying a spectral weight $1-Z$. These are associated to the atomic-like transitions
corresponding to the addition or removal of one electron on an atomic site, which
broaden into Hubbard bands in the solid. As a result, the $\vk$-integrated spectral function
$A(\omega)=\sum_\vk A(\vk,\omega)$ (density of states d.o.s) of the strongly correlated metal
is predicted \cite{georges_kotliar_dmft}
to display a three-peak structure, made of a quasiparticle band close to the Fermi energy surrounded
by lower and upper Hubbard bands (Fig.~\ref{fig:spectra_ipt} and inset of
Fig.~\ref{fig:res_and_dos}).
The quasiparticle part of the d.o.s has a reduced width of
order $ZD\sim\est$. The lower and upper Hubbard bands are separated by
an energy scale $\Delta$.
\begin{center}
\begin{figure}
\includegraphics[width=10 cm]{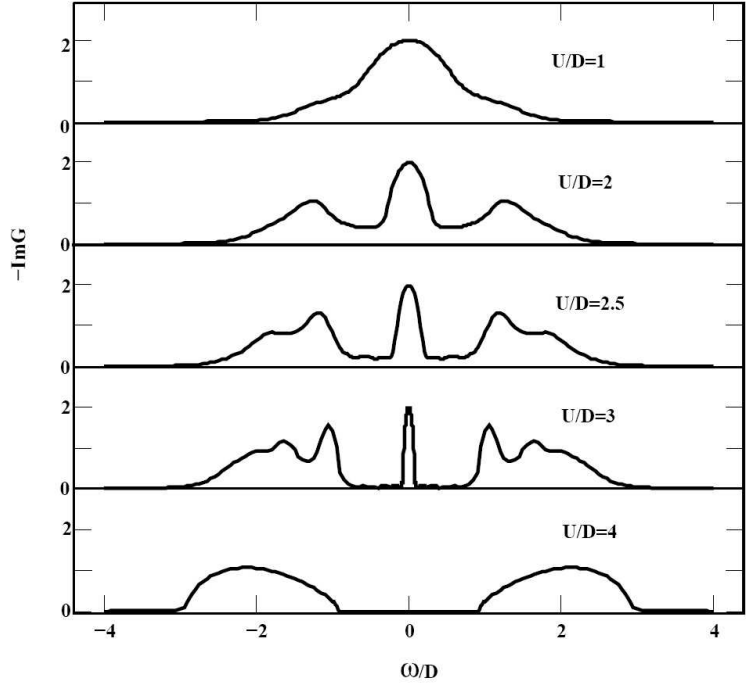}
\caption{Local spectral function for several values of the interaction strength in DMFT.
These results have been obtained using the IPT approximation, for the half-filled
Hubbard model with a semi-circular d.o.s (from Ref.~\cite{georges_review_dmft}).
Close to the transition, the separation of scales between the quasiparticle coherence
energy ($\est$) and the distance between Hubbard bands ($\Delta\,$) is clearly seen.}
\label{fig:spectra_ipt}
\end{figure}
\end{center}
\paragraph{The insulating phase: local moments, magnetism and frustration}
At strong enough coupling (see below), the paramagnetic solution of the DMFT equations
is a Mott insulator, with a gap $\Delta$ in the one-particle spectrum. This phase is characterized
by unscreened local moments,
associated with a Curie law for the local susceptibility $\sum_q\chi_q \propto 1/T$, and
an extensive entropy. Note however that the uniform susceptiblity $\chi_{q=0}$ is finite, of
order $1/J\sim U/D^2$. As temperature is lowered, these local moments order
into an antiferromagnetic phase~\cite{jarrell_1992_qmc_prl,georges_krauth_mott_prb}.
The N\'eel temperature is however strongly dependent on frustration~\cite{georges_review_dmft}
(e.g on the ratio $t'/t$ between the next nearest-neighbour and nearest-neighbour hoppings) and can
be made vanishingly small for fully frustrated models.

\paragraph{Separation of energy scales, spinodals and transition line}

Within DMFT, a separation of energy scales holds close to the Mott transition. The mean-field
solution corresponding to the paramagnetic metal at $T=0$ disappears at a critical coupling
$U_{c2}$. At this point, the quasiparticle weight vanishes ($Z\propto 1-U/U_{c2}$) as in
Brinkman-Rice theory\footnote{Since the self-energy only depends on frequency within DMFT, this
also implies that quasiparticles become heavy close to the transition, with $m^\star/m=1/Z$.
In real materials, we expect however that magnetic exchange will quench out
the spin entropy associated with local moments, resulting in a saturation of the
effective mass close to the Mott transition.
In the regime where $\est\ll J$, the effective mass is then expected to be of order $J$, as
found e.g in slave-boson theories. Describing this effect requires extensions of DMFT in
order to deal with short-range spatial correlations}
On the other hand, a mean-field insulating solution is found
for $U>U_{c1}$, with the Mott gap $\Delta$
opening up at this critical coupling (Mott-Hubbard transition).
As a result, $\Delta$ is a finite energy scale for $U=U_{c2}$ and
the quasiparticle peak in the d.o.s is well separated from the Hubbard bands
in the strongly correlated metal.

These two critical couplings extend at finite temperature into two spinodal lines
$U_{c1}(T)$ and $U_{c2}(T)$, which delimit a region of the $(U/D,T/D)$ parameter space
in which two mean-field solutions (insulating and
metallic) are found (Fig.~\ref{fig:phasediag}). Hence, within DMFT, a first-order
Mott transition occurs at finite temperature even in a purely electronic model.
The corresponding critical temperature $T_c^{el}$ is of order $T_c^{el}\sim \Delta E/\Delta S$, with
$\Delta E$ and $\Delta S\sim \ln (2S+1)$ the energy and entropy differences between the metal
and the insulator. Because the energy difference is small ($\Delta E\sim (U_{c2}-U_{c1})^2/D$), the
critical temperature is much lower than $D$ and $U_c$ (by almost two orders of magnitude).
Indeed, in V$_2$O$_3$ as well as in the organics, the critical temperature corresponding to the
endpoint of the first-order Mott transition line is a factor of 50 to 100 smaller than the
bare electronic bandwith.
\begin{figure}
\includegraphics[width=8 cm]{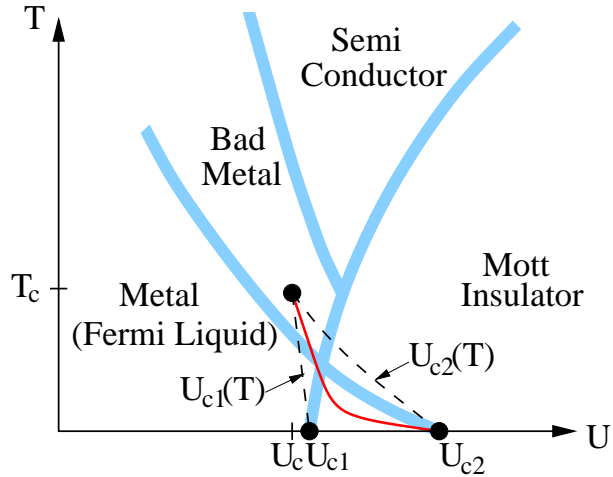}
\caption{
Paramagnetic phases of the Hubbard model within DMFT,
displaying schematically the spinodal lines of the Mott insulating and
metallic mean-field solutions (dashed),
the first-order transition line (plain) and the critical endpoint. The shaded crossover
lines separating the different transport regimes discussed in Sec.3 are
also shown. The Fermi-liquid to ``bad metal'' crossover line corresponds to the
quasiparticle coherence scale and is a continuation of the spinodal $U_{c2}(T)$
above $T_c$. The crossover into the insulating state corresponds to the continuation
of the $U_{c1}$ spinodal.
Magnetic phases are not displayed and depend on the degree of frustration.
Figure from Refs.~\cite{florens_phd} and \cite{georges_mott_iscom}.}
\label{fig:phasediag}
\end{figure}

\subsection{Physical properties of the correlated metallic state: DMFT confronts experiments}

\subsubsection{Three peaks: evidence from photoemission}

In Fig.~\ref{fig:pes_oxides}, we reproduce the early photoemission spectra of some
$d^1$ transition metal oxides, from the pioneering work of
Fujimori and coworkers~\cite{fujimori_pes_oxides}.
This work established experimentally, more than ten years ago, the existence of
well-formed (lower) Hubbard bands in correlated metals, in addition to low-energy
quasiparticles.
\begin{figure}
\includegraphics[width=9 cm]{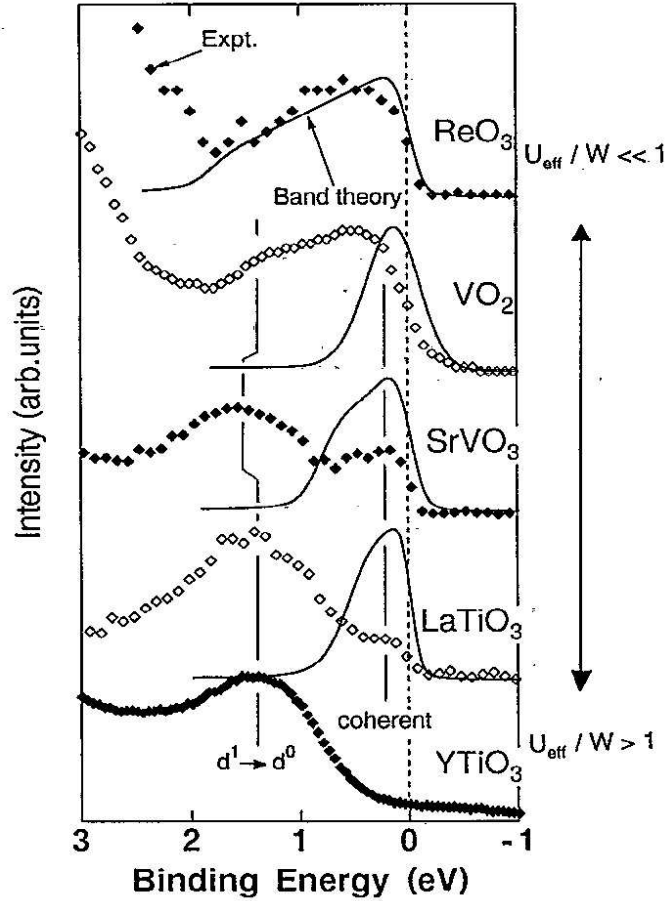}
\caption{Photoemission spectra of several $d^1$ transition metal oxides, reproduced
from Ref.~\cite{fujimori_pes_oxides}. The effects of correlations increases from ReO$_3$ (a weakly
correlated metal) to YTiO$_3$ (a Mott insulator). The plain lines are the d.o.s obtained from
band structure calculations. A lower Hubbard band around $-1.5$ eV is clearly visible in the
most correlated materials, both in the metallic and insulating case.
}
\label{fig:pes_oxides}
\end{figure}
This experimental study and the theoretical prediction of a 3-peak structure
from DMFT~\cite{georges_kotliar_dmft} came independently around the same time.
However, back in 1992, the existence of a narrow quasiparticle peak in $A(\omega)$
resembling the DMFT results was, to say the least, not obvious from these early data.
Further studies~\cite{inoue_casrvo3_1995_prl} on \casr therefore aimed at studying
the dependence of low-energy quasiparticle spectral features upon the degree of
correlations.
One of the main difficulty raised by these photoemission results
is that the weight $Z$ of the low-energy quasiparticle peak estimated from these early data
is quite small (particularly for CaVO$_3$), while specific heat measurements do not
reveal a dramatic mass enhancement.
This triggered
some discussion~\cite{inoue_casrvo3_1995_prl,rozenberg_casrvo3_1996_prl,imada_mit_review}
about the possibility of a strong $k$-dependence of the self-energy.
A decisive insight into this question came from
further experimental developments by Maiti and coworkers~\cite{maiti_2001,maiti_phd}
in which it was demonstrated
that the photoemission spectra are actually quite sensitive to the photon energy. Studies
at different photon energies allowed these authors to extract the estimated spectra
corresponding to the
bulk and the surface of the material. Surface and bulk spectra were found to be very different
indeed: the surface of CaVO$_3$ being apparently insulating-like while the bulk spectrum
did show a much more pronounced quasiparticle peak. Very recently, high resolution, high-photon energy
photoemission studies~\cite{sekiyama_casrvo3,sekiyama_casrvo3_2} clarified considerably this
issue. The high photon-energy spectrum reproduced on Fig.~\ref{fig:pes_casrvo3} displays a
clear quasiparticle d.o.s at low-energy (with a weight in good agreement with $m/m^\star$ and
a height comparable to the LDA d.o.s), as well as a lower Hubbard band carrying the rest of the
spectral weight. Moreover,
recent calculations~\cite{nekrasov_casrvo3_condmat1,pavarini_d1oxides_prl,sekiyama_casrvo3_2}
combining electronic structure
methods and DMFT (see next section)
compare favorably to the experimental spectra, on a quantitative level.
\begin{figure}
\includegraphics[width=8 cm]{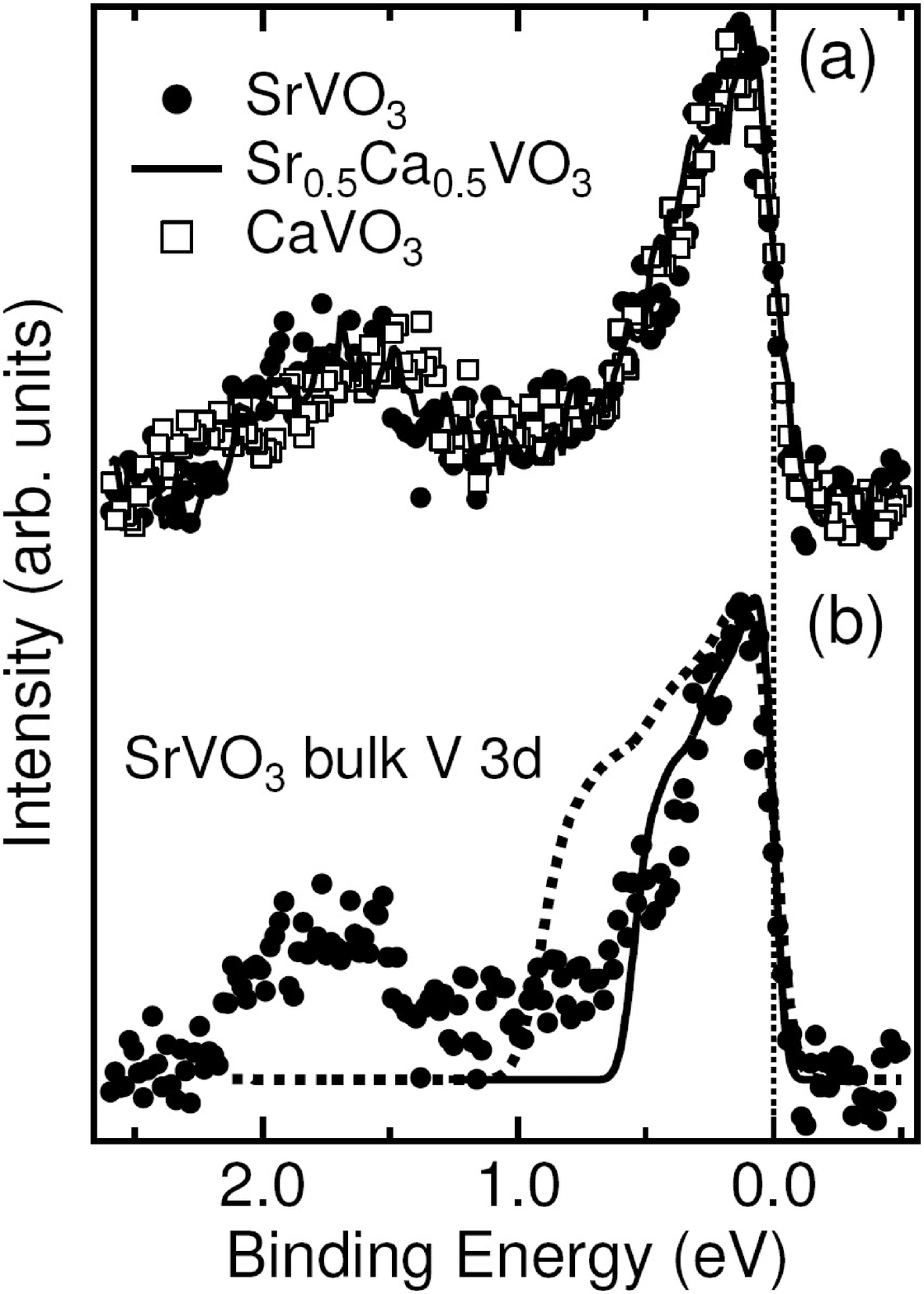}
\caption{(a) Bulk V $3d$ spectral functions of SrVO$_3$
(closed circles), Sr$_{0.5}$Ca$_{0.5}$VO$_3$ (solid line) and CaVO$_3$
(open squares).
(b) Comparison of the experimentally obtained bulk V $3d$ spectral
function of SrVO$_3$ (closed circles) to the V $3d$ partial density
of states for SrVO$_3$ (dashed curve) obtained
from the band-structure calculation,
which has been broadened by the experimental resolution of 140 meV.
The solid curve shows the same V $3d$ partial density of states
but the energy is scaled down by a factor of 0.6. Figure and caption
from Ref.~\cite{sekiyama_casrvo3} (see also \cite{sekiyama_casrvo3_2}).
}
\label{fig:pes_casrvo3}
\end{figure}

In the case of NiS$_{2-x}$Se$_x\,$, angular resolved photoemission have revealed
a clear quasiparticle peak, with strong spectral weight redistributions as a function of
temperature~\cite{matsuura_NiSSe_pes}. For the metallic phase of \v2o3, high photon energy photoemission proved
to be an essential tool in the recent experimental finding of the
quasiparticle peak (Fig.~\ref{fig:pes_v2o3}) by Mo et al.~\cite{mo_V2O3_prominent_peak}.
\begin{figure}
\includegraphics[width=9 cm]{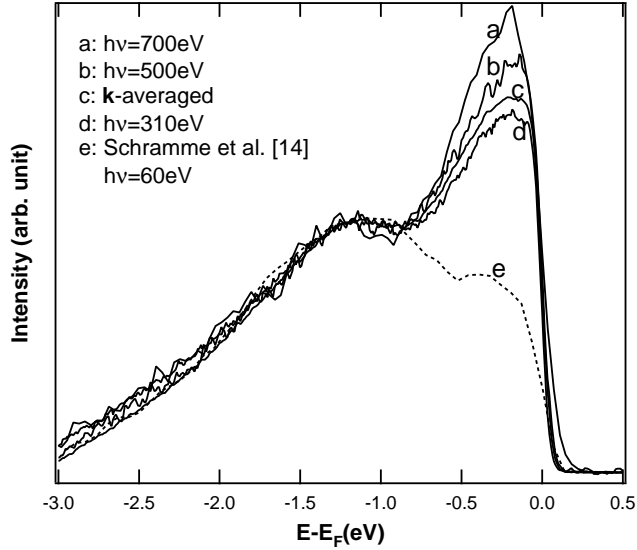}
\caption{Photoemission spectra of \v2o3, for various photon energies, from
Ref.~\cite{mo_V2O3_prominent_peak}. The highest photon energy spectrum, corresponding
to the greatest bulk sensitivity, reveals a prominent quasiparticle peak.
}
\label{fig:pes_v2o3}
\end{figure}

\subsubsection{Spectral weight transfers}

The quasiparticle peak in the d.o.s is characterized by an extreme sensitivity to
changes of temperature, as shown in the inset of Fig.~\ref{fig:res_and_dos}.
Its height is strongly reduced as $T$ is increased, and the peak
disappears altogether as $T$ reaches $\est$, leaving a pseudogap at the Fermi energy. Indeed,
above $\est$, long-lived coherent quasiparticles no longer exist. The corresponding spectral
weight is redistributed over a very large range of energies, of order $U$
(hence much larger than temperature itself).
This is reminiscent of Kondo systems \cite{liu_allen_cerium_spectro},
and indeed DMFT establishes a formal and physical connection \cite{georges_kotliar_dmft} between
a metal close to the Mott transition and the Kondo problem. The local moment present at short time-scales
is screened through a self-consistent Kondo process involving the
low-energy part of the (single- component) electronic fluid itself.

These spectral weight
transfers and redistributions are a distinctive feature of strongly correlated systems.
As already mentioned, they have been observed in the photoemission spectra
of \nisse. They are also commonly observed in optical spectroscopy of correlated materials,
as shown on Fig.~\ref{fig:exp_optics} for metallic \v2o3 \cite{rozenberg_optics_prl}
and the $\kappa$-BEDT organics \cite{eldridge_optics_bedt}.
DMFT calculations give a good description of the optical spectral weight
transfers for these materials, at least on a qualitative
level~\cite{rozenberg_optics_prl,Mer00}.
\begin{figure}
\begin{tabular}{cc}
\includegraphics[width=8 cm]{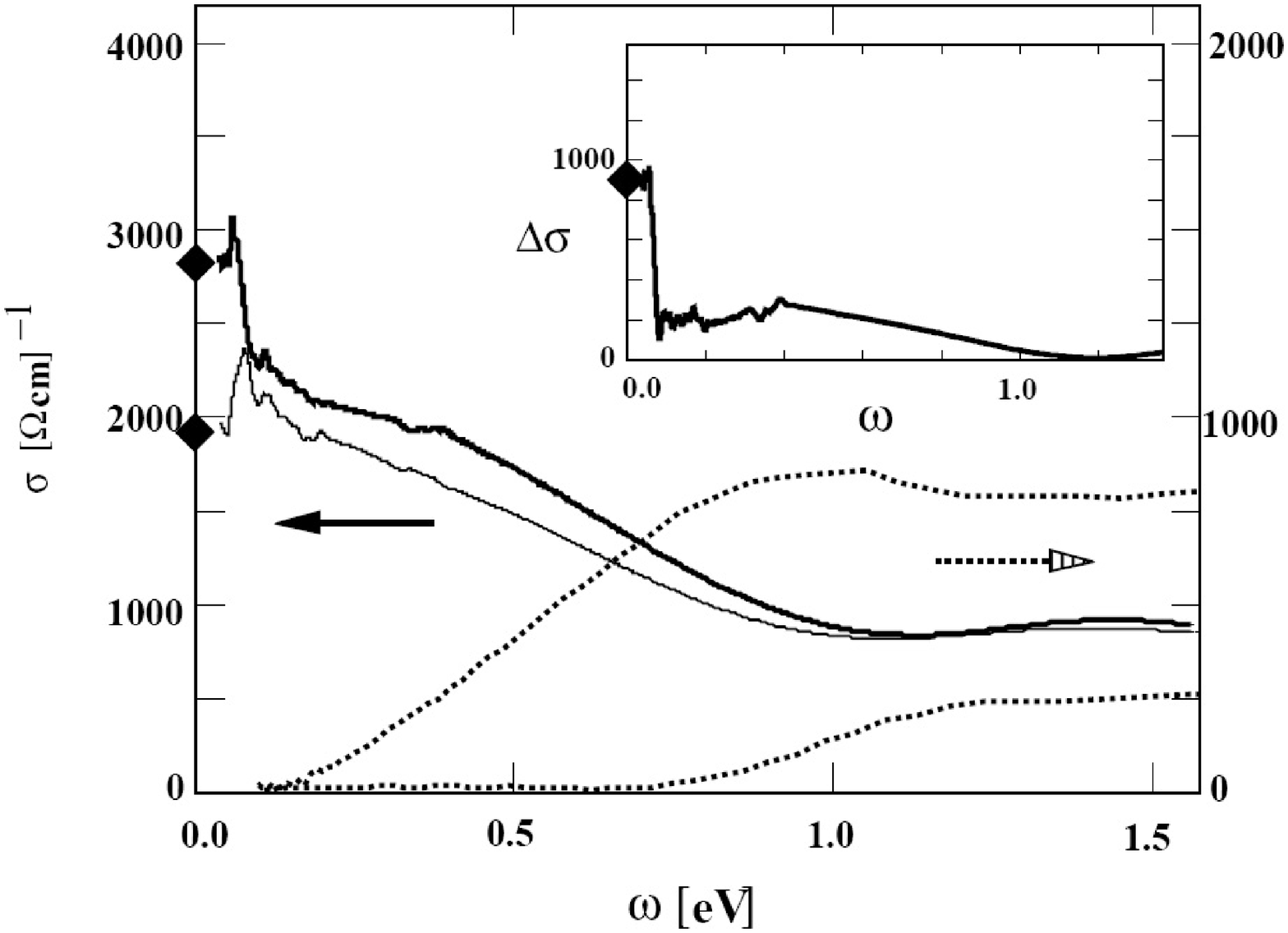} &
\includegraphics[width=7 cm]{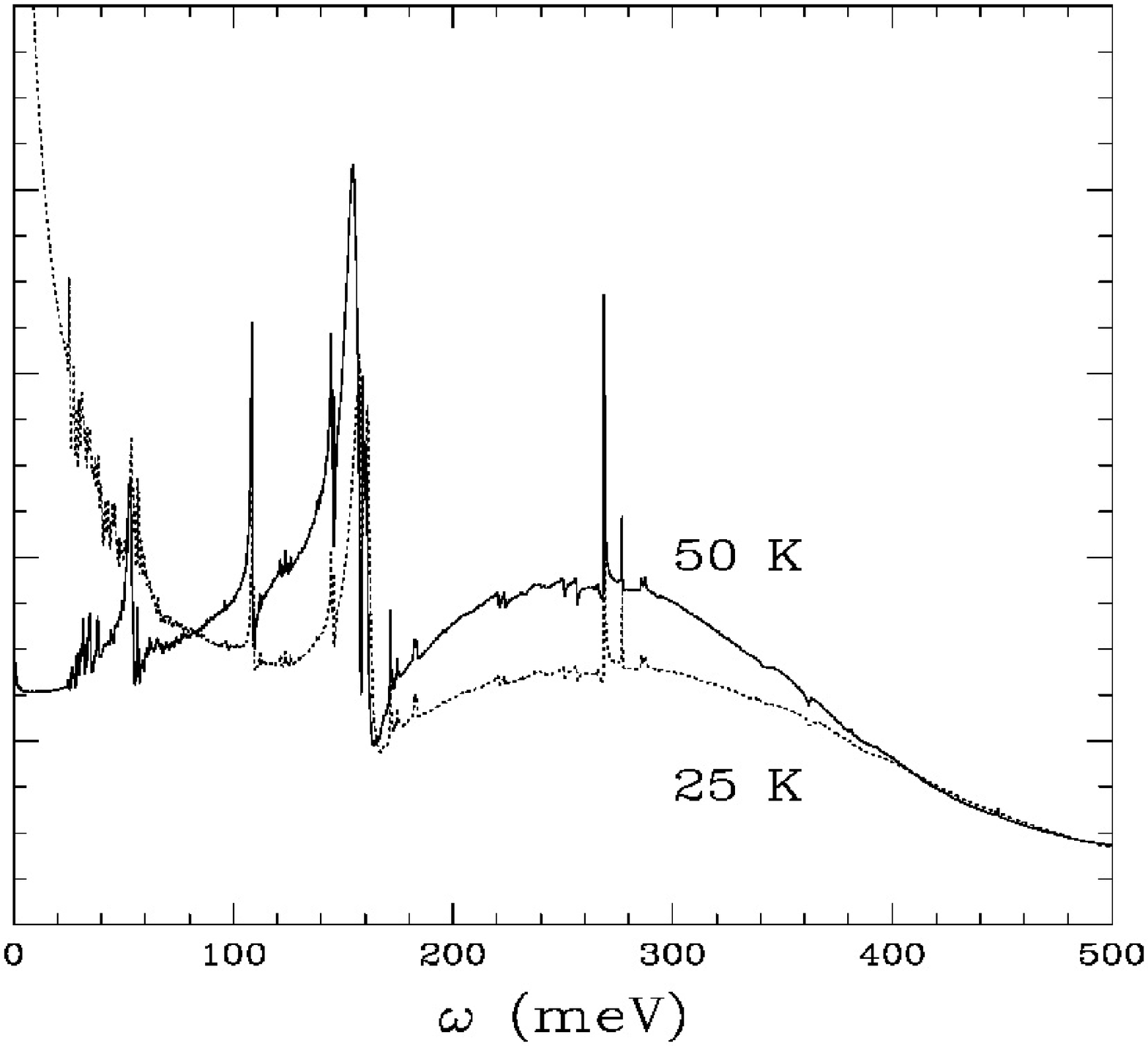}
\end{tabular}
\caption{Left: Optical conductivity of metallic
$V_2O_3$~\cite{rozenberg_optics_prl}
at $T=170K$ (thick line) and $T=300K$ (thin line)).
The inset contains the difference of the two spectra
$\Delta\sigma(\omega)=\sigma_{170K}(\omega)-\sigma_{300K}(\omega)$.
Diamonds indicate the measured dc conductivity $\sigma_{dc}$.
Dotted lines are for the insulating compounds
$V_{2-y}O_3$ with $y=.013$ at $10K$ (upper)
and $y=0$ at $70K$ (lower).
Right: Optical conductivity of $\kappa$-(BEDT-TTF)$_2$Cu[N(CN)$_2$]Br at
ambiant pressure~\cite{eldridge_optics_bedt}, for $T=25$K and $T=50$K.
For both materials, transfer of spectral weight from high energies to the Drude peak
is clearly visible as temperature is lowered.}
\label{fig:exp_optics}
\end{figure}

\subsubsection{Transport regimes and crossovers}

The disappearance of coherent quasiparticles, and associated spectral weight transfers, results
in three distinct transport
regimes~\cite{rozenberg_optics_prl,majumdar_transport,Mer00,limelette_bedt_prl,georges_mott_iscom}
for a correlated metal close to the Mott transition,
within DMFT (Figs.~\ref{fig:phasediag} and \ref{fig:res_and_dos}):

\begin{itemize}

\item In the {\it Fermi-liquid regime} $T\ll \est$, the resistivity obeys a $T^2$
law with an enhanced prefactor:
$\rho = \rho_M\,(T/\est)^2$. In this expression, $\rho_M$
is the Mott-Ioffe-Regel resistivity $\rho_M \propto ha/e^2$ corresponding
to a mean-free path of the order of a single lattice spacing in a Drude picture.

\item For $T\sim\est$, an {\it ``incoherent'' (or ``bad'') metal} regime is entered.
The quasiparticle lifetime shortens dramatically, and the quasiparticle peak is
strongly suppressed (but still present). In this regime, the resistivity is metallic-like (i.e
increases with $T$) but reaches values considerably larger than the Mott ``limit'' $\rho_M$.
A Drude description is no longer applicable in this regime.

\item Finally, for $\est\ll T \ll \Delta$, quasiparticles are gone altogether and the d.o.s
displays a pseudogap associated with the scale $\Delta$ and filled with thermal
excitations. This yields an insulating-like regime of transport,
with the resistivity decreasing upon heating ($d\rho/dT<0$). At very low
temperature, the resistivity follows an activated behaviour, but deviations from
a pure activation law are observed at higher temperature (these two regimes are
depicted as the ``insulating'' and ``semi-conducting'' ones on Fig.~\ref{fig:phasediag}).

\end{itemize}
\begin{figure}
\begin{tabular}{cc}
\includegraphics[width=8 cm]{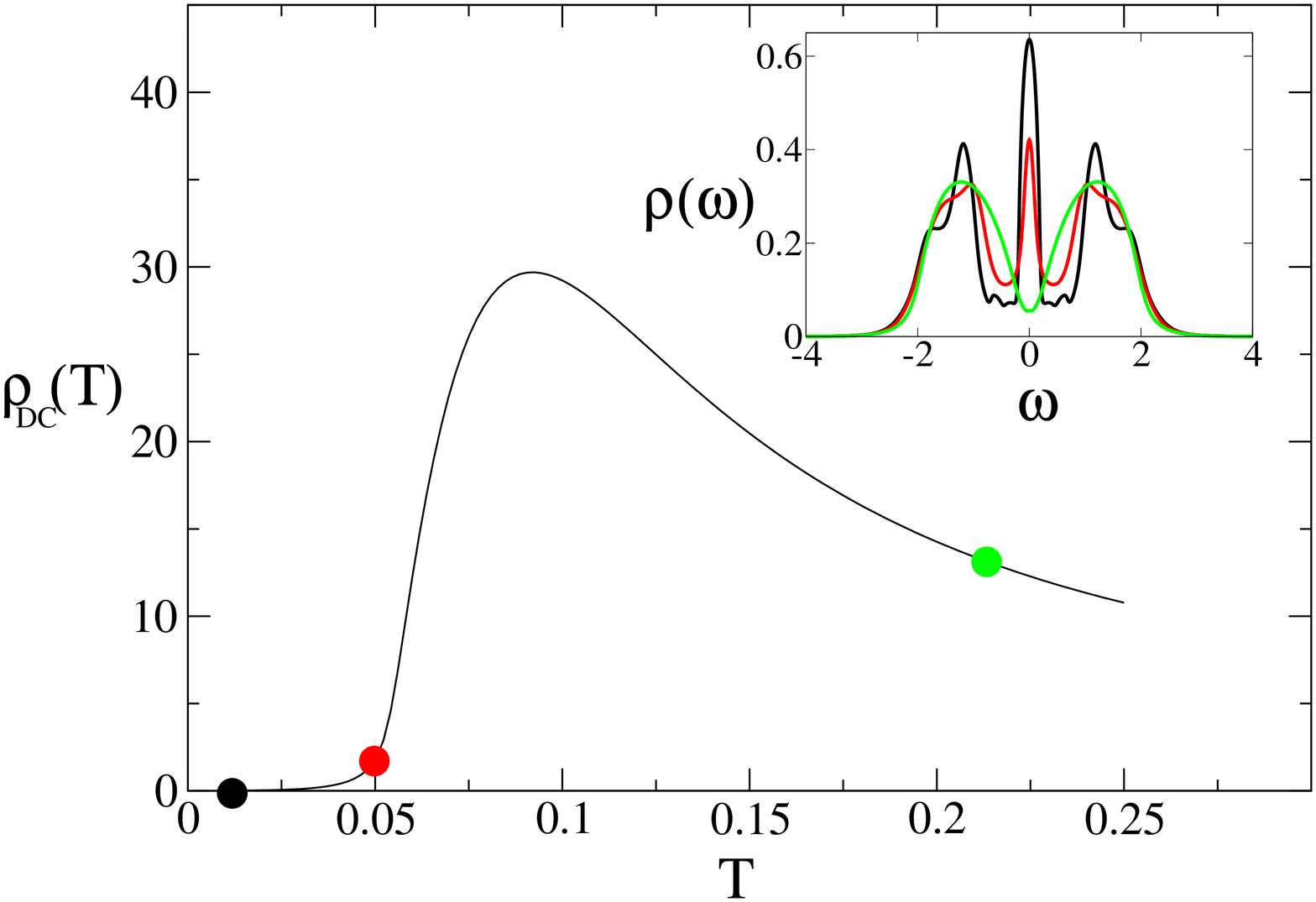} &
\includegraphics[width=8 cm]{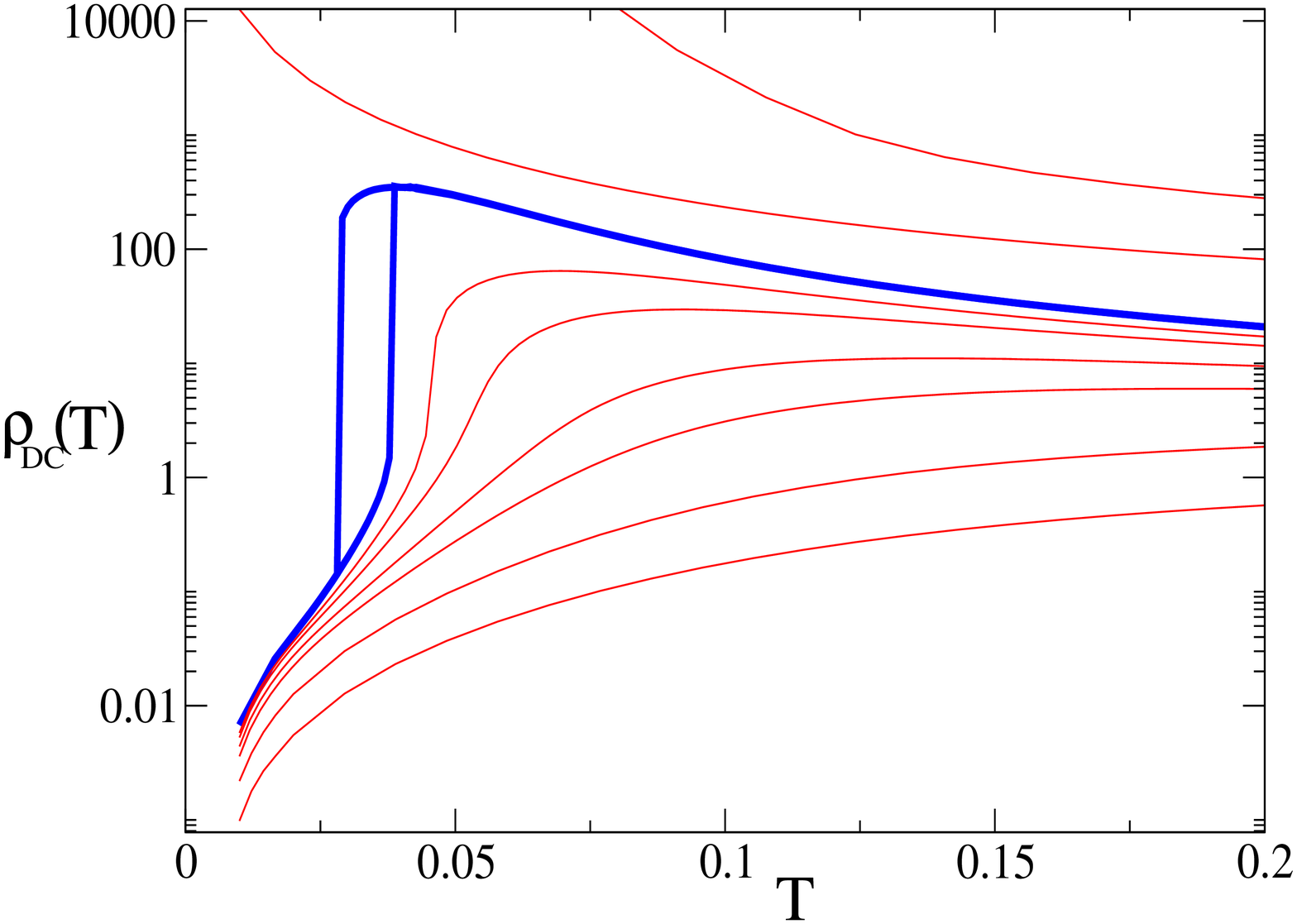}
\end{tabular}
\caption{
Left: Resistivity in the metallic
phase close to the Mott transition ($U=2.4 D$), as a function of temperature, calculated within
DMFT using the IPT approximation. For three
selected temperatures, corresponding to the three regimes discussed in the
text, the corresponding spectral density is displayed in the inset.
Right: IPT results for the resistivity for values of $U$ in the metallic regime (lower curves),
the coexistence region (bold curve) and the insulating regime (upper two curves).
From Ref.~\cite{florens_phd,georges_mott_iscom}.}
\label{fig:res_and_dos}
\end{figure}
These three regimes, and the overall temperature dependence
of the resistivity obtained within DMFT are illustrated by Fig.~\ref{fig:res_and_dos}.
A distinctive feature is the resistivity maximum, which
occurs close to the Mott transition. This behaviour is indeed observed experimentally in both
Cr-doped V$_2$O$_3$ and the organics. In the latter case, the transport data obtained
recently in the Orsay group are depicted on Fig.~\ref{fig:bedt_resist}, and compared to
DMFT model calculations~\cite{limelette_bedt_prl,georges_mott_iscom}.

\begin{center}
\begin{figure}
\begin{tabular}{cc}
\includegraphics[width=8 cm]{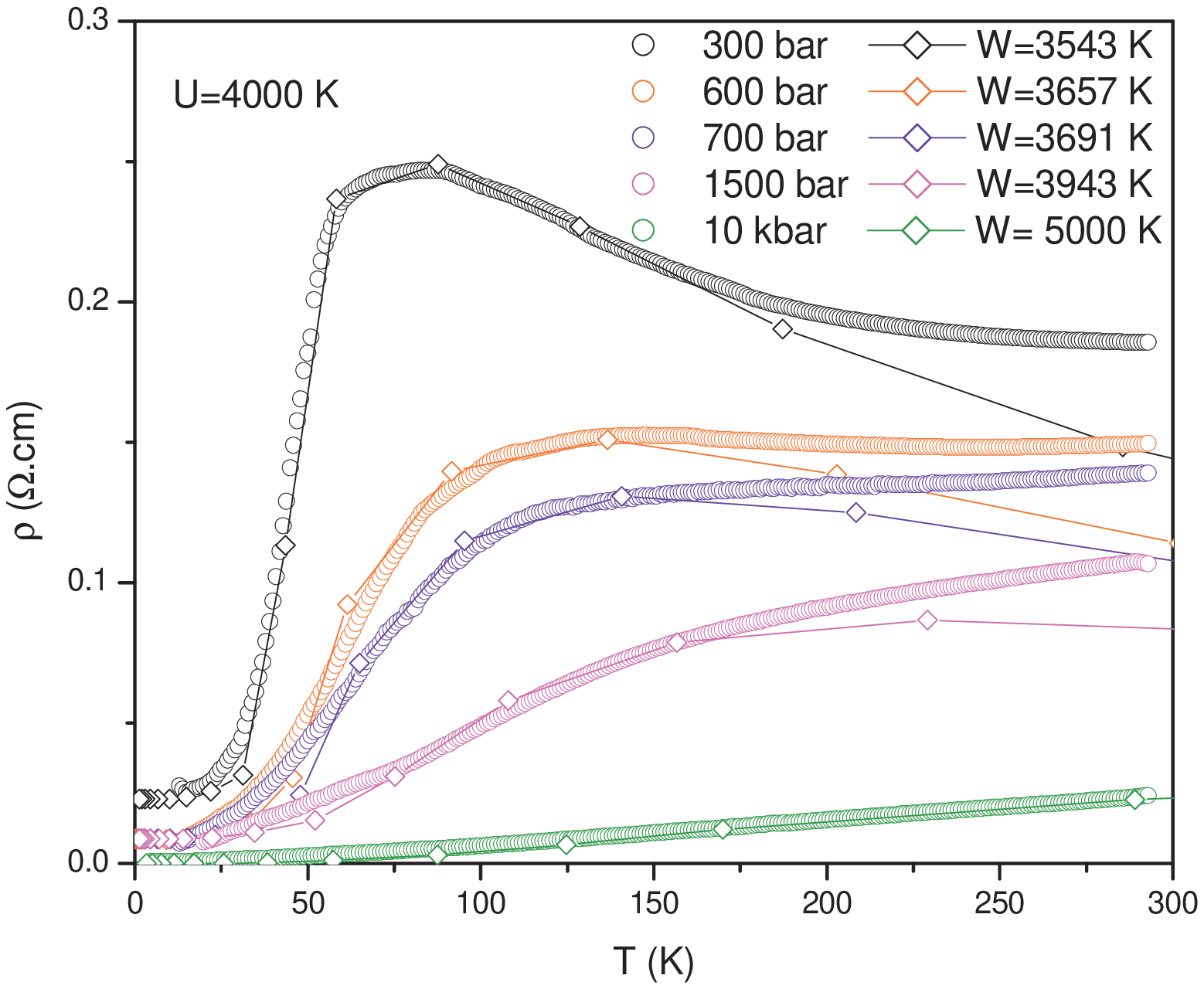} &
\includegraphics[width=8 cm]{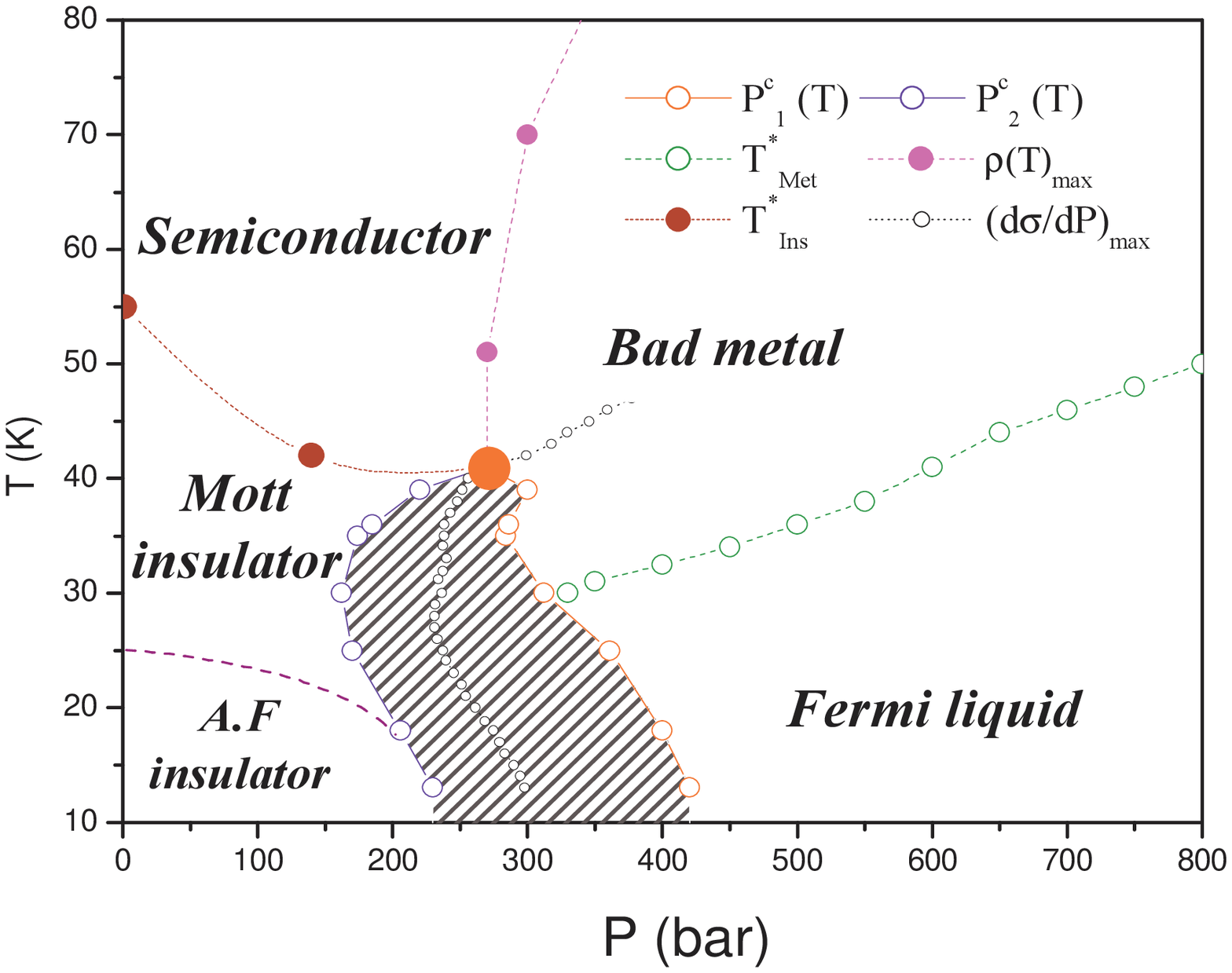}
\end{tabular}
\caption{Left: Temperature-dependence of the resistivity at different pressures, for
$\kappa$-(BEDT-TTF)$_{2}$Cu[N(CN)$_{2}$]Cl.
The data (circles) are compared to a DMFT-NRG calculation (diamonds), with a
pressure dependence of the bandwidth as indicated. The measured residual
resistivity $\rho_0$ has been added to the theoretical curves.
Right: Transport regimes and crossovers for this compound.
Figures reproduced from Limelette {\it et al.}~\cite{limelette_bedt_prl}.}
\label{fig:bedt_resist}
\end{figure}
\end{center}

Within DMFT, the conductivity can be simply obtained from a calculation of the
one-particle self-energy since vertex corrections are absent~\cite{khurana_vertex,georges_review_dmft}.
However, a precise
determination of both the real and imaginary part of the real-frequency self-energy is
required. This is a challenge for most ``impurity solvers''. In practice, early
calculations\cite{majumdar_transport,rozenberg_optics_prl,Mer00}
used the iterated perturbation theory (IPT) approximation\cite{georges_kotliar_dmft}.
The results displayed in
Fig.~\ref{fig:res_and_dos} have been obtained with this technique, and the overall shape of the
resistivity curves are qualitatively reasonable. However, the IPT approximation does a poor
job on the quasiparticle lifetime in the low-temperature regime, as
shown on Fig.~\ref{fig:compare_ipt_nrg}. Indeed, we expect on general grounds
that, close to the transition, $D\,\mbox{Im}\Sigma$ becomes a scaling function~\cite{moeller_projective}
of $\omega/\est$ and $T/\est$, so that for $T\ll\est$ it behaves as:
$\mbox{Im}\Sigma(\omega=0)\propto D (T/\est)^2\propto T^2/(Z^2D)$ which leads to an
enhancement of the $T^2$ coefficient of the resistivity by $1/Z^2$ as mentioned above.
The IPT approximation does not capture this enhancement and yields the incorrect result
$\mbox{Im}\Sigma_{IPT}(\omega=0)\propto U^2 T^2/D^3$, as illustrated
For this reason, the numerical renormalization group (NRG)
has been used recently~\cite{limelette_bedt_prl,georges_mott_iscom}
in order to
perform accurate transport calculations within DMFT.
This method is very appropriate in this context,
since it is highly accurate at low energies and
yields real-frequency data\cite{bulla_mott_NRG}. DMFT-NRG calculations
compare favorably to transport data on organics, as shown on Fig.~\ref{fig:bedt_resist}.
%
%
\begin{center}
\begin{figure}
\includegraphics[width=8 cm]{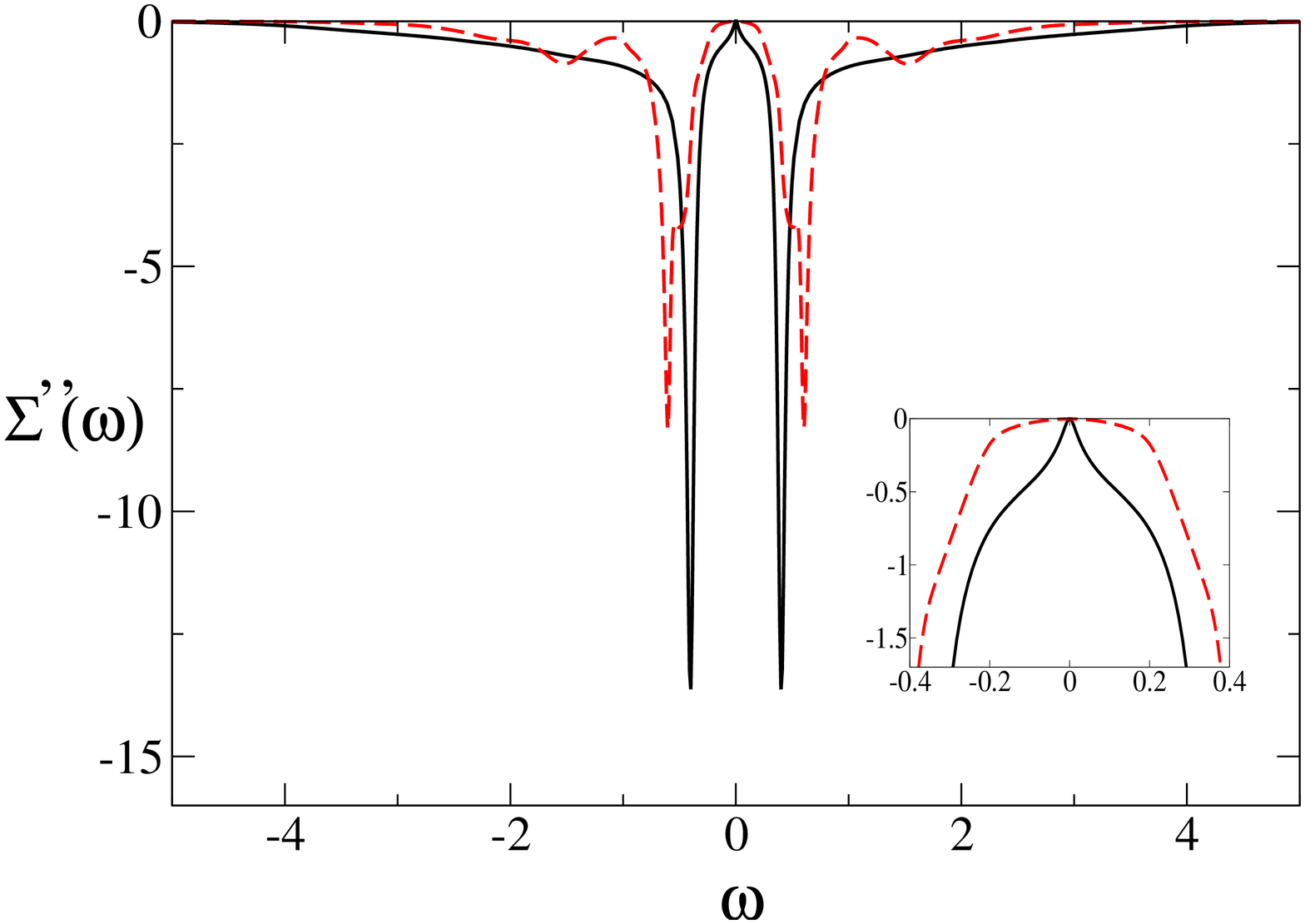}
\caption{
Comparison between the IPT (dashed lines) and NRG methods (plain lines), reproduced
from Ref.~\cite{georges_mott_iscom}.
The low-frequency behaviour of the inverse lifetime $\mbox{Im}\Sigma$
clearly displays a critically enhanced curvature, which is not reproduced by IPT.}
\label{fig:compare_ipt_nrg}
\end{figure}
\end{center}

The crossovers described here in electrical transport also have consequences for
thermal transport. The thermopower, in particular, displays a saturation in the
incoherent metal regime~\cite{palsson_thermo_1998_prl,Mer00}. This is
presumably relevant for the cobalt- based thermoelectric oxides such as
\naxcoo2 . Finally, let me emphasize that an interesting experimental
investigation of the correlations between transport crossovers (both
ab-plane and c-axis) and the loss
of quasiparticle coherence observed in photoemission has been performed by
Valla \etal~\cite{valla_coherence_2002_nature} for several layered materials.
This study raises
intriguing questions in connection with DMFT, and particularly its
$\vk$-dependent extensions.

\subsection{Critical behaviour: a liquid-gas transition}

Progress has been made recently in identifying the critical behaviour at the Mott
critical endpoint, both from a theoretical and experimental standpoint.
It was been pointed
early on by Castellani {\it et al.}\cite{Cast79} (see also \cite{Jaya70})
that an analogy exists with the liquid-gas transition in a classical fluid.
This is based on a qualitative picture illustrated on Fig.~\ref{fig:cartoon}.
The Mott insulating phase has few double occupancies (or holes) and
corresponds to a low-density ``gas'', while
the metallic phase corresponds to a high-density ``liquid'' with many
double-occupancies and holes (so that the electrons can be itinerant).
\begin{center}
\begin{figure}
\begin{tabular}{cc}
\includegraphics[width=9 cm]{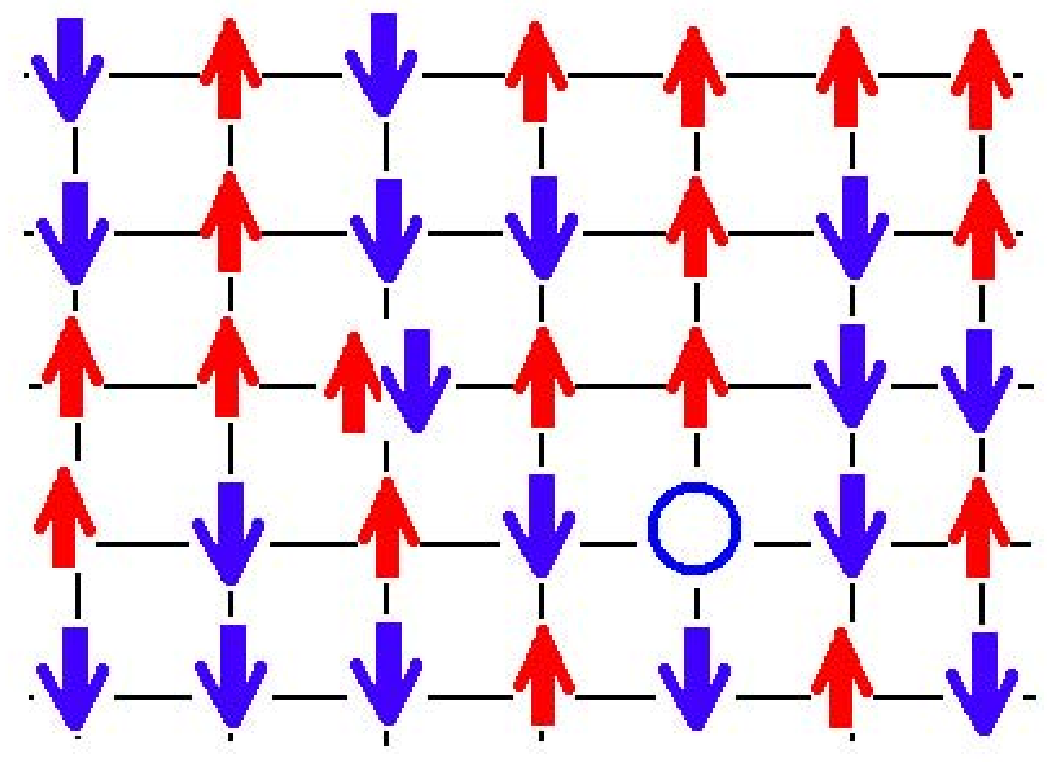} &
\includegraphics[width=9 cm]{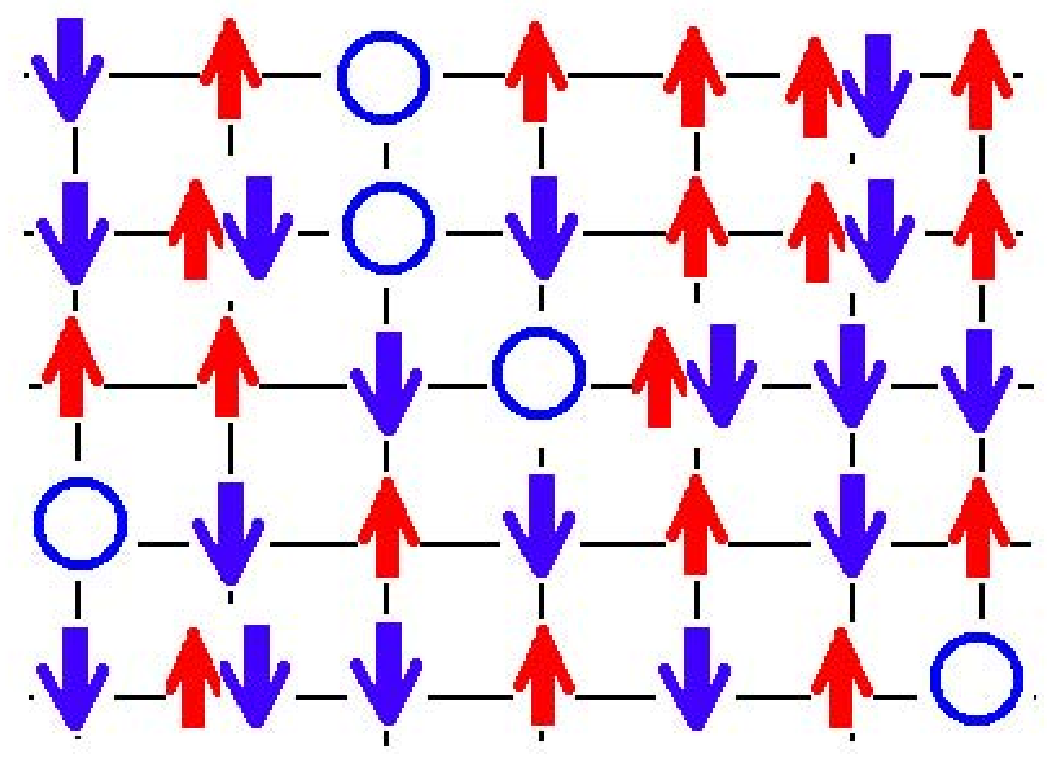}
\end{tabular}
\caption{Cartoon of a typical real-space configuration of electrons in
the Mott insulator (left) and metallic (right) phase. The insulator has few
double-occupancies or holes, and corresponds to a gas of these excitations.
Fluctuating local moments exist in this phase. The metal has many double-occupancies and holes,
corresponding to a dense ``liquid''. Electrons are itinerant in the metallic phase,
and the local moments are quenched. Within DMFT this quenching is akin to a (self-consistent) local
Kondo effect.
 }
\label{fig:cartoon}
\end{figure}
\end{center}
Recently, this analogy has been
given firm theoretical foundations within the framework of a
Landau theory~\cite{kotliar_landau_functional_mott,kotliar_landaumott_prl,rozenberg_finiteT_mott}
derived from DMFT by Kotliar and coworkers.
In this framework, a scalar order parameter $\phi$  is associated with
the low-energy electronic degrees of freedom which build up the quasiparticle
resonance in the strongly correlated metallic phase close to the transition.
This order parameter couples to the singular part of the double occupancy (hence
providing a connection to the qualitative picture above), as well as to other
observables such as the Drude weight or the dc-conductivity.
Because of the scalar nature of the order parameter, the transition falls in
the Ising universality class. In Table 1, the correspondence between the Ising
model quantities, and the physical observables of the liquid-gas transition and of
the Mott metal-insulator transition is summarized.
\begin{center}
\begin{table}
\begin{tabular}{|c|c|c|c|} \hline
{\bf Hubbard model} & {\bf Mott MIT} & {\bf Liquid-gas} & {\bf Ising model} \\ \hline
 & & & \\
  $D-D_c$ & $p-p_c$ & $p-p_c$ & Field $h$ \\
  & & & (w/ some admixture of $r$) \\ \hline
  &  &  & Distance to \\
  $T-T_c$ & $T-T_c$ & $T-T_c$ & critical point $r$ \\
  & & & (w/ some admixture of $h$) \\ \hline
  Low-$\omega$ & Low-$\omega$ & $v_g-v_L$ & Order parameter \\
  spectral weight & spectral weight & & (scalar field $\phi$) \\ \hline
\end{tabular}
\caption{Liquid-gas
description of the Mott critical endpoint.
The associated Landau free-energy density reads
$r\phi^2+u\phi^4-h\phi$ (a possible $\phi^3$ can be eliminated by an appropriate change
of variables and a shift of $\phi$).}
\end{table}
\end{center}
In Fig.~\ref{fig:sigma_vs_D}, the dc-conductivity obtained from DMFT in the half-filled Hubbard model
(using IPT) is plotted as a function of the half-bandwith $D$, for several different temperatures.
The curves qualitatively resemble those of the Ising model order parameter as a function of
magnetic field (in fact, $D-D_c$ is a linear combination of the field $h$ and of the mass term $r$
in the Ising model field theory). Close to the critical point, scaling implies that
the whole data set can be mapped onto a universal form of the equation of state:
\beq
\langle\phi\rangle\,=\, h^{1/\delta}\,f_{\pm}\left(h/|r|^{\gamma\delta/(\delta-1)}\right)
\label{eq:eq_of_state}
\eeq
In this expression, $\gamma$ and $\delta$ are critical exponents associated with the
order parameter and susceptibility, respectively:
$\langle\phi\rangle\sim h^{1/\delta}$ at $T=T_c$ and
$\chi=d\langle\phi\rangle/dh \sim |T-T_c|^{-\gamma}$.
$f_{\pm}$ are universal scaling functions associated with $T>T_c$ (resp. $T<T_c$).
A quantitative study of the critical behaviour of the double occupancy within DMFT was made
in Ref.~\cite{kotliar_landaumott_prl}, with the expected mean-field values of the
exponents $\gamma=1,\delta=3$.
\begin{center}
\begin{figure}
\includegraphics[width=8 cm]{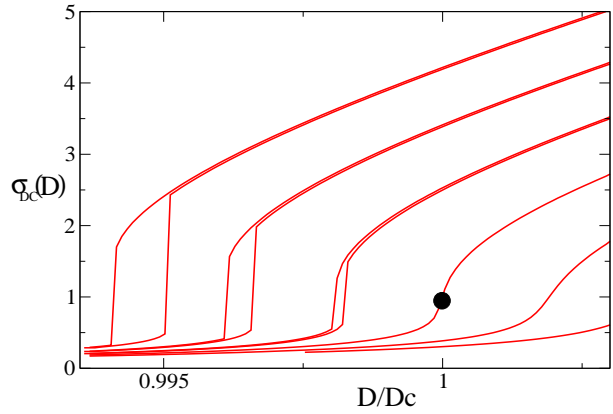}
\caption{
IPT calculation of the
dc-conductivity as a function of the half-bandwith for the half-filled Hubbard model within DMFT,
for several different temperatures.
Increasing $D$ drives the system more metallic. The curve at $T=T_c$ displays a singularity
(vertical slope: dot), analogous to the non-linear dependence of the order parameter upon the
magnetic field at a second-order magnetic transition.
Hysteretic behaviour is found for $T<T_c$.
}
\label{fig:sigma_vs_D}
\end{figure}
\end{center}

Precise experimental studies of the critical behaviour at the Mott critical
endpoint have been performed very recently, using a variable pressure technique,
for Cr-doped \v2o3 by Limelette {\it et al.}~\cite{limelette_v2o3_science}
(Fig.~\ref{fig:v2o3_sigma_critique})
and also for the $\kappa$-BEDT organic compounds by Kagawa {\it et al.}~\cite{kagawa_bedt}.
These studies provide the first experimental demonstration of the liquid-gas
critical behaviour associated with the Mott critical endpoint, including a
a full scaling~\cite{limelette_v2o3_science} onto the universal
equation of state (\ref{eq:eq_of_state}).
\begin{center}
\begin{figure}
\includegraphics[height=6cm]{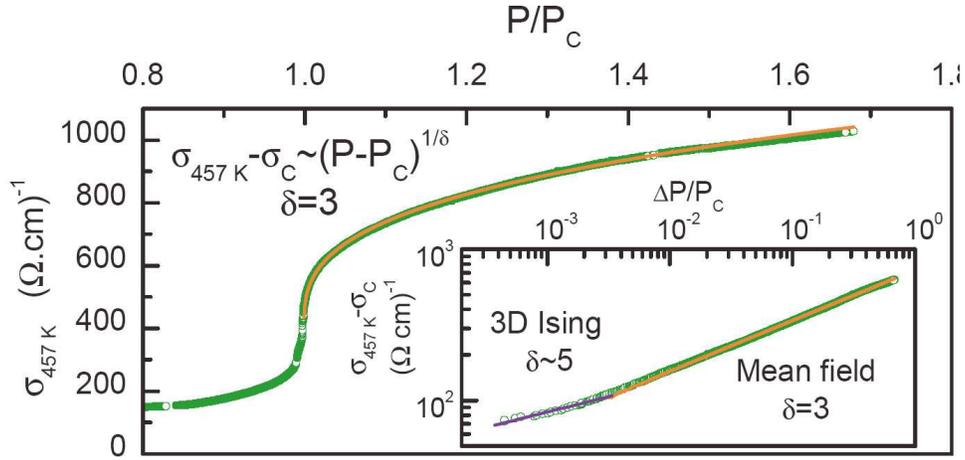}
\caption{Conductivity of Cr-doped \v2o3, at the critical endpoint $T=T_c$,
measured as a function of pressure $P/P_c$ (Limelette \etal~\cite{limelette_v2o3_science}).
A characteristic sigmoidal form is found, which is well fitted by
$\sigma-\sigma_c\sim |P-P_c|^{1/\delta}$ (plain line). Inset: log-log scale.
See Ref.~\cite{limelette_v2o3_science} for a full experimental study of the critical behaviour,
including scaling onto the universal equation of state.
}
\label{fig:v2o3_sigma_critique}
\end{figure}
\end{center}

\subsection{Coupling to lattice degrees of freedom}
\label{sec:lattice}

Lattice degrees of freedom do play a role at the Mott transition in real materials,
e.g the lattice spacing changes discontinuously through the first-order
transition line in (V$_{1-x}$ Cr$_{x}$)$_2$O$_3$, as displayed in
Fig.~\ref{fig:v2o3_lattice}. In the metallic
phase, the d-electrons participate in the cohesion of the solid,
hence leading to a smaller lattice spacing than in the insulating phase.
\begin{center}
\begin{figure}
\begin{tabular}{cc}
\includegraphics[width=8 cm]{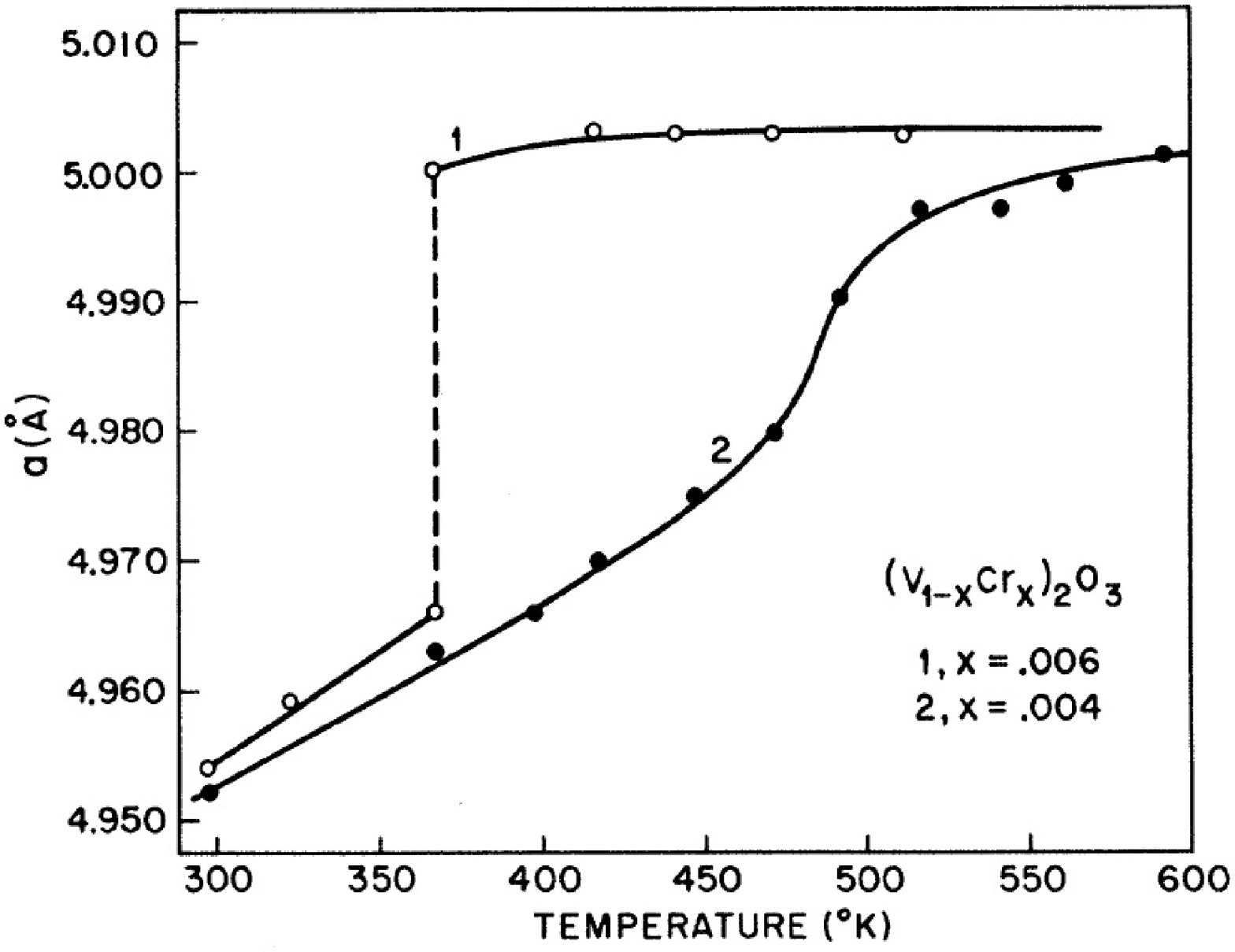} &
\includegraphics[width=8 cm]{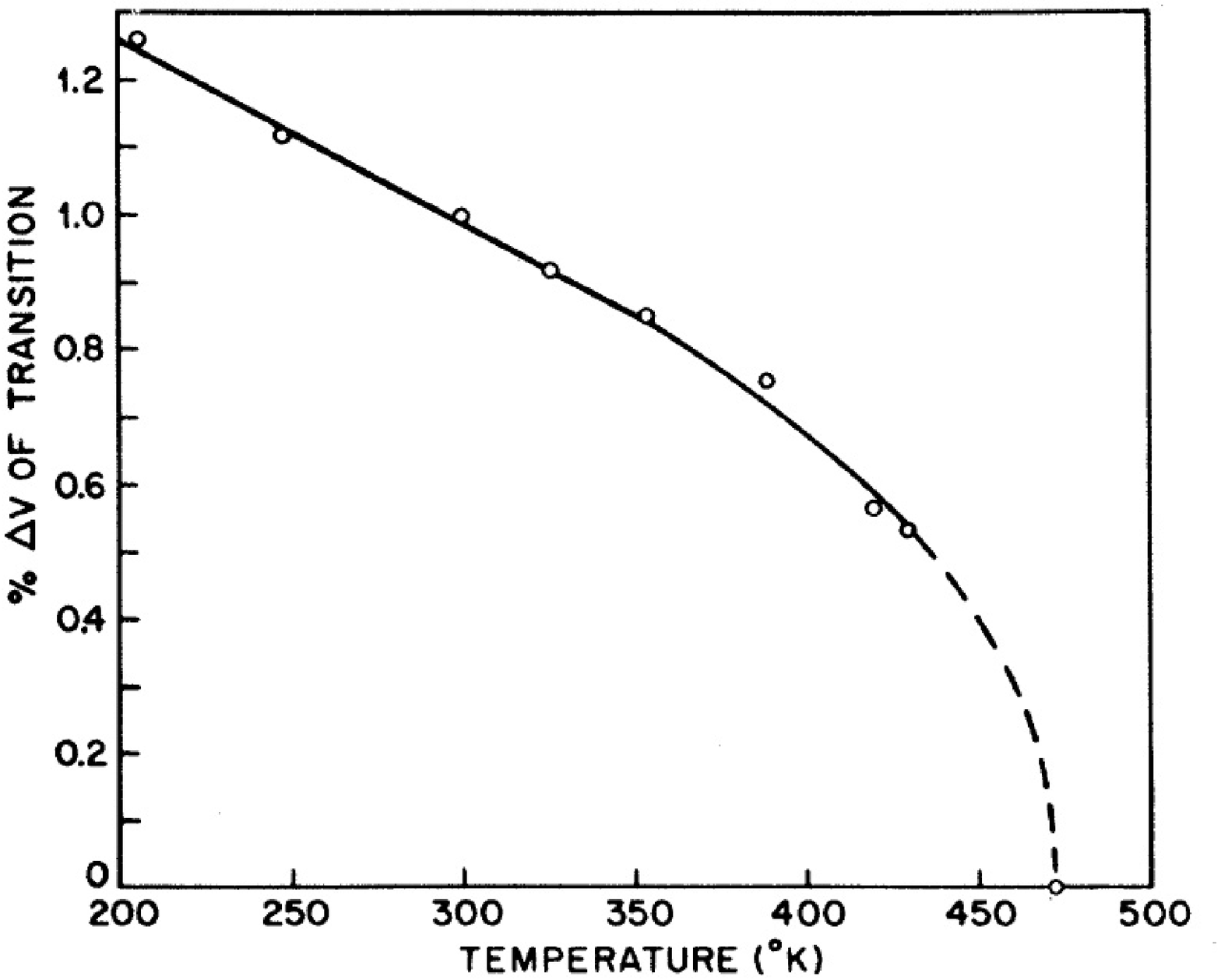}
\end{tabular}
\caption{Left: Change of the lattice constant as a function of temperature for
two samples of \vcr2o3 with different Cr-concentrations. The discontinuous change
in the lattice constant through the first-order transition transition line is clearly
seen for $x=.006$, while the sample with $x=.004$ is slightly to the
right of the critical point. Right: Percentage volume change of the
unit-cell volume close to the critical line, reflecting the critical behaviour of the
order parameter. Reproduced from Ref.~\cite{Jaya70}}
\label{fig:v2o3_lattice}
\end{figure}
\end{center}
Both the electronic degrees of freedom and the ionic positions must be retained
in order to describe these effects.
In Ref.~\cite{majumdar_compress} (see also~\cite{cyrot_1972_compress}),
such a model was treated in the simplest approximation
where all phonon excitations are neglected. The free energy then reads:
\beq
F= \frac{1}{2}\,B_0\,\frac{(v-v_0)^2}{v_0} + F_{el}\left[D(v)\right]
\eeq
In this expression, $v$ is the unit-cell
volume, $B_0$ is a reference elastic modulus and the electronic
part of the free-energy $F_{el}$ depends on
$v$ through the volume-dependence of the bandwith. In such a model, the critical
endpoint is reached when the electronic response function:
\beq
\chi=-\frac{\partial^2 F_{el}}{\partial D^2}
\eeq
is large enough (but not infinite), and hence the critical temperature $T_c$ of the
compressible model is larger than
$T_c^{el}$ (at which $\chi$ diverges in the Hubbard model). The compressibility
$\kappa = \left(v \partial^2 F/\partial v^2\right)^{-1}$ diverges at $T_c$. This implies
an anomalous lowering of the sound-velocity at the
transition~\cite{merino_sound,hassan_sound_2004}, an effect that has been
experimentally observed in the $\kappa$-BEDT compounds
recently \cite{fournier_sound_bedt}, as shown on Fig.~\ref{fig:bedt_sound}.

We emphasize that, within DMFT, an unambiguous answer is given to the
``chicken and egg'' question: is the first-order Mott transition driven
by electronic or lattice degrees of freedom ? Within DMFT, the transition
is described as an electronic one, with lattice degrees of freedom
following up. In fact, it is aremarkable finding of DMFT that
a purely electronic model can display
a first-order Mott transition and a finite-T critical endpoint (associated with a diverging
$\chi$), provided that magnetism is frustrated enough so that ordering
does not preempt the transition. Whether this also holds for the finite-dimensional
Hubbard model beyond DMFT is to a large extent an open question
(see \cite{onoda_finiteT_mott} for indications supporting this conclusion
in the 2D case).
\begin{figure}
\includegraphics[width=9 cm]{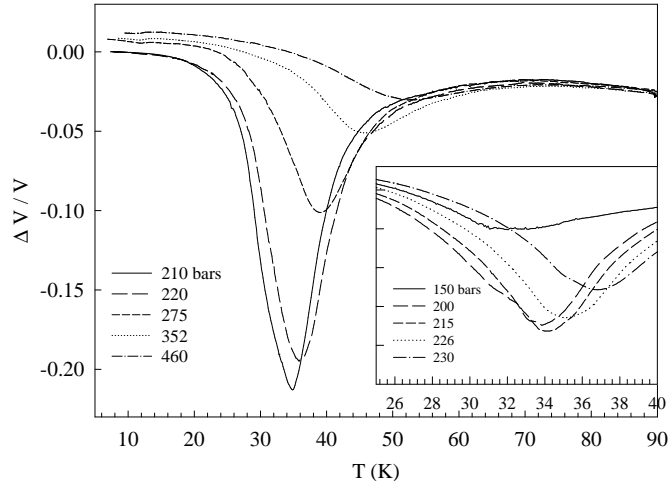}
\caption{Relative change in the sound velocity of \kappacl
as a function of temperature, at various pressures.
The velocity variation is relative to
the value at 90 K. Inset: position and amplitude of the anomaly
below 230 bars. (Figure and caption from Ref.~\cite{fournier_sound_bedt}).}
\label{fig:bedt_sound}
\end{figure}

\subsection{The frontier: $\vk$-dependent coherence scale,
cold and hot spots}

A key question, still largely open, in our theoretical understanding of the Mott transition
is the role of spatial correlations (inadequately treated by DMFT).
This is essential in materials like cuprates, in which short-range spatial
correlations play a key role (in particular magnetic correlations due to superexchange,
leading to a strong tendency towards the formation of singlet bonds, as well
as pair correlations).
In the regime where the quasiparticle coherence scale $\est$ is small as compared
to the (effective strength of the) superexchange $J$, the DMFT picture
is certainly deeply modified.
There is compelling experimental evidence that the quasiparticle coherence
scale then has a strong variation as the momentum $\vk$ is varied along the Fermi
surface, leading to the formation of ``cold spots'' and ``hot regions''.
Such effects have been found in recent studies using
cluster extensions of the DMFT framework
(\cite{parcollet_hot}, see also \cite{biermann_chain_2001_prl,giamarchi_orga_iscom}).
\section{Electronic structure and Dynamical Mean-Field Theory}
\label{esc}

The possibility of using DMFT in combination with electronic
structure calculation methods, in order to overcome some of the
limitations of DFT-LDA for strongly correlated materials, was pointed out
early on~\cite{georges_review_dmft}.
In the last few years, very exciting developments
have taken place, in which theorists from the electronic structure and many-body
communities joined forces and achieved concrete implementations of
DMFT within electronic structure calculations.
The first papers~\cite{anisimov_lda+dmft_1997,lichtenstein_lda+dmft_1998}
implementing this combination appeared in 1997-1998, and the field has been
extremely active since then. For reviews of the early developments in
this field\footnote{This section is merely a brief introduction to the field
and certainly not as an exhaustive review.},
see Refs.~\cite{held_review_lda+dmft_1,held_review_lda+dmft_2,
kotliar_savrasov_newton,lichtenstein_magnetism_2002}.
For on-line material presented at
recent workshops, see Refs.~\cite{kitp_cem02_workshop,kitp_cem02_conference,ictp_conference_2003}

\subsection{Limitations of DFT-LDA for strongly correlated systems}

In Sec.~\ref{sec:dft}, I briefly presented the basic principles of density functional theory (DFT).
In practice, the local density approximation (LDA) to the exchange-correlation energy, and
its extensions (such as the generalised gradient approximation) have been remarkably
successful at describing ground-state properties of many solids from first principles. This is
also the state of the art method for band structure calculations, with the additional
assumption that Kohn-Sham eigenvalues can be interpreted as single-particle excitations.
For strongly correlated materials however, DFT-LDA has severe limitations, which we
now briefly review.

\paragraph{Issues about ground-state properties} Ground-state properties, such as equilibrium unit-cell volume,
are not accurately predicted from LDA (or even GGA) for the most strongly correlated materials. This is
particularly true of materials in which some electrons are very localized, such as the 4f electrons
of rare-earth elements at ambiant or low pressure (Sec.~\ref{sec:felec}).
If these orbitals are treated as valence orbitals, the LDA
leads to a much too itinerant character, and therefore overestimates the contribution of these
orbitals to the cohesive energy of the solids, hence leading to a too small unit-cell volume.
If instead the f-orbitals are treated as core states, the equilibrium volume
is then overestimated (albeit closer to experimental value, in the case of rare earth),
since binding is underestimated. Phenomena such as the volume-collapse transitions,
associated with the partial delocalization of the f-electrons,
(and associated structural changes) under pressure~\cite{mcmahan_collapse_review}
are simply out of reach of standard methods.
At high pressures however, the f-electrons recover itinerant
character and DFT-LDA(GGA) does better, as expected.
In some particular cases, the electrons are just on the verge of the itinerant/localized
behaviour. In such cases, standard electronic structure methods
perform very poorly. A spectacular example is the $\delta$-phase of metallic plutonium
in which the unit-cell volume is underestimated (compared to the experimental value)
by as much as $35\%$ by standard
electronic structure methods (Fig.~\ref{fig:pu_energy}) !
All these examples illustrate the need of a method which is able to handle
intermediate situations between fully localized and fully itinerant electrons.
I emphasize that this issue may depend crucially on energy scales, with localized character
most pronounced at high-energy (short time) scales, and itinerant quasiparticles forming
at low-energy (long time scales).

\paragraph{Excitation spectra}
Even though the Kohn-Sham eigenvalues and wavefunctions are, strictly speaking,
auxiliary quantities in the DFT formalism used to represent the local density,
they are commonly interpreted as energy bands in electronic structure
calculations. This is very successful in many solids, but does fail
badly in strongly correlated ones. The most spectacular difficulty is that
Mott insulators are found to have metallic Kohn-Sham spectra. This is documented,
e.g by Fig.~\ref{fig:d1oxides_dmft}, in which the LDA density of states
of two Mott insulators, \latio3 and \ytio3 are shown. I emphasize that, in
both compounds (as well as in many other Mott insulators),
the Mott insulating gap has nothing to do with the magnetic ordering in the
ground-state. Even though magnetic long-range order is
found at low-enough temperatures in both materials (below $T_N\simeq 140$~K in
\latio3 and $T_{\,C}\simeq 30$~K in \ytio3), the insulating behaviour and
Mott gap ($\simeq 1\eV$ for \ytio3)
are maintained well above the ordering temperature. In other cases (such as
\vo2), the insulating phase is a paramagnet and the LDA spectrum is
again metallic.

In strongly correlated metals, e.g close to Mott insulators, the
LDA bandstructure is also in disagreement with experimental
observations. The two main discrepancies are the following. (i)
LDA single-particle bands are generally too broad. Correlation
effects lead to band-narrowing, corresponding to a (Brinkman-Rice)
enhancement of the effective masses of quasiparticles. This
becomes dramatic in f-electron materials, where the large
effective mass is due to the Kondo effect, a many-body process
which is beyond the reach of single-particle theories. (ii) The
spectral weight $Z$ associated with quasiparticles is reduced by
correlations, and the corresponding missing spectral weight $1-Z$
is found in intermediate or high-energy incoherent excitations. In
correlated metals, as well as in Mott insulators, lower and upper
Hubbard bands are observed, which are absent in the LDA density of
states (e.g for \srvo3 and \cavo3 in Fig.~\ref{fig:pes_casrvo3}
and Fig~\ref{fig:d1oxides_dmft}).

Related correlation effects are observed also for
pure transition metals, such as nickel, in which the LDA spectrum is unable to
account for: the $\sim -6\eV$ photoemission satellite, and for the correct values of
the occupied bandwidth and exchange splitting between the majority and minority band in the
ferromagnetic ground-state.

\subsection{Marrying DMFT and DFT-LDA}

In this section, I briefly describe the (happy) marriage of electronic structure methods
and dynamical mean-field theory.
I first give a simple practical formulation in terms of a
realistic many-body hamiltonian, and keep for the next section the construction of
energy functionals.

The first issue to be discussed is the choice of the basis set for the valence electrons.
Since DMFT emphasizes local correlations, we need a localised basis set, i.e basis
functions which are centered on the atomic positions $\vR$ in the crystal lattice.
Up to now, most implementations have used basis sets based on
linear muffin-tin orbitals~\cite{andersen_lmto_1975_prb,skriver_book} (LMTOs)
$\chi_{L\vR}(\vr)=\chi_L(\vr-\vR)$
(in which $L=\{l,m\}$ stands for the angular momentum quantum number of the valence
electrons). These basis sets offer the advantage to carry over the physical intuition
of atomic orbitals from the isolated atoms to the solid. In the
words of their creator, O.K. Andersen, LMTO- based electronic structure methods
are ``intelligible'' because they are based on a minimal and flexible basis set of
short-range orbitals~\cite{andersen_nmto_psik_2000}.
There are several possible choices of basis even within the LMTO
method. Basically, a compromise has to be made between the degree of localisation and
the orthogonality of the basis set. The most localised basis set (the so-called ``screened''
or $\alpha$-basis) is not orthogonal and will therefore involve\footnote{In the following, we
assume an orthogonal basis set to simplify the formalism. The overlap matrix
can be easily reintroduced where it is appropriate}
an overlap matrix
$O_{LL'}=\bra \chi_L|\chi_{L'}\ket$. Since DMFT neglects non-local correlations, they
may be the best one to choose. However, a non-orthogonal basis set may not be simple
to implement, for technical reasons, when using some impurity solvers (e.g QMC).
Orthogonal LMTOs basis sets are somewhat more extended.

Another possibility is to use basis sets made of Wannier functions. This has been little
explored yet in combination with DMFT. Wannier functions can in fact be constructed
starting from the LMTO formalism by using the ``downfolding'' procedure (the so-called
third-generation LMTO~\cite{andersen_nmto_psik_2000,andersen_nmto_2000_wshop}).
Recently, DMFT has been implemented within
a downfolded (NMTO) Wannier basis, and successfully
applied to transition metal oxides with non-cubic structures. Other routes to Wannier functions
(such as the Marzari-Vanderbilt construction of
maximally localised Wannier functions~\cite{marzari_wannier_1997_prb}) might be worth pursuing.
Given a basis set, the electron creation operator at a point $\vr$ in the solid can be
decomposed as:
\beq
\psi^\dagger(\vr) = \sum_{\vR,L} \chi^*_{L\vR}(\vr)\,c_{L\vR}^\dagger
\eeq
The decomposition of the full Green's function in the solid:
$G(\vr,\vr',\tau-\tau')\equiv - \bra T\psi(\vr,\tau)\psi^\dagger(\vr',\tau')\ket$
(as well as of any other
one-particle quantity) thus reads:
\beq
G(\vr,\vr',i\omega) = \sum_{\vR\vR'}\sum_{LL'} \chi_{L\vR}(\vr)
G_{LL'}(\vR-\vR',i\omega) \chi_{L'\vR'}(\vr')^*
\label{eq:green_basis}
\eeq

The simplest combination of DMFT and electronic structure methods uses a starting point which
is similar to that of the LDA$+$U approach~\cite{anisimov_lda+u_1991_prb,anisimov_lda+u_review_1997_jpcm}.
Namely, one first separates the valence electrons
into two groups: those for which standard electronic structure methods are sufficient on one hand
(e.g $l=s,p$ in an oxide or $l=s,p,d$ in rare-earth compounds),
and on the other hand the subset of orbitals which will feel strong correlations
(e.g $l=d$ or $l=f$). This separation
refers, of course, to the specific choice of basis set which has been made. In the following,
I denote the orbitals with $l$ in the correlated subset by the index $a\equiv\{m,\sigma\}$ (and
$b,\cdots$).
Let us then consider the one-particle hamiltonian:
\beq
H_{KS} = \sum_\lambda \epsilon^{KS}_\lambda |\lambda\ket\bra\lambda|
= \sum_{\vk L} h^{KS}_{LL'}(\vk) c^\dagger_{\vk L}c_{\vk L'}
\eeq
obtained from solving the
Kohn-Sham equations for the material under consideration.
The Kohn-Sham potential we have in mind is, in the simplest implementation,
the one obtained within a standard DFT-LDA (or GGA) electronic structure calculation of
the local density.
In a more sophisticated implementation, one may also correct
the local density by correlation effects
and use the associated Kohn-Sham potential (i.e modify the self-consistency
cycle over the local density in comparison to standard LDA, see below).
A many-body hamiltonian is then constructed as follows:
\begin{equation}
H = H_{KS} - H_{DC} + H_U
\label{eq:mb_ham}
\eeq
In this expression, $H_U$ are many-body terms acting in the subset of correlated orbitals
only. They correspond to matrix elements of the Coulomb interaction, and will in general
involve arbitrary 2-particle terms $U_{abcd}c^\dagger_ac^\dagger_bc_dc_c$.
In practice however, one often
makes a further simplification and keep only density-density interactions (for technical
reasons, this is always done when using QMC as a solver). To simplify notations, we
shall limit ourselves here to this case, and use:
\beq
H_U =\frac{1}{2}\sum_\vR\sum_{ab\sigma} U_{ab}\, \hn_{\vR a}\hn_{\vR b}
\label{eq:HU}
\eeq
with:
\beq
U_{mm'}^{\spinup\spindown}=U_{mm'}\,\,\,,\,\,\,
U_{m\neq\,m'}^{\spinup\spinup}=U_{m\neq\,m'}^{\spindown\spindown}=U_{mm'}-J_{mm'}
\label{eq:U_matrix}
\eeq
In this expression, $J_{mm'}$ is the Hund's coupling. For a more detailed discussion
of the choice of the matrix of interaction parameters, see
e.g Ref.~\cite{anisimov_lda+u_review_1997_jpcm}.

The ``double-counting'' term $H_{DC}$ needs to be introduced,
since the contribution of interactions between the correlated orbitals to the
total energy is already partially included in the exchange-correlation
potential.
Unfortunately, it is not possible to derive this term explicitly, since the energy
within DFT is a functional of the total electron density, which combines all orbitals
in a non-linear manner. In practice, the most commonly used form of the double-counting term
is (for other choices, see e.g \cite{lichtenstein_magnetism_dmft_2001_prl}):
\begin{eqnarray}\nonumber
H_{DC} = \sum_{\vR\sigma ab} V^{DC}_{ab\sigma} c^\dagger_{a\sigma}c_{b\sigma}\\
V^{DC}_{ab\sigma} = \delta_{ab}\left[U (N-\frac{1}{2})-J(N^\sigma-\frac{1}{2})\right]
\end{eqnarray}
The many-body hamiltonian (\ref{eq:mb_ham}) is then soved using the DMFT approximation.
This means that a local self-energy matrix is assumed, which acts in the subset of correlated
orbitals only:
\beq
\Sigma^{\vR\vR'}_{LL'}(i\omega)\,=\,\delta_{\vR,\vR'}\,
\left(%
\begin{array}{cc}
  0 & 0 \\
  0 & \Sigma_{ab}(i\omega) \\
\end{array}%
\right)
\label{eq:sigma_matrix}
\eeq
In the DMFT framework, the local Green's function in the correlated subset:
\beq
G_{ab}(\tau-\tau') \equiv - \bra Tc^\dagger_a(\tau)c_b(\tau')\ket
\eeq
is represented as the Green's function of the multi-orbital impurity model:
\begin{equation}
S=
- \int^{\beta}_{0} d\tau \int^{\beta}_{0} d\tau'\, \sum_{ab}
c^{\dagger}_{a}(\tau) [{\cal G}_0^{-1}]_{ab}(\tau-\tau') c_{b}(\tau')
+\frac{1}{2}\sum_{ab}\,U_{ab}\,\int^{\beta}_{0}d\tau\,
n_{a}(\tau)n_{b}(\tau)
\label{Smulti}
\end{equation}
The Weiss function (or alternatively the dynamical mean-field, or effective
hybridisation function $\Delta_{ab}=(\iomn+\mu)\delta_{ab}-[{\cal G}_0^{-1}]_{ab}$)
is determined, as before, from the self-consistency condition requesting that the
on-site Green's function in the solid coincides with the impurity model Green's function.
The components of the Green's function of the solid in the chosen basis set read:
\beq
[G^{-1}]_{LL'}(\vk,\iomn) = (\iomn+\mu)\delta_{LL'} - h^{KS}_{LL'}
+ V^{DC}_{LL'} - \Sigma_{LL'}(\iomn)
\label{eq:gll'}
\eeq
In this expression, the self-energy matrix $\Sigma_{LL'}$ is constructed by
using the components of the
impurity self-energy $\Sigma_{ab}\equiv [{\cal G}_0^{-1}]_{ab}-[G_{imp}^{-1}]_{ab}$
into (\ref{eq:sigma_matrix}). The self-consistency condition relating implicitly
${\cal G}_0$ and $G_{imp}$ finally reads:
\beq
G(\iomn)_{ab}\,=\,\sum_{\vk}\,
\left[(\iomn+\mu)\delta_{LL'} - h^{KS}_{LL'}
+ V^{DC}_{LL'} - \Sigma_{LL'}(\iomn)\right]^{-1}_{ab}
\label{eq:scc_esc_dmft}
\eeq
Note that this involves a matrix inversion at each $\vk$-point, as well as a $\vk$-summation
over the Brillouin zone (which does not, in general, reduces to an integration over
the band density of states, in contrast to the single-band case).
Also let us emphasize that, even though the self-energy matrix has only components in the
subspace of correlated orbitals, the components of the Green's function corresponding to all
valence orbitals ($s,p,d,\cdots$) are modified due to the matrix inversion.
Correlation effects encoded in the self-energy affect the local electronic density,
which can be calculated from the full Green's function as:
\beq
\rho(\vr)\,=\, \sum_\vk\,\chi_{L\vk}(\vr)\,
G_{LL'}(\vk,\tau=0^{-})\, \chi_{L'\vk}^*(\vr)
\eeq
In a complete implementation, self-consistency over the local density
should also be reached~\cite{savrasov_kotliar_pu_nature_2001,kotliar_savrasov_spectral_short}.
The general structure of the combination of DMFT with electronic structure calculations, as well
as the iterative procedure used in practice to solve the DMFT equations, is summarised on
Fig.~\ref{fig:lda+dmft_loop}.
\begin{center}
\begin{figure}
\includegraphics[width=10 cm]{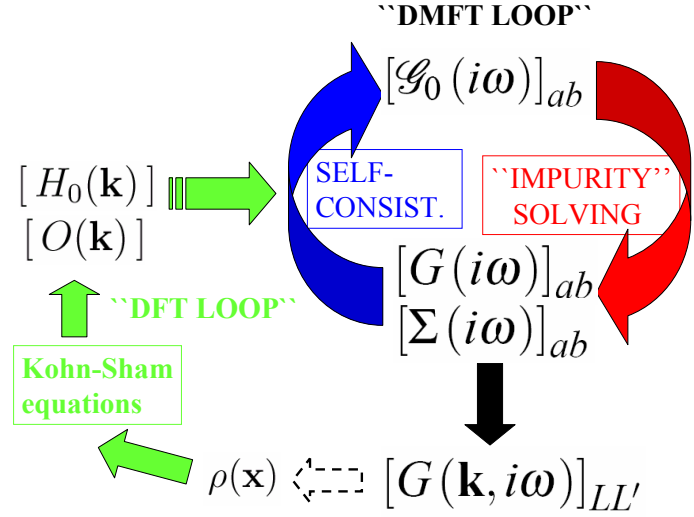}
\caption{DMFT combined with electronic structure calculations. Starting from a local electronic
density $\rho(\vr)$, the associated Kohn-Sham potential is calculated and the Kohn-Sham equations
are solved. The Kohn-Sham hamiltonian $H^{KS}_{LL'}(\vk)$ is expressed in a localised basis set
(e.g LMTOs).
A double-counting term is substracted to obtain the
one-electron hamiltonian $H_0\equiv H^{KS}-H^{DC}$. The local self-energy matrix for the subset of
correlated orbitals is obtained through the iteration of the DMFT loop: a multi-orbital impurity
model for the correlated subset is solved (red arrow), containing as an input the dynamical mean-field
(or Weiss field ${\cal G}_0$). The self-energy $\Sigma_{ab}$ is combined with $H_0$ into
the self-consistency condition Eq.~(\ref{eq:scc_esc_dmft}) in order to update the Weiss field (blue arrow).
At the end of the DMFT loop, the components of the full, $\vk$-dependent,
Green's function in the local basis set can be calculated and thus also an updated local
density $\rho(\vr)$. This is used (dashed arrow)
as a new starting density for the Kohn-Sham calculation until
a converged local density is also reached .
Alternatively, in a simplified implementation of this full scheme, the DFT-LDA calculation can
be converged first and the corresponding $H_0$ injected into the DMFT loop without attempting
to update $\rho(\vr)$.
}
\label{fig:lda+dmft_loop}
\end{figure}
\end{center}

\subsection{An application to $d^1$ oxides}
\label{sec:esc_d1oxides}

On Fig.~\ref{fig:d1oxides_dmft}, I show the spectral functions recently
obtained in Ref.~\cite{pavarini_d1oxides_prl} for
\srvo3, \cavo3, \latio3 and \ytio3.
These oxides have the same formal valence of the d-shell ($d^1$).
The single electron sits in the $t_{2g}$ multiplet, and the (empty)
$e_g$ doublet is well separated in energy.
They have a perovskite structure with perfect cubic symmetry
for the first one (Fig.~\ref{fig:perovskite})
and increasing degree of structural distortion for the three others
(corresponding mainly to the GdFeO$_3$-like tilting
of oxygen octahedra).
These calculations were performed in
a downfolded (NMTO) basis set, including the off-diagonal components of the
self-energy matrix. The latter are important for the compounds with the largest
structural distortions. For comparison, the LDA density of states are shown on
the same plot. For an independent DMFT calculation of the Ca/SrVO$_3$ compounds,
see Ref.~\cite{nekrasov_casrvo3_condmat1,sekiyama_casrvo3_2} and
Ref.~\cite{anisimov_lda+dmft_1997,nekrasov_lasrtio3_2000_epjb} for early calculations of
the doped system \lasrtio3. The spectra in Fig.~\ref{fig:d1oxides_dmft} have features
which should be familiar to the reader at this point, namely:
\begin{itemize}
\item \srvo3 and \cavo3 are correlated metals with lower ($\sim -1.5$~eV)
and upper ($\sim 2.5$~eV) Hubbard bands, as well
as a relatively moderate narrowing of the quasiparticle bandwith. The calculated
spectra compare favorably to the recent photoemission experiments of
Fig.~\ref{fig:pes_casrvo3} (see \cite{sekiyama_casrvo3_2} for a comparison).

\item \latio3 and \ytio3 are Mott insulators, with quite different values of the Mott gap
($\sim 0.3$~eV and $\sim 1$~eV, respectively) as observed experimentally. It was
emphasized in \cite{pavarini_d1oxides_prl} that the main reason for this difference is that the
orbital degeneracy of the $t_{2g}$ multiplet is lifted to a greater degree in \ytio3 than in
\latio3 due to the larger structural distortion. Indeed, reducing orbital degeneracy is known
to increase the effect of correlations (for comparable
interaction strength)~\cite{gunnarsson_orbital_1997_prb,koch_filling_1999_prb,
florens_orbital_2002_prb,manini_orbital_2002_prb}.
It was also found in Ref.~\cite{pavarini_d1oxides_prl} that both compounds develop
a very pronounced orbital polarization, of a quite different nature in each compound
(see \cite{mochizuki_orbital_rtio3_2003_prl} for a discussion of orbital
ordering in these materials and \cite{cwik_orbital_latio3_2003_prb} for a recent
experimental investigation).
\end{itemize}
This example, as well as several other recent studies, demonstrate that the embedding of DMFT
within electronic structure calculations yields a powerful quantitative tool for understanding the
rich interplay between correlation effects and material-specific aspects.
\begin{center}
\begin{figure}
\includegraphics[angle=270,width=14 cm]{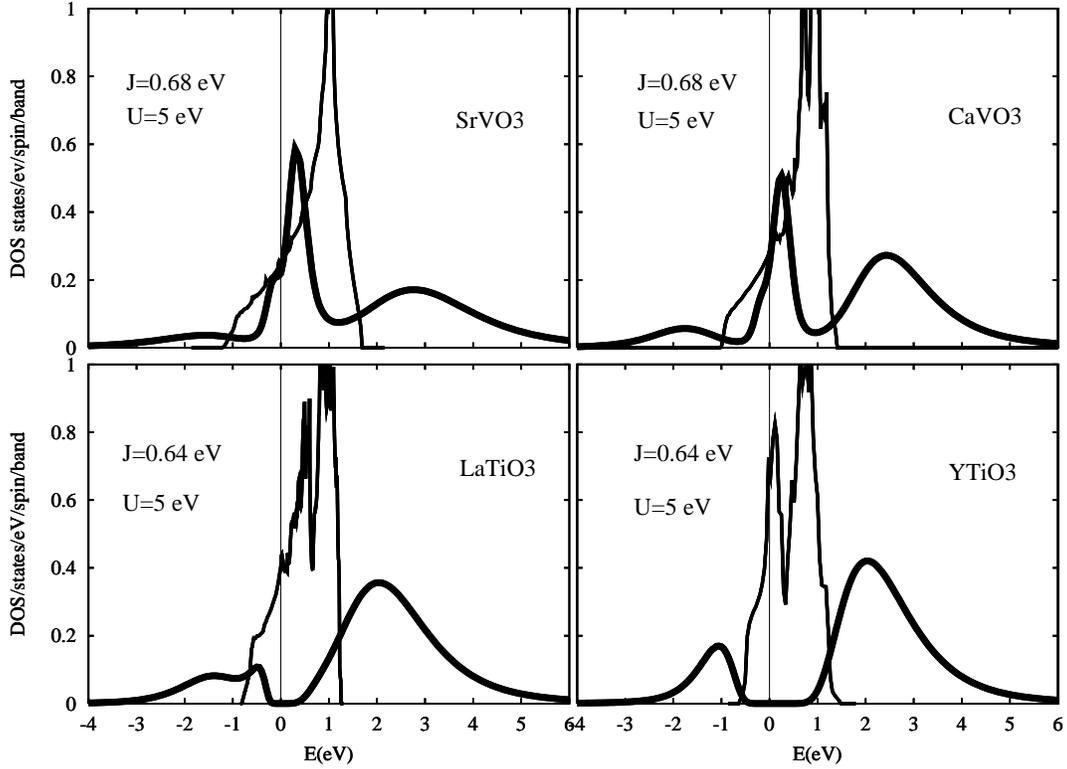}
\caption{LDA+DMFT spectral densities of the transition-metal oxides
discussed in the text, from Ref.~\cite{pavarini_d1oxides_prl}.
The (QMC) calculations were
made at $T=770$K. For comparison, the LDA d.o.s are also displayed (thin lines).
}
\label{fig:d1oxides_dmft}
\end{figure}
\end{center}

\subsection{Functionals and total- energy calculations}
\label{sec:energy_lda+dmft}

In order to discuss total energy calculations in the LDA$+$DMFT
framework\footnote{I acknowledge a collaboration with B. Amadon and
S. Biermann~\cite{amadon_energy} on the topic of this section.},
it is best to use a formulation of this scheme in terms of
a (free-) energy functional.
Kotliar and Savrasov~\cite{kotliar_savrasov_spectral_short,savrasov_kotliar_spectral_long}
have introduced for this
purpose a (``spectral-density-'') functional of both the total local electron density $\rho(\vr)$
and the on-site Green's function in the correlated subset:
$G_{ab}^{\vR\vR}$ (denoted $G_{ab}$ for simplicity in the following).
Let us emphasize that these quantities are independent, since $G_{ab}$ is
restricted to local components and to
a subset of orbitals so that $\rho(\vr)$ cannot be reconstructed from it.
The functional is
constructed by introducing (see Section.~3) source terms
$\l(\vr)=v_{KS}(\vr)-v_c(\vr)$
and $\Delta\Sigma_{ab}(\iomn)$
coupling to the operators $\psi^\dagger(\vr)\psi(\vr)$ and to
$\sum_{\vR}\chi^*_a(\vr-\vR)\psi(\vr,\tau)\psi^\dagger(\vr',\tau')\chi_b(\vr'-\vR)
=c_{a\vR}(\tau)c^\dagger_{b\vR}(\tau')$, respectively.
Furthermore, the Luttinger-Ward part of the functional is approximated by
that of the on-site
local many-body hamiltonian $H_U-H_{DC}$ introduced above.
This yields:
\begin{eqnarray}\nonumber
&\Omega[\rho(\vr),G_{ab};v_{KS}(\vr),\dS_{ab}]_{LDA+DMFT}
\,=\\ \nonumber
&-\t\ln[\iomn+\mu+\frac{1}{2}\nabla^2-v_{KS}(\vr)-\chi^*.\dS.\chi]
-\int d\vr\,(v_{KS}-v_c)\rho(\vr) -\t [G.\dS] + \\ \nonumber
&+\frac{1}{2}\int d\vr\,d\vr' \rho(\vr) U(\vr-\vr') \rho(\vr')
+ E_{xc}[\rho(\vr)]
+\sum_\vR\left(\Phi_{imp}[G^{\vR\vR}_{ab}]-\Phi_{DC}[G^{\vR\vR}_{ab}]\right)
\end{eqnarray}
In this expression, $\chi^*.\dS.\chi$ denotes the ``upfolding'' of the local
quantity $\dS$ to the whole solid:
$\chi^*.\dS.\chi=\sum_\vR\sum_{ab}\chi^*_a(\vr-\vR)\Sigma_{ab}(\iomn)\chi_b(\vr'-\vR)$.
Variations of this functional with respect to the sources
$\delta\Omega/\delta\,v_{KS}=0$ and $\delta\Omega/\delta\Sigma_{ab}=0$ yield the
standard expression of the local density and local Green's function in terms of the
full Green's function in the solid:
\beq
\rho(\vr) = \bra\vr|\hat{G}|\vr\ket\,\,\,,\,\,\,
G_{ab}(\iomn) =
\bra\chi_{a\vR}|\hat{G}|\chi_{b\vR}\ket
\label{eq:rho_and_G}
\eeq
with:
\beq
\hat{G} = \left[\iomn+\mu+\frac{1}{2}\nabla^2-v_{KS}(\vr)-\chi^*.\dS.\chi\right]^{-1}
\eeq
or, in the local basis set (see (\ref{eq:gll'})):
\beq
\hat{G} = \sum_{\vk,LL'} |\chi_{L\vk}\ket
\left[(\iomn+\mu).1 - \hat{h}^{KS}(\vk)- \Delta\hat{\Sigma}(\iomn)\right]^{-1}_{LL'}
\bra\chi_{L'\vk}|
\eeq
From these relations, the Legendre multiplier
functions $v_{KS}$ and $\dS$ could be eliminated in terms of
$\rho$ and $G_{ab}$, so that a functional
of the local observables only is obtained:
\beq
\Gamma_{LDA+DMFT}[\rho,G_{ab}]=
\Omega_{LDA+DMFT}\left[\rho(\vr),G_{ab};\l[\rho,G],\dS[\rho,G]\right]
\eeq
Extremalisation of this functional with respect to $\rho$
($\delta\Gamma/\delta\rho=0$)
and $G_{ab}$ ($\delta\Gamma/\delta\,G_{ab}=0$) yields the expression of the
Kohn-Sham potential and self-energy correction at self-consistency:
\beq
v_{KS}(\vr) = v_c(\vr) + \int d\vr' U(\vr-\vr')\rho(\vr') + \frac{\delta E_{xc}}{\delta\rho(\vr)}
\eeq
\beq
\dS_{ab} = \frac{\delta\Phi_{imp}}{\delta\,G_{ab}}-\frac{\delta\Phi_{DC}}{\delta\,G_{ab}}
\equiv \Sigma^{imp}_{ab}-V^{DC}_{ab}
\eeq
Hence, one recovers from this functional the defining equations of the LDA$+$DMFT
combined scheme, including self-consistency over the local density (\ref{eq:rho_and_G}).
Using (\ref{eq:total_functional}) and (\ref{A0DFT}), one notes that the free-energy
can be written as:
\begin{eqnarray}\nonumber
&\Omega_{LDA+DMFT} = \Omega_{DFT}+\t\ln\,G_{KS}(\vk,\iomn)^{-1}
-\t\ln\,G(\vk,\iomn)^{-1}
-\t[G_{imp}\Sigma^{imp}]+\sum_\vR\Phi_{imp}+\\
&+\t[G_{imp}V^{DC}]-\sum_\vR\Phi_{DC}
\label{eq:omega_lda+dmft_2}
\end{eqnarray}
In this expression, $\Omega_{DFT}$ is the usual density-functional theory
expression (\ref{eq:total_functional}), while $G_{KS}$ is the Green's function
corresponding to the Kohn-Sham hamiltonian, i.e without the self-energy correction:
\beq
G_{KS}^{-1}\equiv \iomn+\mu - \hat{h}_{KS}(\vk)
\eeq
A careful examination of the zero-temperature limit of
(\ref{eq:omega_lda+dmft_2})
leads to the following expression of the total energy~\cite{amadon_energy}:
\begin{eqnarray}
&E_{LDA+DMFT} \,=\, E_{DFT} -\sum'_\lambda \e^{KS}_\l+ \bra H_{KS}\ket + \bra H_U \ket - E_{DC}
\label{eq:energy_lda+dmft_1}\\
&= E_{DFT} +\sum_{\vk,LL'} h^{KS}_{LL'}
[\bra c^\dagger_{L\vk}c_{L'\vk} \ket_{DMFT} - \bra c^\dagger_{L\vk}c_{L'\vk} \ket_{KS}]
+ \bra H_U \ket - E_{DC}
\label{eq:energy_lda+dmft_2}
\end{eqnarray}
The first term, $E_{DFT}$ is the energy
found within DFT(LDA), using of course the
local density obtained at the end of the LDA$+$DMFT convergence cycle, namely:
\beq
E_{DFT} = \sum'_\lambda \e^{KS}_\l + \int d\vr [v_c(\vr)-v_{KS}(\vr)]\rho(\vr)
+ \frac{1}{2}\int d\vr d\vr' \rho(\vr) u(\vr-\vr') \rho(\vr') + E_{xc}[\rho]
\eeq
Hence the total energy within LDA$+$DMFT is made of several terms.
Importantly, it does {\it not} simply reduce to the expectation value
$\bra H\ket$ of the many-body hamiltonian (\ref{eq:mb_ham}) introduced in
the previous section.
Furthermore, $\bra H_{KS}\ket =\t[H_{KS}\hat{G}]$
must be evaluated with the full Green's function including the self-energy correction.
Therefore, this quantity does not coincide with
the sum of the (occupied) Kohn-Sham eigenvalues
$\sum'_\l\epsilon^{KS}_\l=\t H_{KS}G_{KS}$.
Eq.~(\ref{eq:energy_lda+dmft_1}) expresses that the latter has to be removed from
$E_{DFT}$, in order to correctly take into account the change of energy
coming from the Kohn-Sham orbitals. This change can also be written
$\bra H_{KS} \ket_{DMFT}-\bra H_{KS} \ket_{KS}=\t[(G-G_{KS})H_{KS}]$.
This is used in the second
expression for the energy, which emphasizes the modification of the
density matrix $\bra\,c^\dagger_{L\vk}c_{L'\vk}\ket$ by correlations.
Finally, the double-counting correction to the energy is the zero-temperature
limit of $-\bra H_{DC}\ket+\t[G\Sigma^{DC}]-\Phi_{DC}$.
The simplest form of double-counting correction (neglecting $J$ for simplicity)
corresponds to: $\Phi_{DC}[G_{ab}]=U\,N(N-1)/2$ with
$N=\sum_a n_a = \sum_a\t G_{aa}$.
Hence $V^{DC}_{ab}=\delta\Phi_{DC}/\delta\,G_{ab}=U(N-1/2)n_a\delta_{ab}$, and
$\bra H_{DC}\ket=\t[G\Sigma^{DC}]=UN(N-1/2)$ so that, finally:
$E_{DC}=U\,N(N-1)/2$.

Another formula for the total energy within LDA+DMFT has been used by
Held \etal in their investigation of the volume collapse
transition of Cerium~\cite{held_cerium_2001_prl,held_cerium_2003_prb}.

Total energy calculations within LDA$+$DMFT,
with full self-consistency on the local density
have been performed by Savrasov, Kotliar and
Abrahams~\cite{savrasov_kotliar_pu_nature_2001,kotliar_savrasov_spectral_short,
savrasov_kotliar_spectral_long}
for metallic plutonium
with fcc structure, corresponding to the $\delta$-phase.
The results are reproduced in Fig.~\ref{fig:pu_energy}, in which the
total energy is plotted as a function of the unit-cell volume
(normalised by the experimental value), for different
values of the parameter $U$. It is seen that the GGA calculation underestimates
the volume by more than $30\%$. As $U$ increases, the minimum is pushed to higher volumes,
and good agreement with experiments is reached for $U$ in the range $3.8-4\eV$.
Interestingly, in the presence of correlations, the energy curve develops
a metastable shallow minimum at a lower volume, which can be interpreted as a
manifestation of the $\alpha$-phase (which has a more complicated crystal structure
however). For the corresponding spectra, see \cite{savrasov_kotliar_spectral_long}.
In these DMFT calculations, the $\delta$-phase of plutonium is
described as a paramagnetic metal, in agreement with experiments.
In contrast, a static LDA$+$U treatment\cite{savrasov_kotliar_pu_prl_2000,bouchet_pu_lda+u_2000_jpcm}
also corrects the equilibrium volume,
but at the expense of introducing an unphysical spin
polarization\footnote{For an alternative description of the $\delta$-phase of plutonium,
in which a subset
of the f-electrons are viewed as localised while the others are itinerant,
see \cite{eriksson_pu_1999}}.
\begin{center}
\begin{figure}
\includegraphics[width=8 cm]{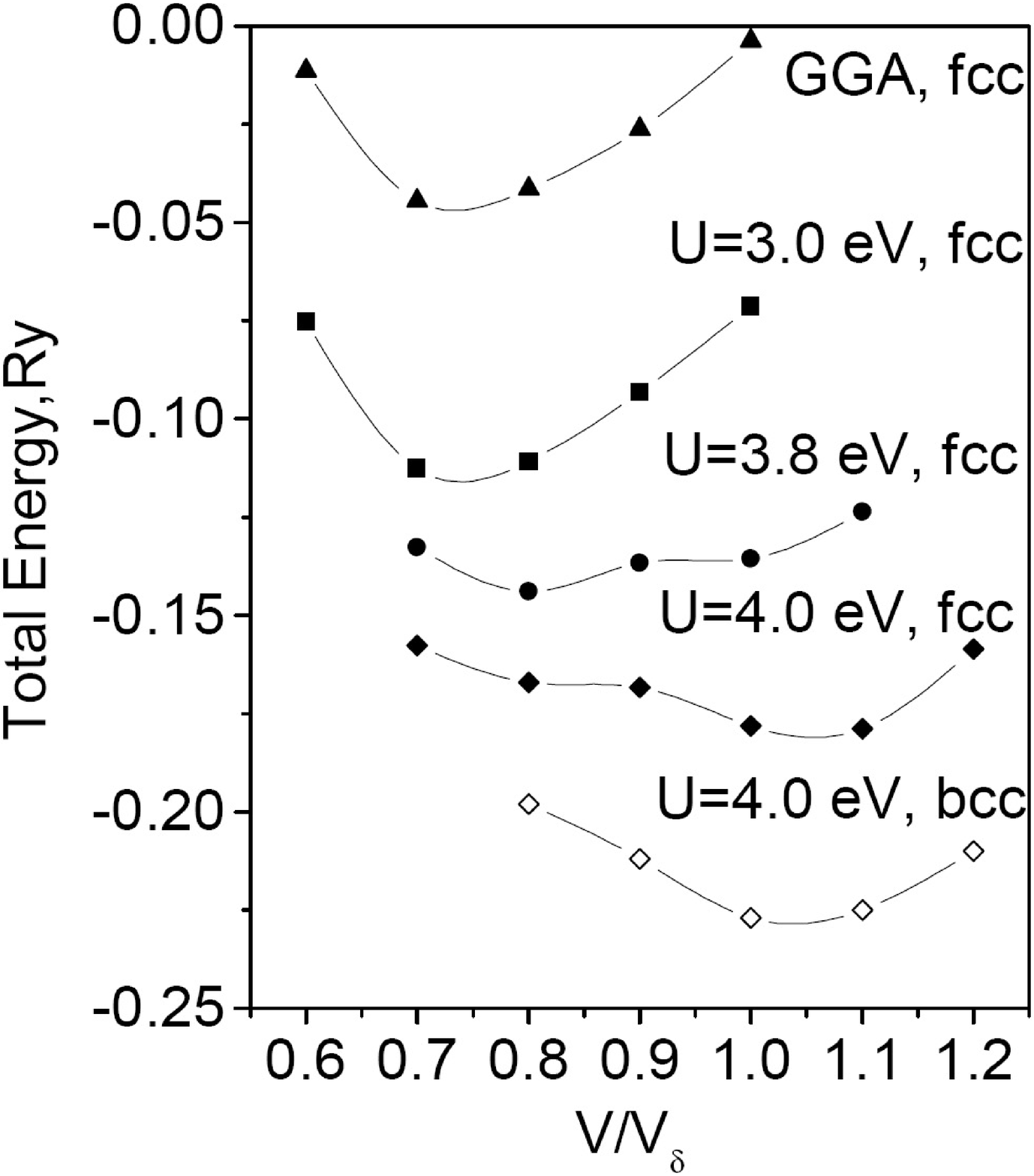}
\caption{Total energy of fcc plutonium as a function of unit-cell
volume (normalised by the experimental volume of the $\delta$-phase),
reproduced from Ref.~\cite{savrasov_kotliar_pu_nature_2001,savrasov_kotliar_spectral_long}.
The upper curve is the GGA result. Other curves are from LDA+DMFT with different values
of $U$. The lower curve is for the bcc structure.
}
\label{fig:pu_energy}
\end{figure}
\end{center}

\subsection{A life without U: towards {\it ab-initio} DMFT}
\label{sec:abinitio_dmft}

The combination of DMFT with electronic structure methods described in the previous
section introduces a matrix $U$ of local interaction parameters acting in
the subset of correlated orbitals, as in the LDA$+$U scheme. Some of these parameters
can be determined from constrained LDA calculations, or instead they can be viewed
as adjustable. Furthermore, introducing these interactions implies the need for a
``double-counting'' correction in order to remove the contribution to the total energy
already taken into account in the (orbital-independent) exchange correlation potential.
As such, this theory has great practical virtues. However, going
beyond this framework and being able to treat the electron-electron interaction entirely
from first- principles is a tempting and challenging project.
Work in this direction have appeared
recently~\cite{biermann_gwdmft_prl,kotliar_savrasov_newton,sun_gwdmft_2002_prb,sun_beyond_gw_2003,
biermann_gwdmft_baku_2004,biermann_gwdmft_ringberg_2004,ferdi_uomega_2004_condmat}.

Physically, the Hubbard interaction is associated with the screened Coulomb interaction
as seen by a given atom in the solid. {\it Screening} is essential for estimating the
order of magnitude of this parameter correctly. The naive view that $U$ is simply the
on-site matrix element of the Coulomb potential in the local basis- set would lead to values
on the scale of tens of electron-volts, while the appropriate value in the solid is
a few eV's ! This immediately points towards a key notion: that, in fact, the Hubbard
$U$ is a concept which {\it depends on the energy-scale}. At high energies (say, above
the plasma frequency in a metal), it has a very large value associated with the bare,
unscreened, matrix element, while at low energy screening takes place and it is considerably
reduced. For first-principle RPA studies of the frequency dependence of the screened local
interaction, see \cite{springer_womega_1998_prb,ferdi_uomega_2004_condmat}.

In fact, the screened effective interaction in a solid can be related, quite generally,
to the density-density correlation function. Let us start
from the first-principles hamiltonian:
\begin{eqnarray}\nonumber
&H = -\sum_i \frac{1}{2} \nabla^2_i + \sum_i v(\vr_i) +
\frac{1}{2} \sum_{i\neq j} u(\vr_i-\vr_j)\\
&= -\frac{1}{2}\int d\vr \psi^\dagger\nabla^2\psi
+ \int d\vr\, v(\vr)\hn(\vr)
+ {1\over 2} \int d\vr d\vr'\,u(\vr-\vr')\, :\hn(\vr)\hn(\vr'):
\end{eqnarray}
in which $u(\vr-\vr')=e^2/|\vr-\vr'|$ is the bare Coulomb interaction,
$\hn\equiv\psi^\dagger(\vr)\psi(\vr)$ and $:(\,):$ denotes normal ordering. The
(connected) density-density correlation function is defined as:
\beq
\chi(\vr,\vr';\tau-\tau') = \bra T\,(\hn(\vr,\tau)-\rho(\vr))\,(\hn(\vr',\tau')-\rho(\vr')\ket
\label{eq:def_density_corr}
\eeq
with $\rho(\vr)=\bra\hn(\vr)\ket$ the local density.
The screened effective interaction reads:
\beq
W(\vr,\vr',i\omega) = u(\vr-\vr') -
\int d\vr_1\,d\vr_2\, u(\vr-\vr_1)\chi(\vr_1-\vr_2;i\omega)u(\vr_2-\vr')
\label{eq:def_W}
\eeq
This can also be expressed in terms of the polarization
$P\equiv -\chi.[1-u.\chi]^{-1}$ as $W=u.[1-P.u]^{-1}$ (the dot is an
abbreviation for spatial convolutions). We emphasize that in this expression, $P$ is
the exact polarization operator, not its RPA approximation. The screened
interaction $W$ can be interpreted as the correlation function of the local scalar potential
field conjugate to $\hn(\vr)$, as can be shown from a Hubbard-Stratonovich transformation.

Armed with this precise formal definition of the screened interaction in the solid
(and, naturally, also of the full Green's function
$G(\vr,\vr';\tau-\tau')\equiv -\bra T\psi(\vr,\tau)\psi^\dagger(\vr',\tau')\ket$), we
would like to adopt now a local picture in which we focus on a given atom.
This is done,as before, by specifying a complete basis set of functions
$\chi_{L\vR}(\vr)$ localised around the atomic positions $\vR$. There is of course
some arbitrariness in this choice, as already discussed.
Adopting a local point of view, we focus on the matrix elements of the Green's function
and of the screened effective interaction {\it on a given atomic site}:
\beq
G_{ab}(i\omega) = \bra\chi_{a\vR}|G|\chi_{b\vR}\ket\,\,\,,\,\,\,
W_{a_1a_2a_3a_4}(i\omega) =
\bra\chi_{a_1\vR}\chi_{a_2\vR}|W|\chi_{a_3\vR}\chi_{a_4\vR}\ket
\eeq
In this expression, the indices $a,b,\cdots$ can run over the full set
of valence orbitals, or alternatively over a subset corresponding to the
more strongly correlated ones. This is a matter of choice of the local
quantities we decide to focus on.
Following the point of view developed in the third section of these lectures,
the key idea is again to introduce an {\it exact representation} of these local
quantities as the solution of
an atomic problem coupled to an effective bath. Because we want to represent the
local components of both $G$ and $W$, this effective problem now involves two Weiss
functions, both in the one-particle and two-particle sectors. This is an extended
form of dynamical mean-field theory (EDMFT). The action of the local problem reads:
\begin{eqnarray} \nonumber
&S  =\int d\tau d\tau^{\prime}\left[-\sum c_{a}^{+}(\tau)
{\cal G}_{ab}^{-1}(\tau-\tau^{\prime})c_{b}(\tau^{\prime})+\right.\\
&\left.+\frac{1}{2}\,\sum:c_{a_{1}}^{+}(\tau)c_{a_{2}}(\tau):\mathcal{U}%
_{a_{1}a_{2}a_{3}a_{4}}(\tau-\tau^{\prime}):c_{a_{3}}^{+}(\tau^{\prime
})c_{a_{4}}(\tau^{\prime}):\right]
\label{eq:S_edmft}
\end{eqnarray}
The local screened interaction is calculated from this effective action as:
$W_{imp}=\cU-\cU \chi_{imp} \cU$ with $\chi_{imp}$ the 2-particle impurity
correlation funcition. The two Weiss fields $\cG$ and $\cU$ are adjusted in such a
way that $G_{imp}=G_{ab}$ and $W_{imp}=W_{abcd}$, the local quantities in the solid.
The impurity model (\ref{eq:S_edmft}) can be viewed as an atom hybridised with
an effective bath of non-interacting fermions and also coupled to a bath of
fluctuating electric scalar potentials.

This construction provides an unambiguous definition of the Hubbard interactions
$\cU_{abcd}(i\omega)$ in the solid (as well as of the usual dynamical mean-field
$\cG$), assuming of course that the local components
of the screened interaction $W$ and of the Green's function $G$ are known.
Frequency- dependence of $\cU$ is essential in a proper definition of these Hubbard
interactions, at least when a wide range of energy scale is considered.
Naturally, one degree of arbitrariness remains, associated with the choice of
the basis set: $\cU$ will change when a different basis set is considered,
keeping the same form of the effective interaction $W(\vr,\vr';i\omega)$ in the full solid.

To proceed from these formal considerations to a practical scheme, we need to
decide how $W$ and $G$ will actually be calculated, and this of course will involve
approximations. Again, a free-energy functional is an excellent guidance and indeed
such a functional of the full $\fG$ and $\fW$ has been introduced by
Almbladh et al.\cite{almbladh_functionals}, generalizing the Baym-Kadanoff
construction (see also \cite{chitra_bk} for independent work). The functional
reads:
\begin{equation}
\Gamma(G,W)=Tr\ln G-Tr[(G_{H}^{-1}-G^{-1})G]-\frac{1}{2}Tr\ln W+\frac{1}%
{2}Tr[(u^{-1}-W^{-1})W]+\Psi\lbrack G,W] \label{LW}%
\end{equation}
$G_{H}^{-1}=i\omega_{n}+\mu+\nabla^{2}/2-v_{H}$ corresponds to the Hartree
Green's function with $v_{H}$ being the Hartree potential.
For a derivation of (\ref{LW}) using a Hubbard-Stratonovich transformation
and a Legendre transformation with respect to both $G$ and $W$,
see~\cite{chitra_bk}.
The functional
$\Psi\lbrack G,W]$ is a generalization of the Luttinger-Ward functional
$\Phi\lbrack G]$, whose derivative with respect to $G$ gives the self-energy.
Here we have, similarly (from $\delta\Gamma/\delta\,G=\delta\Gamma/\delta\,W=0$):
\begin{equation}
G^{-1}=G_{H}^{-1}-\Sigma^{xc}\,\,\,,\,\,\,
\Sigma^{xc}=\frac{\delta\Psi}{\delta G}\,\,\,;\,\,\,
W^{-1}=u^{-1}-P\,\,\,,\,\,\,P=-2\frac
{\delta\Psi}{\delta W}
\label{dGamma}%
\end{equation}
A well established electronic structure calculation method, which offers in part an
alternative to DFT-LDA, is the so-called GW approach~\cite{hedin_gw}
(see \cite{ferdi_gw_review} for a review). This corresponds to the following
approximation to the $\Psi$-functional:
\beq
\Psi_{GWA}\,=\,
-\,\frac{1}{2}\,\int d\vr d\vr' \int d\tau d\tau' \,
G(\vr,\vr',\tau-\tau')W(\vr,\vr',\tau-\tau')G(\vr',\vr,\tau'-\tau)
\label{eq:psi_gwa}
\eeq
which yields the RPA-like approximation to the polarisation and exchange-correlation
self-energy:
$P=G\star\,G$ and $\Sigma^{xc}=-G\star\,W$. The GW approximation to the
$\Psi$-functional is easily written in terms of the components of $G$ and $W$ in the
chosen basis set:
\beq
\Psi_{GWA}=
-\frac{1}{2}\int d\tau \sum_{L_1\cdots L_2'}
\sum_{\vR\vR'}G^{\vR\vR'}_{L_1L'_1}(\tau)
W^{\vR\vR'}_{L_1L_2L'_1L'_2}(\tau)
G^{\vR'\vR}_{L_2'L_2}(-\tau)
\label{eq:psi_gwa_comp}
\eeq
This can be separated into a contribution
$\Psi_{GWA}^{non-loc}$
from non-local components (corresponding to
the terms with $\vR\neq\vR'$ in (\ref{eq:psi_gwa_comp})) and a contribution
$\Psi_{GWA}^{loc}[G^{\vR\vR},W^{\vR\vR}]$ from local components only ($\vR=\vR'$).

The GW approximation does treat the screened Coulomb interaction from first-principles,
but does not treat successfully strong correlation effects.
Recently, it has been suggested to improve on the GWA for the local contributions
by using the DMFT framework~\cite{biermann_gwdmft_prl,sun_gwdmft_2002_prb}
(see also \cite{kotliar_savrasov_newton,sun_beyond_gw_2003,
biermann_gwdmft_baku_2004,biermann_gwdmft_ringberg_2004}).
One can think of different approximations to the $\Psi$-functional in this
context, depending on whether the DMFT approach is used for all the valence
orbitals $L=s,p,d,\cdots$, or for a subset
(corresponding to the index $a,b,...$)
of correlated orbitals only. The corresponding $\Psi$-functional
reads:
\beq
\Psi_{GW+DMFT}[G^{\vR\vR'}_{L_1L'_1},W^{\vR\vR'}_{L_1L_2L'_1L'_2}]\,=\,
\Psi_{GWA}^{non-loc} + [\Psi_{GWA}^{loc}-\Delta\Psi] +
\sum_\vR\Psi_{imp}[G^{\vR\vR}_{ab},W^{\vR\vR}_{abcd}]
\label{eq:psi_gw_dmft}
\eeq
In this expression, $\Psi_{imp}$ is the $\Psi$-functional corresponding to the
local effective model (\ref{eq:S_edmft}), while $\Delta\Psi$ removes the
components from $\Psi_{GWA}^{loc}$ which will be taken
into account in $\Psi_{imp}$, namely:
\beq
\Delta\Psi = -\frac{1}{2}\sum_\vR\int d\tau
\sum_{abcd} G^{\vR}_{ab}(\tau)
W^{\vR\vR}_{abcd}(\tau)
G^{\vR\vR}_{DC}(-\tau)
\label{eq:deltapsi}
\eeq
If all valence orbitals are included in the DMFT treatment, the second term in the
r.h.s of (\ref{eq:psi_gw_dmft}) is absent altogether.
If only a correlated
subset is treated with DMFT, $\Delta\Psi$ can be thought of as a term preventing double-counting
of interactions in the correlated subset.
In this context however, in contrast to LDA$+$DMFT, the form of this double-counting
correction is known explicitly.

Taking derivatives of this functional with respect to the components of $G$ and $W$,
one sees that, in the GW$+$DMFT approach, the non-local components of the self-energy
and of the polarization operator keep the same form as in the GWA, while the local
components are replaced by the ones from the effective impurity model (possibly
in the correlated subset only). The GW$+$DMFT theoretical framework is
fully defined by (\ref{eq:psi_gw_dmft}) and the form of the impurity model
(\ref{eq:S_edmft}). As before, an interative self-consistent process
must be followed in order to obtain the self-energy and screened effective
interaction, as well as the dynamical mean-field $\cG$ and effective Hubbard
interactions $\cU$. This is described in more details
in Refs.~\cite{biermann_gwdmft_prl,biermann_gwdmft_baku_2004,biermann_gwdmft_ringberg_2004}.
Concrete implementations of this scheme to electronic structure (and to model
hamiltonians as well) is currently being pursued by several groups.
For early results, see \cite{biermann_gwdmft_prl,sun_gwdmft_2002_prb,
sun_beyond_gw_2003,
biermann_gwdmft_baku_2004,biermann_gwdmft_ringberg_2004,ferdi_uomega_2004_condmat}.
\section{Conclusion and perspectives}
\label{sec:concl}

In these lectures notes, I have tried to give an introduction to
some aspects of the physics of strong electron correlations in
solids. Naturally, only a limited number of topics could be
covered. The field is characterized by a fascinating diversity of
material-dependent properties. It is, to a large extent,
experimentally driven, and new discoveries are undoubtedly yet to
come. Also, new territories outside the traditional boundaries of
solid-state physics are currently being explored, such as
correlation effects in nano-electronic devices or the condensed
matter physics of cold atoms in optical lattices.

On the theory side, these lectures are influenced by the author's
prejudice that (i) physics on intermediate energy scale matters
and may be a key to the unusual behaviour of many strongly
correlated materials and that (ii) quantitative theoretical
techniques are essential to the development of the field, in combination
with phenomenological considerations and experimental investigations.

Dynamical mean-field theory is a method of choice for treating these
intermediate energy scales. The basic principles of this approach have been
reviewed in these lectures. On the formal side, analogies with classical
mean-field theory and density-functional theory have been emphasized, through
the construction of free-energy functionals of local observables.
A distinctive aspect of DMFT is that it treats quasi-particle excitations
and higher energy incoherent excitations, on equal footing. As a result, it is
able to describe transfers of spectra weight between quasiparticle and incoherent features as
temperature, coupling strength, or some other external parameter (doping, pressure,...)
is varied. I have emphasized that the {\it quasiparticle coherence scale}
plays a key role in the physics of a strongly correlated metal.
Above this scale, which can be dramatically reduced by correlations,
unusual (non-Drude) transport and spectroscopic properties are observed,
corresponding to an incoherent metallic regime.
This is the case, in particular, for metals which are close to a Mott insulating
phase. I have briefly reviewed the DMFT description of these effects in these
lectures, in comparison to experiments, as well as the detailed
theory of the Mott transition which has been one of the early successes
of this approach. I have also provided an (admittedly quite succinct)
introduction to the recent combination of DMFT with electronic structure
calculations. These developments have been made possible by researchers
from two communities joining forces towards a common goal. It provides
us with a powerful quantitative tool for investigating material-dependent aspects
of strong electron correlations.

Despite these successes, some key open questions in the physics of strongly
correlated electron systems remain out of reach of the simplest
version of DMFT. Indeed, in materials like cuprates, short-range spatial
correlations play a key role (in particular magnetic correlations due to superexchange,
leading to a strong tendency towards the formation of singlet bonds, as well
as pair correlations). These
correlations deeply affect the nature of quasiparticles.
There is compelling experimental evidence that the quasiparticle coherence
scale has thus a strong variation as the momentum $\vk$ is varied along the Fermi
surface, leading to the formation of ``cold spots'' and ``hot regions''.
Extending the DMFT framework in order to take these effects into account may well be
the most important frontier in the field.

\begin{theacknowledgments}

The content of these lecture notes has been greatly influenced by all
the colleagues with whom I recently collaborated in this field, both theorists
and experimentalists:
B. Amadon, O.K. Andersen,
F. Aryasetiawan,
S. Biermann,
S. Burdin,
T.A. Costi,
L. de Medici,
S. Florens,
T. Giamarchi,
M. Grioni,
S.R. Hassan,
M. Imada,
D. J\'erome,
G. Kotliar,
H.R. Krishnamurthy,
F. Lechermann,
P. Limelette,
A. Lichtenstein,
S. Pankov,
O. Parcollet,
C. Pasquier,
E. Pavarini,
L. Perfetti,
A. Poteryaev,
M. Rozenberg,
S. Sachdev,
R. Siddharthan,
P. Wzietek,
as well as by all other colleagues with whom I have had profitable discussions
over the last few years.
I would like to thank particularly the members of the Ecole Polytechnique group for the
friendly and stimulating atmosphere, and for the lively daily discussions.
I am grateful to F.~Mila for the invitation to lecture on this subject
in the ``Troisi\`eme cycle de la Suisse Romande'' in may, 2002,
to C.~Berthier, G.~Collin, C.~Simon and the other organizers of the school on
``Oxydes \`a propri\'et\'es remarquables'' (GDR 2069) at
Aussois in june, 2002~\cite{gdr_ox_school_2002},
to W.Temmermann and D.Szotek for organizing lectures in Daresbury in june, 2003,
to F.Mancini and A. Avella for the organisation of the training course at Vietri
in october, 2003, and to C.~Ortiz and A.~V\'anyolos for help with the notes of
my lectures at Vietri.
Hospitality of the KITP (Santa Barbara) and of ICTP (Trieste) is acknowledged.
Support for research has been
provided by CNRS, Ecole Polytechnique, the European Union (through the Marie Curie
and RTN programs), and the Indo-French program of IFCPAR.
                                                                         
\end{theacknowledgments}

\end{document}